\documentclass[fleqn,usenatbib]{mnras}

\usepackage[T1]{fontenc}
\usepackage{ae,aecompl}

\usepackage{graphicx,color}	
\usepackage{amsmath}	
\usepackage{amssymb}	

\newcommand\mj{{\,{\rm M}_{\rm J}}}
\newcommand\msun{{\,{\rm M}_{\odot}}}
\newcommand\Lsun{{\,{\rm L}_{\odot}}}
\newcommand\simlt{\la}
\newcommand\simgt{\ga}

\newcommand\p{\partial}

\title[Dust accretion onto planets]{Changes in the metallicity of gas giant planets due to pebble accretion}

\author[]{R. J. Humphries\thanks{rjh73@le.ac.uk}  \& S. Nayakshin
\\
Department of Physics and Astronomy, University of
  Leicester, Leicester LE1 7RH, UK.
}

\date{Accepted XXX. Received YYY; in original form ZZZ}

\pubyear{2018}

\begin{document}
\label{firstpage}
\pagerange{\pageref{firstpage}--\pageref{lastpage}}
\maketitle

\begin{abstract}
We run numerical simulations to study the accretion of gas and dust grains onto gas giant planets embedded into massive protoplanetary discs. The outcome is found to depend on the disc cooling rate, planet mass, grain size and irradiative feedback from the planet. If radiative cooling is efficient, planets accrete both gas and pebbles rapidly, open a gap and usually become massive brown dwarfs. In the inefficient cooling case, gas is too hot to accrete onto the planet but pebble accretion continues and the planets migrate inward rapidly. Radiative feedback from the planet tends to suppress gas accretion. Our simulations predict that metal enrichment of planets by dust grain accretion inversely correlates with the final planet mass, in accordance with the observed trend in the inferred bulk composition of Solar System and exosolar giant planets. 
To account for observations, however, as much as $\sim 30-50$\% of the dust mass should be in the form of large grains.

\end{abstract}


\section{Introduction}

The formation mechanisms of gas giant planets can be constrained by studying their metallicities \citep[e.g.,][]{Guillot05,MillerFortney11,HelledEtalPP62014}.
Gravitational instability (GI) \citep{Kuiper51b,Boss97} is an alternative planet formation mechanism to the classical Core Accretion (CA) \citep{PollackEtal96} theory. In GI, the outer regions of gravitationally unstable protoplanetary discs may fragment  into Jupiter mass and larger clumps that were previously believed to share the composition of their parent discs \citep[e.g.,][]{Boss97}. These clumps may then undergo further collapse and cooling and evolve into gas giant planets.
More recent work has shown that the initial metallicities of these planets may be enhanced by large grains (decimetres to metres) that are efficiently concentrated into spiral arms \citep{ForganRice11} from within which the fragments are born \citep{BoleyDurisen10}. Planetesimal accretion can also alter the composition of such planets \citep{BoleyEtal11a}.

In this paper we study pebble accretion onto gas giant planets embedded in massive protoplanetary discs at large ($\sim$ 100 AU) separations from the host star. Pebbles are defined as grains with radius $a \sim 0.1 - 10 $ cm. The stopping time of such grains due to aerodynamic friction \citep{Weidenschilling77} is comparable to the local dynamical time in the massive discs necessary for GI.  Pebbles are dynamically decoupled to a certain degree from the gaseous disc \citep{OrmelKlahr10,JohansenLacerda10} and may therefore accrete onto embedded massive objects such as planetesimals and planets \citep{LambrechtsJ12,LJ14} even whilst gas accretion is suppressed. 

The planetary mass clumps can be born by gravitational instability only at large radii \citep[$R \simgt 50$ AU][]{Rafikov05,Clarke09}, however observations show that the occurence rate of planets at these radii is low, at around 1-5\% \citep{ViganEtal17}.
Detailed simulations find that clumps born in the outer disc will migrate into the inner disc in $\sim 10^3 - 10^4$ years \citep[e.g.,][]{BoleyEtal10,MachidaEtal10,BaruteauEtal11} and even migrate closer to the host star \citep[e.g.,][]{GalvagniEtal12,NayakshinFletcher15}. This may explain why it is rare to observe these planets at large radii. 

These simulations imply that GI planets may end up at separations $\sim 0.1-10$ AU due to interactions with their gas disc. Although less robustly, planet-planet and planet-secondary star scatterings may also put a fraction of wide separation planets into the inner disc \citep{RiceEtal15}. There is arguably direct evidence that such large scale migration does take place for both massive planets and brown dwarfs. It has been recently shown that, unlike hot Jupiters, planets more massive than $\sim 4$~Jupiter masses \citep{Nayakshin_Review,SantosEtal17} and brown dwarfs \citep{TroupEtal16} observed to orbit their primary stars in the inner $\simlt 1$~AU do not correlate with host metallicity, similar to the stellar mass companions of these stars. 
It is therefore important to consider to what degree pebble accretion onto migrating gas giants at tens of AU may affect the dust distribution in the disc and contribute to the seemingly metal rich compositions of hot and warm Jupiters \citep[e.g.,][]{MillerFortney11}.

To constrain the range of possible outcomes, we try to sample the important parameter space of the problem. Naturally, this includes the pebble size and the planet mass. Additionally, even though we are focused on pebble not gas accretion onto the planet, the latter process is crucial in determining the fate of the forming planet \citep[e.g.,][]{ZhuEtal12a,KratterL16}, and therefore we must also consider the physics controlling gas accretion. 

\cite{Gammie01} showed that the rate of radiative cooling is key to determining whether self-gravitating gas fragments grow in mass by accreting gas from the disc quickly. 
This result was explored by one of us in recent 3D SPH simulations \citep{Nayakshin17a}.
The rate of radiative cooling is a crucial parameter since gas entering the planet domain (its Hill sphere) from the disc has a net positive energy with respect to the planet. To become gravitationally bound to the planet, the gas needs to radiate this excess energy away. For a Keplerian shear flow, the time spent by the gas within the Hill sphere is of the order  of the dynamical time, $\Omega_K^{-1}$, where $\Omega_K$ is the local angular frequency. If the gas cooling time is expressed as  $t_{\rm rad} = \beta \Omega_K^{-1}$, where $\beta > 0$ is a dimensionless parameter, then for $\beta \ll 1$,  radiative cooling is efficient. The excess energy is lost by the gas rapidly and the planet enters a phase of runaway accretion during which it typically enters the brown dwarf regime.

In the opposite limit, when $\beta \gg 1$,  gas entering the Hill sphere is unable to lose its energy rapidly enough. It simply remains too hot to be bound to the planet and hence exits the Hill sphere before it can be accreted\footnote{In the classical Core Accretion scenario \citep{PollackEtal96}, growing embryos migrate through the parent disc slowly. Thus they have a considerable amount of time for their envelopes to cool down and contract. Planets created by gravitational instability migrate inward on the time scale of a few orbits. If they do not accrete gas on time scales approaching dynamical then gas accretion is negligible and they evolve at nearly constant gas mass.}. In this situation the planets do not grow in mass quickly and instead migrate inward rapidly in the Type I regime \citep[e.g.,][]{BaruteauEtal11,MichaelEtal11}. Fragment evolution in this case is drastically different. Instead of evolving into a brown dwarf they may be torn apart by tidal forces of the host star if they migrate too close \citep[e.g.,][]{VB06,BoleyEtal10,Nayakshin10c,MachidaEtal10}; if they collapse by hydrogen molecule dissociation \citep{Bodenheimer74} then they remain planetary mass and may migrate much closer to the host star.
In this paper we therefore sample the two opposite limits for gas accretion by setting $\beta = 0.1$  or $\beta=10$ for most of the simulations presented below, although we also explore an intermediate case, $\beta=1$ and radiative feedback from the planet. 

Note that although protoplanetary discs are expected to fragment at $\beta \sim 3-6$ \citep{Rice05}, it would seem that these are the most interesting values of the dimensionless cooling time to consider. However, there is a debate about the appropriate value for the critical fragmentation $\beta$ \citep{MeruBate11a}. In addition, $\beta$ in realistic discs is a strongly decreasing function of radius \citep{Clarke09}. Due to planet migration or planet-planet scatterings, a planet can find itself far from the region where it was born, where $\beta$ could be either considerably larger or smaller than the critical fragmentation value. For these reasons we choose to explore a wide range of the dimensionless cooling parameter.

In addition to this, there is a non trivial question of how pebbles are distributed in the disc prior to introduction of the planet. As shown by \cite{RiceEtal04,BoleyDurisen10}, large grains can collect into spiral arms very effectively, resulting in a highly non uniform dust-to-gas ratio distribution throughout the protoplanetary disc. Here we wish to avoid these modelling uncertainties for the sake of first understanding the basics of pebble accretion onto massive planets. To this end we choose to work with massive discs, $M_{\rm d} =0.2\msun$, which are however passively irradiated by their host stars and have \cite{Toomre64} parameter $Q$ somewhat larger than unity so that no strong spiral density waves form in the disc. The even more interesting problem of pebble accretion in a strongly self-gravitating disc is thus left for a near future work pending understanding of the present simpler situation.

\section{Analytical expectations}\label{sec:analytics}

Consider a planet migrating in the Type I regime, e.g., when the planet Hill radius satisfies the constraint

\begin{equation}
R_{\rm H} = R \left(\dfrac{M_{\rm p}}{3 M_*}\right)^{1/3} \le H\;.
\label{type1}
\end{equation}

The \textit{maximum} gas accretion rate onto the planet can be estimated from

\begin{equation}
\dot M_{\rm acc} = 2\pi \Sigma \Omega_{\rm K} R_{\rm cap}^2\;,
\label{dotm_gas0}
\end{equation}

where $\Sigma$ is the disc surface density and $R_{\rm cap} = (2/3) (G M_{\rm p}/c_s^2)$. Note that Equation \ref{dotm_gas0} is exactly equivalent to Equation 3 in \cite{Nayakshin17a} and also assumes that $R_{\rm cap} \le R_{\rm H}$. For the pebble accretion rate, the maximum is given by 

\begin{equation}
\dot M_{\rm peb} = 2 R_{\rm H}^2 \Omega_{\rm K} \Sigma_{\rm peb} \;,
\label{dotm_peb0}
\end{equation}

where $\Sigma_{\rm peb}$ is the pebble surface density, which we express as

\begin{equation}
\Sigma_{\rm peb} = f_{\rm peb} Z_0 \Sigma\;,
\label{fpeb}
\end{equation}

where $Z_0$ is the mass fraction of metals in the gas disc ($Z_0 = 0.02$ for a Solar metallicity disc), and $0 \le f_{\rm peb} < 1 $ is the fraction of the metal mass in pebbles. 

It is then convenient to define the doubling time scales for the planet gas and metal mass, assuming that the initial metallicity of the planet is equal to that of the parent disc:

\begin{equation}
t_{\rm acc} = \dfrac{M_{\rm p}}{\dot M_{\rm acc}} \qquad \text{and} \qquad t_{\rm Z} = \dfrac{Z_0 M_{\rm p}}{\dot M_{\rm peb} }\;.
\label{times0}
\end{equation}

The planet will not be in the Type I migration regime indefinitely. Since the disc is geometrically thinner in the inner regions, the gap opening planet mass is always an increasing function of radial separation \citep[e.g.][]{MalikEtal15}. Usually, the planet will open a gap when reaching separation of a few to a few tens of AU, at which point accretion rates of both gas and pebbles onto the planet drop very strongly. Therefore, the planet migration time scale is a rough measure of how long the planet can continue accreting pebbles and gas. The Type I migration time scale is approximately

\begin{equation}
t_{\rm mig} \sim \dfrac{1}{ \Omega_{\rm K}} \dfrac{M_*^2}{M_{\rm p} M_{\rm d}} \left(\dfrac{H}{R}\right)^2\;.
\label{tI_approx}
\end{equation}

A more detailed analytical estimate for $t_{\rm mig}$ is given in Equation \ref{eq:tanakamig} \citep{Tanaka02}.

Now, via simple algebra one can show that

\begin{equation}
\dfrac{t_{\rm acc}}{t_{\rm Z}} \sim \dfrac{f_{\rm peb}}{\pi} \left(\dfrac{H}{R}\right)^4 \left(\dfrac{M_*}{M_{\rm p}}\right)^{4/3}
\label{gas_to_peb}
\end{equation}

and

\begin{equation}
\dfrac{t_{\rm mig}}{t_{\rm Z}} \sim \dfrac{f_{\rm peb}}{\pi} \left(\dfrac{H}{R}\right)^2 \left(\dfrac{M_*}{M_{\rm p}}\right)^{4/3}\;.
\label{mig_to_peb}
\end{equation}

These two relations of course also imply that

\begin{equation}
\dfrac{t_{\rm acc}}{t_{\rm mig}} \sim \left(\dfrac{H}{R}\right)^2 \ll 1\;.
\label{acc_to_mig}
\end{equation}

We now see that if gas accretion onto the clump is efficient, that is, proceeds at around the estimate given by Equation \ref{dotm_gas0}, then 
gas accretion is the most rapid process. The clump will runaway in mass, becoming a massive brown dwarf or even a low mass star, at an approximately fixed separation (This is the origin of the nearly vertical tracks in the clump mass-separation plane seen in Figure 15 in \cite{Nayakshin17a}\footnote{note that the logarithmic-linear scales on the quoted figure somewhat hide the nearly vertical motion of the planets in those coordinates.}). Accretion of pebbles in this case is a secondary, usually negligible, effect. Assuming that $f_{\rm peb}\ll 1$, the bulk metallicity of massive objects built by an effective gas accretion from the disc should hence be expected to be close to that of the parent disc. Indeed, microscopic grains in the disc, carrying most of the disc metallic mass, will be accreted together with the gas, keeping the object metal abundance close to $Z_0$.

If gas inside the Hill sphere is however unable to cool rapidly, and/or is heated by radiative feedback from the planet, then gas accretion onto the clump is suppressed strongly compared with the estimate given in Equation \ref{dotm_gas0} \citep[see][]{NayakshinCha13,Stamatellos15,Nayakshin17a}. In this situation we have

\begin{equation}
t_{\rm acc} \gg t_{\rm Z} \qquad \text{and} \qquad t_{\rm acc} \gg t_{\rm mig}\;,
\end{equation}

so that we can neglect accretion of gas onto the clump. In this case we obtain the nearly horizontal tracks in the planet mass-separation diagram \citep[see Figure 15 in][again]{Nayakshin17a}: the clump migrates inward very rapidly at nearly a constant mass. From Equation \ref{mig_to_peb} we deduce that

\begin{equation}
\dfrac{t_{\rm mig}}{t_{\rm Z}} \propto \left(\dfrac{M_*}{M_{\rm p}}\right)^{4/3}\;,
\label{peb_dom}
\end{equation}

implying that pebble accretion is the more efficient the less massive the planet is. In particular, if metal content of planets is dominated by the pebbles accreted during the Type I migration phase, then Equation \ref{peb_dom} suggests that bulk metallicity of gas giant planets will scale with planet mass as $Z\propto M_{\rm p}^{-4/3}$. This prediction is qualitatively correct but is much steeper than the  $Z \propto M_{\rm p}^{-0.45}$ trend for exoplanets deduced by modelling of the planet mass-radius correlations for giants \citep{ThorngrenEtal15}. We shall  discuss in Section \ref{sec:discussion} many caveats in applying Equation \ref{peb_dom} to the final bulk composition of planets.

Physically, the scaling in Equation \ref{peb_dom} comes about because lower mass planets take longer to migrate into the inner disc ($t_{\rm mig} \sim M_{\rm p}^{-1}$), yet their specific pebble accretion rate (that is, per planet mass) is actually higher \citep[see results by][which are relevant to the problem at hand too]{LambrechtsJ12}. In other words, lower mass planets accrete pebbles more rapidly and also have more time to accrete them before they are plunged into the inner disc where they open a gap in the disc and stop accreting pebbles. This trend is further amplified by the fact that more massive planets open gaps at larger radii than the less massive ones \citep[this can be glimpsed from the purple curves in Figure 15 of][]{Nayakshin17a}. 

Finally, even in the `grey area' where gas accretion is not completely suppressed, i.e., $t_{\rm acc} \sim t_{\rm Z}$ or $t_{\rm acc} \sim t_{\rm mig}$, we see that $t_{\rm acc}/t_{\rm Z} \propto (M_*/M_{\rm p})^{4/3}$, so that the less massive the planet is, the more likely pebble accretion is to be important.

\section{Numerical methods}\label{sec:num}
\subsection{General}\label{sec:general}
Our numerical methods build on earlier work by \cite{ChaNayakshin11a}, \cite{NayakshinCha13} and \cite{Nayakshin17a}. The gas disc with embedded dust particles is modelled via the coupled hydrodynamical and N-body code \textsc{gadget-3} \citep[see][]{Springel05}. The code uses the lagrangian Smoothed Particle Hydrodynamics (SPH)  algorithm \citep{Gingold77,Lucy77,Monaghan92,Springel10} to describe dynamics of gas and is well suited for irregular gas geometries and self-gravitating systems. Dust grain particles are treated as a set of individual particles interacting with gas/SPH particles (see below). Gravitational force from all the components in the system is calculated using an N-body tree algorithm. 

The star is a sink particle with an initial mass of $M_* = 1\msun$ and an accretion (sink) radius, $R_{\rm sink}$, set to about 1 AU (the exact value varies slightly between the simulations as described below). Any gas or dust particle that is separated from the star by a distance $R$ less than $R_{\rm sink}$ is accreted, with its mass and momentum added to that of the star. The disc is irradiated by the central star, so that the equilibrium gas temperature ($T_{eq}$) is given by the function

\begin{equation}
T_{\rm eq} = 20 \;\hbox{K } \left(\frac{100 \hbox{ AU}}{R}\right)^{1/2}\;.
\label{Teq0}
\end{equation}

This temperature floor prevents the fragmentation of low $\beta$ discs and allows us to easily explore the rapid cooling regime without worrying about additional effects due to fragmentation. An ideal equation of state  with adiabatic index $\gamma = 5/3$ is used. The radiative cooling prescription is the simple $\beta$-cooling law \citep{Gammie01} and the gas specific energy $u$ is evolved according to

\begin{equation}
\dfrac{\mathrm{d}u}{\mathrm{d}t} = - \dfrac{u - u_{\rm eq}}{ t_{\rm cool}}\;,
\label{beta0}
\end{equation}

where $u_{\rm eq} = k_B T_{\rm eq}/(\mu (\gamma-1))$, with $k_B$ and $\mu = 2.45 m_p$ being the Boltzmann constant and the mean molecular weight for gas of Solar composition, respectively. The radiative cooling time $t_{\rm cool}$  is a function of radius $R$ only:

\begin{equation}
t_{\rm cool}(R) = \beta \Omega_K^{-1}\;,
\label{beta_def}
\end{equation}

where $\beta$ is a positive constant and $\Omega_K = (GM_*/R^3)^{1/2}$ is the local Keplerian angular frequency at $R$. 

\subsection{Gas-grain interactions and grain dynamics}\label{sec:grain_dyn}
A dust grain particle equation of motion is given by

\begin{equation}
\dfrac{\mathrm{d} \mathbf{v_a}}{\mathrm{d}t} = \mathbf{g} - \dfrac{\mathbf{v_a - v_g}}{ t_{\rm st}}\;,
\label{dust_v1}
\end{equation}

where ${\bf g}$ is the total gravitational force on the grain particle,  ${\mathbf v_a}$ and $\mathbf v_g$ are the grain and the gas velocities, respectively, and $t_{\rm st}$ is the grain stopping time defined by 

\begin{equation}
t_{\rm st} = \dfrac{m_a \Delta v_a }{ F_a} \dfrac{\rho_{\rm gas} + \rho_{\rm p} }{ \rho_{\rm gas}}\;,
\label{tstop1}
\end{equation}

where $m_a = (4\pi/3) \rho_a a^3$ is the dust particle mass, $\Delta v_a = |{\bf v_a - v_g}|$, and $F_a$ is the aerodynamical friction force \citep{Epstein24}. Here $\rho_{\rm gas}$ and $\rho_{\rm p}$ are the gas and particle volume densities, with $\rho_{\rm p} \ll \rho_{\rm gas}$, usually. The local gas density $\rho_{\rm gas}$ (at the grain particle location) is calculated with the usual SPH formalism \citep[e.g.,][]{Price12}, with the number of SPH particle neighbours of the dust particle fixed at 40.

The dust particle {\it material} density is set in this paper to $\rho_a = 5$ g cm$^{-3}$. This is relatively high, being appropriate for a material composed of a mix of silicates and Fe. However, grain dynamics in our paper depends almost exclusively on the product  $\rho_a a$ only, since this quantity enters in the grain stopping time in the Epstein regime,

\begin{equation}
t_{\rm Eps} = \dfrac{\rho_a a}{ \rho_{\rm gas} v_{\rm th}}\;,
\label{tEps0}
\end{equation}

where $v_{\rm th} = (8/\pi)^{1/2} c_s$ and $c_s^2 = k_B T/\mu$ are the typical gas thermal and the isothermal sound speeds for the gas, respectively. The dimensionless grain stopping time $\tau$ \citep[e.g.,][]{Weiden77,Armitage10} is defined as

\begin{equation}
\tau = t_{\rm st} \Omega_{\rm k}(R),
\label{eq:tau}
\end{equation}

and is equivalent to the Stokes number for grains suspended in the gas disc. The Epstein regime is appropriate for grains smaller than about the H$_2$ molecule mean free path, i.e.,

\begin{equation}
a < \dfrac{9}{4} \lambda = \dfrac{1 }{ n\, \sigma_{H2}} \approx 10^5 \hbox{ cm } \rho_{13}^{-1}\;,
\label{eq:eps}
\end{equation}

where $n$ and $\sigma_{H2}$ and the molecule density and the interaction cross section, and $\rho_{13} = 10^{13} \hbox{g cm}^{-3}\, \rho_{\rm gas} \sim 1$ is a scaled dimensional gas density. Note that because gas density is $\sim 3-5$ orders of magnitude lower in the outer $R\sim 100$~AU disc than in the inner few AU discs, we find that even very large grains fall well within the Epstein regime. Grains only start to enter the Stokes regime at scales of $\sim$ $R_H/100$ which is well below the resolution limit of our simulations. For this reason our results can be simply rescaled to less dense grains, e.g., water ice, by keeping the product $a \rho_a$ invariant. We plot the maximum grain size for the Epstein regime across the disc in the Appendix.

The gas equation of motion has a corresponding back reaction term to conserve the total momentum of the fluid. In terms of the grain stopping time, the total Lagrangian time derivative of the gas velocity is:

\begin{equation}
\dfrac{\mathrm{d} \mathbf{v_g}}{\mathrm{d}t} = \mathbf{\tilde{g}} + \dfrac{\mathbf{v_a - v_g}}{ t_{\rm st}} \dfrac{\rho_{\rm p} }{ \rho_{\rm gas}}\;,
\label{gas_v1}
\end{equation}

where $\mathbf{\tilde{g}}$ is the total acceleration acting on gas (the result of all of the forces included in the problem, such as gravity, gas pressure gradient term, radiation pressure force, etc.). 

The usual approach in SPH is to integrate the equations of motion explicitly, that is, assuming that the terms on the right hand side of equations such as \ref{gas_v1} are constant during the time step. A finite differencing scheme is then applied to advance particle velocity and position. Integrating Equation \ref{dust_v1} via an explicit scheme is numerically challenging for dust particles with short stopping times. In general this requires the dust particle timestep $\Delta t$ to be shorter than $t_{\rm st}$, which can grind the code to a halt in regions of high gas density or if grain particle size $a$ is too small. 

One way of dealing with too small dust particle time steps is to use one-fluid formulation for the dust-gas mixture \citep{LP14,PriceLaibe15} in which the dust particles are strongly coupled to the parent SPH particles but can drift from one to another. One then needs to use a separate integration scheme for larger particles. For the problem at hand, this approach would work for small dust particles which are always in the one fluid regime, but not for larger ones. The gas density may change by many orders of magnitude between the self-gravitating fragment, its Hill sphere, and the main body of the disc, so that particles with a large size $a$ can be in the one fluid (short stopping time) in dense regions but in the two fluid (long stopping times) in the lower density regions.
We thus need to find an integration scheme that can treat both short and long $t_{\rm st}$ in the same formulation. We are inspired by ideas of \cite{LAandBate14,LAandBate15b,BoothEtal15}. We assume that dust stopping time is constant during a time step of duration $\Delta t$, which requires gas density $\rho_{\rm gas}$ and its temperature do not change appreciably during $\Delta t$. \cite{BoothEtal15} define the difference $\Delta \mathbf{v}$ between the dust and the gas velocity, $\Delta \mathbf{v = v_a - v_g} $. They show that the evolution equation for this quantity is

\begin{equation}
\dfrac{\mathrm{d} \Delta \mathbf{v}}{\mathrm{d}t} = - \dfrac{\Delta \mathbf{v} }{ t_{\rm st}} + \mathbf{\left( g - \tilde{g}\right)  } + \Delta \mathbf{v} \cdot \nabla \mathbf{v_g}\;,
\label{Booth0}
\end{equation}

where $\nabla \mathbf{v_g}$ is velocity divergence for the gas velocity field. Neglecting the last term in this equation, \cite{BoothEtal15} show that there is an exact solution to the evolution equation for $\Delta \mathbf{v}$ which they then use to advance by a time step $\Delta t$.

We choose a slightly simpler solution in which the gas particle velocity is assumed constant when solving the Equation \ref{dust_v1}. This is justified as following. Assuming that the stopping time is constant across the dust time step automatically implies that the changes in the gas density during the time step are negligible too because $t_{\rm st}$ is usually a strong function of gas density (e.g., cf. Equation \ref{tEps0}). Now, if we assume the gas density to be constant during the dust particle time step $\Delta t$ we may as well assume the gas velocity to be constant because the two are connected via the mass continuity equation, 

\begin{equation}
\dfrac{\p \rho_{\rm gas}}{ \p t} + \nabla \cdot \left(\rho_{\rm gas} \mathbf{v}\right) = 0\;,
\end{equation}

and hence the error committed by assuming $t_{\rm st}$ constant should be of the same order as that by assuming gas velocity to be constant. Therefore we simplify Equation \ref{Booth0} to

\begin{equation}
\dfrac{\mathrm{d} }{ \mathrm{d}t} \left( \mathbf{v_a} - \mathbf{v_g}\right)  = - \dfrac{ \mathbf{v_a} - \mathbf{v_g} }{ t_{\rm st}} + \mathbf{g } \;,
\label{deBooth}
\end{equation}

where the gas velocity $\mathbf{v_g}$ is kept constant but dust velocity $\mathbf{v_a}$ should be solved for implicitly. An exact solution of the above equation yields the following scheme for advancing the dust particle velocity from time $t$ to $t+\Delta t$:

\begin{multline}
\mathbf{v_a}(t+\Delta t) = \mathbf{v_a}(t) \exp \left(-\dfrac{\Delta t }{ t_{\rm st}}\right) + \\
\left(\mathbf{v_g}+\mathbf g t_{\rm st}\right) \left[1 - \exp \left(-\dfrac{\Delta t }{ t_{\rm st}}\right)\right]\;.
\label{vdust_dt1}
\end{multline}

Here $\mathbf v_g$ is evaluated at the beginning of the time step. Equation \ref{dust_v1} is an exact solution as long as $t_{\rm st} = $~const. One can also see that it gives us the correct result in the opposite limits $\Delta t \ll t_{\rm st}$ and $\Delta t \gg t_{\rm st}$. For example, for static gas we recover the terminal velocity solution for grain particle (cf. Equation \ref{dust_v1}), $d \mathbf v_a/ dt=0$, $\mathbf v_a = \mathbf g t_{\rm st}$. The momentum conservation in the dust-gas interactions is enforced by passing the momentum lost by the dust particles to their neighbouring SPH particles, weighted by the SPH kernel.

This approach is sensible as long as the dust time step $\Delta t$ is short enough that the gas velocity indeed does not vary significantly. In \textsc{gadget-3} the dust particle time step is controlled by the condition

\begin{equation}
\Delta t_{\rm grav} \le \left( \dfrac{2 \epsilon_{d} h_{\rm d}}{ g}\right)^{1/2}\;,
\label{dt_grav0}
\end{equation}

where $\epsilon_{d} = 0.003$ is the accuracy parameter for most of the runs in this paper and $h_{\rm d}$ is the gravitational softening length for dust particles \citep{Springel05}. Estimating $g \approx GM_*/R^2$ for gravitational acceleration, and using $h_{\rm d} = 0.01$~AU, we find that at $R = 100$~AU the time step condition (\ref{dt_grav0}) yields

\begin{equation}
\Delta t_{\rm grav} \lesssim 10^{-3} \Omega_K^{-1}\;.
\label{dt_grav1}
\end{equation}

This is quite short compared with the local dynamical time, $\Omega_K^{-1}$, ensuring that SPH particle velocities are very unlikely to change significantly during such a short time interval. From the practical perspective, the time step in Equation \ref{dt_grav1} is shorter than the particle stopping time unless the Stokes number is smaller than $10^{-3}$, so it may appear that our scheme does not really provide much advantage over an explicit integration. However, just integrating gravitational forces for dust accurately requires such short time steps, and we are now free to explore arbitrary short $t_{\rm st}$ at no significant computational extra cost.

There is one further problem with semi-implicit dust treatments, dust grains may become trapped and clump together underneath the SPH smoothing length scale \citep{Tricco17}. This has implications for the regions deep inside the planet Hill sphere where the dust to gas ratio is high.
In this paper however we are concerned with the large scale collection of grains inside the Hill sphere. We leave modelling of specific grain dynamics inside the Hill sphere to later work.

\subsubsection{Weidenschilling radial dust migration test}

\begin{figure}
\includegraphics[width=1.02\columnwidth]{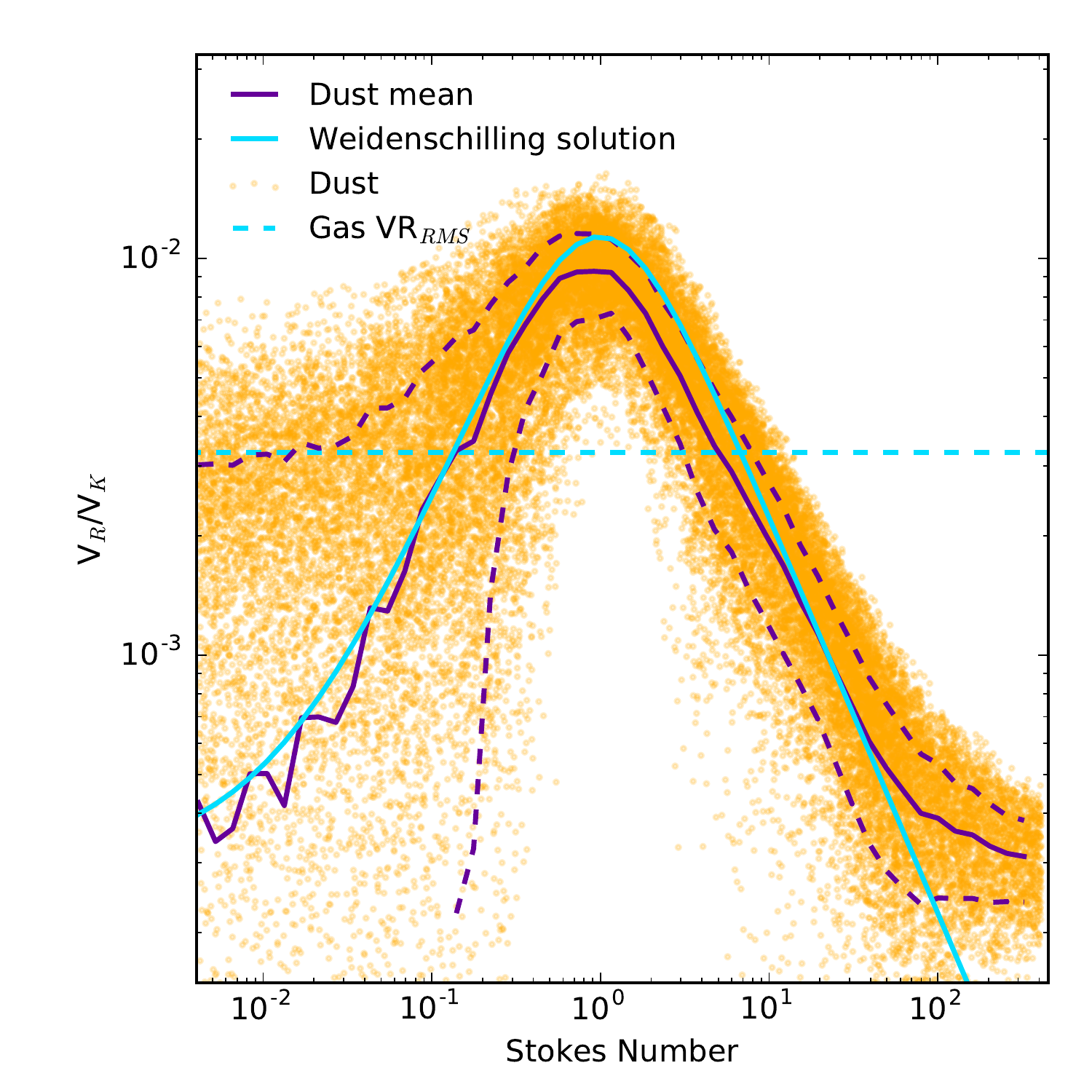}
\caption{The ratio of the radial velocity of dust particles to the local Keplerian velocity (orange scatter) compared to the analytical solution (blue line) given by Equation \ref{urad1}. The purple curve gives binned average radial velocity of dust particles. The dashed purple curves give the one standard deviation for these data. The dashed horizontal line shows the velocity dispersion in the radial velocity of the SPH particles at the location of the dust ring.}
\label{fig:vrad}
\end{figure}

In order to test the implementation of the dust-gas aerodynamical friction in the code, we set up the following test. 
For a non self-gravitating gas disc orbiting a central star of mass $M_*$, there is a well known analytical solution for the steady-state radial velocity of dust particles, $v_{\rm d}$.

\begin{equation}
\dfrac{v_{\rm d} }{ v_{\rm k}} = \dfrac{v_{R}/\tau v_K - \eta}{\tau + \tau^{-1}},
\label{urad1}
\end{equation}

where $\eta= 1 - (v_{\phi}/v_{\rm k})^2 \ll 1$, $\tau$ is the dimensionless grain stopping time defined in Equation \ref{eq:tau}, $v_{R}$ is the radial gas velocity and $v_\phi$ is the azimuthal gas velocity.

To arrive at a setting maximally approaching the analytical test, we turn off gas self-gravity for this test, and relax a gas-only disc with SPH initial particle number $N = 0.5$ million arranged in a disc with surface density profile $\Sigma \propto 1/R$ for $10 \hbox{ AU} \le R \le 160 \hbox{AU}$. Having relaxed the disc for about 10,000 years, we inject 50,000 dust particles with grain size $0.01 \hbox{ cm} \le a \le 1000 \hbox{ cm}$ initially put on a circular Keplerian orbit in a narrow ring around $R_0 = 80$~AU. The dust particle masses are set to negligibly small values so that they do not gravitationally influence the gas disc.

Figure \ref{fig:vrad} shows the dust particle radial velocities versus particle size (orange scatter) at time $t=980$~years after the dust particles were inserted into the gas disc.  The blue curve shows the analytically computed expected solution for $v_{\rm d}$ given by Equation \ref{urad1}. The purple curve shows the dust particle velocity averaged over logarithmic bins in dust particle size. The dashed purple curves show the one standard deviation for these data.

The individual dust particle radial velocities have a wide spread at small grain sizes, however this is simply a reflection of the general particle jitter inherently present in the SPH method \citep[e.g.][]{Lucy77}. This can be seen from the fact that particle average radial velocity curve (purple) fits the analytical result very well. Additionally, the dotted horizontal line shows the magnitude of the kernel weighted SPH particle RMS radial velocity, that is the dispersion in the radial gas velocity, $\sigma_v \approx 0.003 v_k$. We can see that smallest dust particles, tightly coupled to gas, are indeed scattered around the correct mean radial velocity curve with a spread comparable to the gas radial dispersion velocity. The scatter in the dust radial velocity decreases for larger particle sizes. This is expected because larger particles react to changes in the local gas velocity field slower (in proportion to their $t_{\rm st}$), and hence their velocity field reflects averaging over longer times and hence many more SPH particle neighbour interactions. This indicates that it is indeed the instantaneous jitter in the gas velocity field that drives the excursions of the individual dust particle velocities away from the expected solution in Figure \ref{fig:vrad}. This point is discussed further in Appendix \ref{app:v_disp}.

The total pebble mass is small compared to the mass of the gas, and therefore we can pass the back reaction force acting onto the gas via adding (always small) velocity kicks to the SPH neighbour particles interacting with the dust particle. The same approach is employed to account for the gas heating by the dust-gas aerodynamical friction. This ensures both momentum and energy conservation.

\subsubsection{Dust settling tests}
\label{sec:dust_settle}
\begin{figure}
\includegraphics[width=0.49\columnwidth]{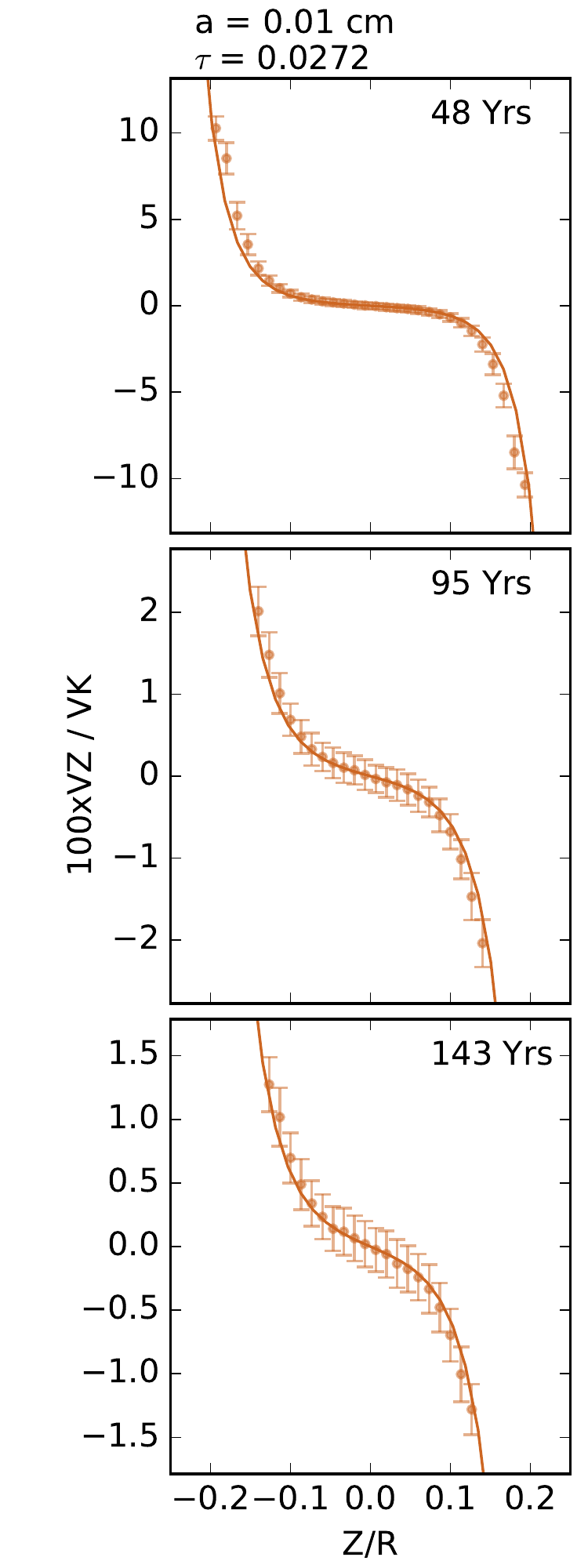}
\includegraphics[width=0.49\columnwidth]{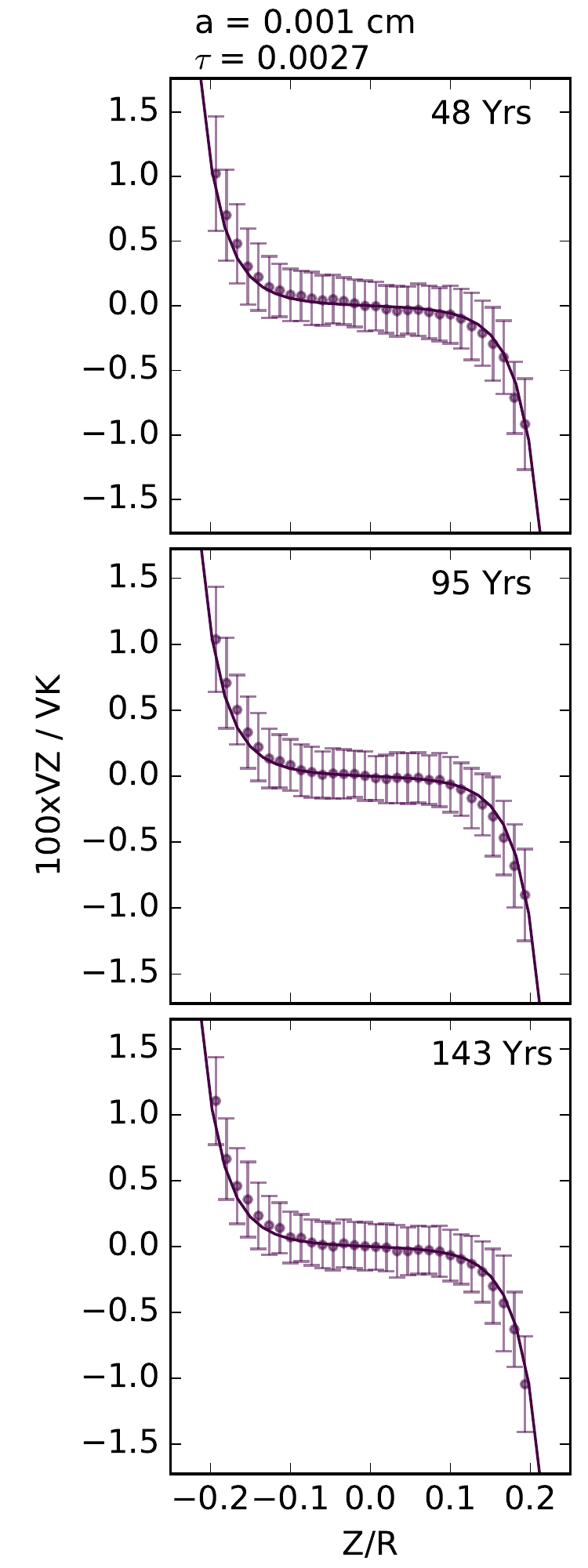}
\caption{Dust settling tests for $a$=0.01cm \& 0.001cm in a 0.01M$_\odot$ gas disc at 50 AU. The solid line marks the analytic settling solution. Filled circles and error bars represent binned particle locations and the 1$\sigma$ spread within each bin. Notice that the $a$=0.01cm grains have a higher Stokes number and so settle rapidly. }
\label{fig:dust_settle}
\end{figure}

In order to test the accuracy of our semi-implicit dust scheme we seek to reproduce the dust settling tests of \cite{LAandBate14}. 
For a non-self gravitating gas disc, the expected vertical density profile is a gaussian with the form 
\begin{equation}
\rho_{gas} (Z) = \rho_{0gas} \mathrm{exp} \left( \dfrac{-Z^2}{2H^2} \right) 
\end{equation}
where $Z$ is the vertical coordinate and $H$ is the vertical scale height of the gas. 
Dust experiences no pressure support and so will settle to the mid-plane of the disc on a timescale determined by its coupling to the gas. For dust in the Epstein regime it is possible to derive the terminal settling velocity ($v_{settle}$) by equating the vertical component of stellar gravity to the drag force on the dust. In terms of the midplane Stokes number ($\tau_0$), 

\begin{equation}
v_{settle} = -\dfrac{\rho_a a}{\rho_{gas}(Z)} \dfrac{Z \Omega_K^2}{v_{th}} = -\tau_0 \Omega_K  Z \mathrm{exp} \left( \dfrac{Z^2}{2H^2} \right).
\label{eq:v_term}
\end{equation}

Since this expression is only valid in the non-self-gravitating case we use a relaxed $M = 0.01 M_\odot$ disc with 1 million gas particles that extends from 2 to 100 AU for these experiments. We seed dust particles in the ratio 1:1 at the same location as the gas particles in the disc but with a much lower mass such that they do not perturb the hydrostatic vertical structure. Over the course of the simulation these particles settle towards the disc mid-plane at a velocity given by Equation \ref{eq:v_term}. We start the particles at an initial $v_Z$=0 and they very quickly accelerate to reach the terminal settling velocity in these tests.
For comparison with \citet{LAandBate14}, their $K_S^E / \hat{m}_D$ = 10 \& 100 correspond to $\tau$ = 0.06 and 0.006 in our set up.

The binned mean values that we find for the $v_{settle}$ of the ensemble of dust particles match the analytic expectation very well. The spread in $v_{settle}$ is caused by noise in the SPH field which is transferred to the dust through Equation \ref{vdust_dt1}. The spread is about 5 \% of the local sound speed in the disc.
We do not expect to reproduce the high accuracy of the tests in \cite{LAandBate14} since theirs were computed in a high resolution 1D SPH simulation. Our tests demonstrate that the bulk behaviour of the dust ensemble in our 3D disc matches the analytic settling expectations very well.  

From Equation \ref{eq:v_term} it can be seen that larger grains are expected to settle faster. This can be seen in Figure \ref{fig:dust_settle}, the population of $a$=0.01cm grains settles more rapidly than the $a$=0.001cm grains.
Tests for $a$=0.1cm \& 0.0001cm grains are presented in the Appendix.

\subsection{Modelling of accretion of gas and dust onto a planet}\label{sec:num_acc}

The accretion rate of gas onto a planet depends on  the cooling rate of the gas entering the Hill sphere, the planet physical size  \citep{Nayakshin17a} and its luminosity through the radiative feedback that the planet exerts on its surroundings \citep{MachidaEtal10,NayakshinCha13,Stamatellos15,MercerStam2017} by pre-heating it. Here we are most interested in the accretion of pebbles onto the planet rather than the detailed physics of gas accretion. We find that the main points of our paper are well demonstrated via a simple sink particle approach \citep{Bate95} to accretion of both gas and dust particles. Therefore, despite certain numerical drawbacks of the approach, discussed below, we opt for it in favour of simplicity and clarity of arguments and results.

We define the sink radius ($r_{\rm s}$) as a spherical region around the planet within which any gas or dust particle is captured by the planet. Operationally, once a dust or gas particle enters the region, it is removed from the simulation, with its mass and momentum added to that of the planet. 

We should expect the sink particle prescription to be least reliable when the gas around the planet is `hot' in the sense that its temperature is comparable to the virial temperature, $T_{\rm vir} = G M_{\rm p}\mu/(2 k_B r)$, where $r$ is distance to the planet centre. In this case the gas pressure gradient may be able to impede gas accretion onto the sink. However, the artificial vacuum within the sphere $r_{\rm s}$, maintained by a constant removal of the gas there, creates an unphysical negative pressure gradient. This effect tends to push gas particles into the sink sphere, probably resulting in an over-estimate of the gas accretion rate onto the sink \citep{Bate95}. Selecting as small a sink radius as numerically practical in this situation is the best solution \citep[see the Appendix in][]{Nayakshin17a}.

Finally, we may expect that pebble accretion is reasonably well modelled with the sink particle approach because pebbles are not directly supported against sink gravity by gas pressure. 
Due to gas drag and lack of pressure support, once pebbles reach the sink radius it is very unlikely that they will be able to escape the fragment and so the exact value of the sink radius should not be too important. We recognise that we do not resolve the planet interiors and so in this paper we often discuss results for both the sink particle and for material within a half Hill radius of the sink which we refer to as the half Hill sphere. This provides a perspective on material gravitationally bound to the planet that is in the process of accreting onto the sink particle and weakens the dependence of results on the artificially imposed sink radius.

\subsection{Initial conditions}\label{sec:ic}

We simulate a gas disc with an initial mass of $M_{\rm d} = 0.2\msun$ with $\Sigma(R) \propto1/R$ between the inner and the outer radii of 10 and 300 AU, respectively, and $\Sigma = 0$ outside of this radial range. 
Since dust grain dynamics are very sensitive to any non-power law features in the gas density distribution \citep[e.g.,][]{DipierroEtal15,DipierroEtal16a}, the gas disc is relaxed for time equal to about 10 orbits at the outer edge. The inner boundary condition at the star inevitably leads to a reduction in the disc surface density around the inner edge due to SPH particle accretion onto the star, so that $\Sigma(R)$ drops and deviates from the initial profile there. Similarly, the outer edge with a free boundary rearranges itself beyond $R \sim 200$~AU. Nevertheless, the relaxed initial condition disc is very close to a perfect $\propto R^{-1}$ surface density profile  and the radial gas velocity approaching zero between radii of $\sim 30$ and $200$~AU. Based on this, it is more appropriate to talk about the inner computational boundary rather than the inner disc edge since we do not resolve the disc properly below around 30 AU.

This relaxed gas only disc is then used to add pebbles and the planet. Unless specified otherwise, pebbles are added in the ratio 1:1 to the gas particles in terms of particle number, but only in the radial range $20 \le R \le 200$~AU. Our planets are never outside this radial range during the simulation time, and so we choose not to follow dust dynamics there, saving computational resources. Inside the radial range $20 \le R \le 200$~AU, pebbles are seeded at the same locations as their `parent' SPH particles except for the vertical coordinate $z$ which is suppressed by a factor of 100 \citep{LambrechtsJ12}. Pebbles are thus seeded in a rather thin disc. However, since pebble particle mass is set to $10^{-3}$ of the SPH particle mass, the effects of pebble self-gravity remain negligible. This implies that local initial abundance of pebbles in our disc is $Z_0 = 10^{-3}$. 

For simplicity, we use pebbles of just one fixed size per simulation even though the code could handle a range of grain sizes at the same time. In a more complete simulation, there would be more metal mass in the disc distributed across a range of grain sizes, smaller and perhaps also larger than the pebbles that we simulate here. Modelling only one grain size neglects the expected dust and gas coupling in areas of high dust density and also neglects indirect coupling of different grain sizes via the gas distribution \citep{Bai10}. 
In this work we aim to identify the regime where metal enrichment due to pebble accretion is significant. Full grain population studies will be necessary to calculate the exact degree of this enrichment. These will lead to additional complications since the population distribution of grain sizes is not well known.

Pebbles are set on local circular orbits. While this deviates from the quasi-steady state velocity profile for dust grains (cf. Equation \ref{urad1}), the difference in results was found to be very small when started from the latter velocity profile.
The planet is initialised on a circular prograde orbit around the star at the initial separation $R_{\rm 0} =120$~AU in order to represent a gas clump recently formed by gravitational fragmentation of a massive disc.

\subsection{Numerical convergence}\label{sec:num_con}
\begin{figure}
\includegraphics[width=0.99\columnwidth]{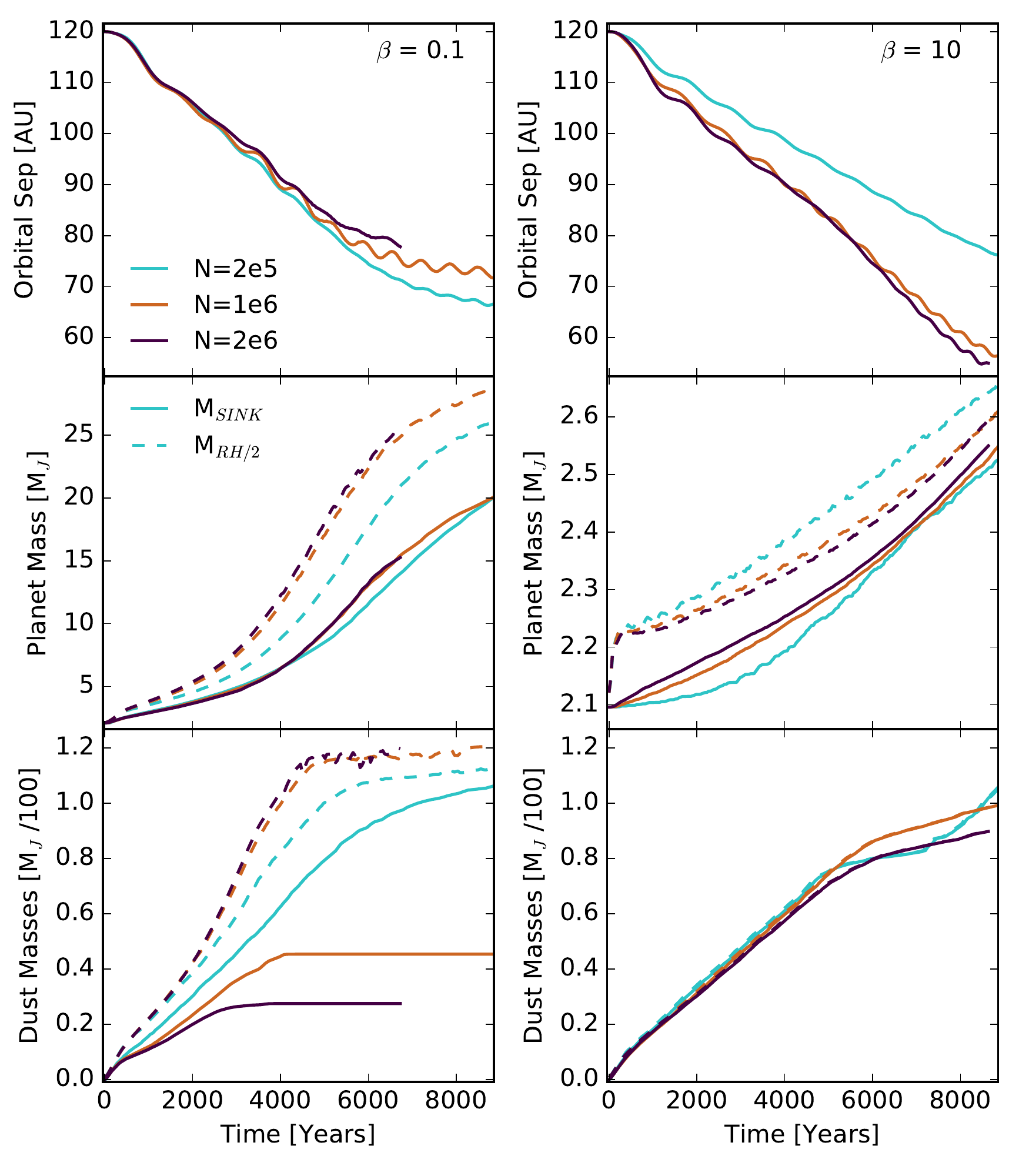}
\caption{Change in orbital separation (top panels), evolution of planet mass (middle panels) and accreted dust mass (bottom panels). Two values of the cooling parameter $\beta$, 0.1 (left panels) and 10 (right panels), are considered. Curves with 0.2, 1 \& 2 million SPH particles are shown. Dashed lines correspond to masses within half the Hill Radius, including the sink particle.}
\label{fig:res_N}
\end{figure}

\begin{figure}
\includegraphics[width=0.99\columnwidth]{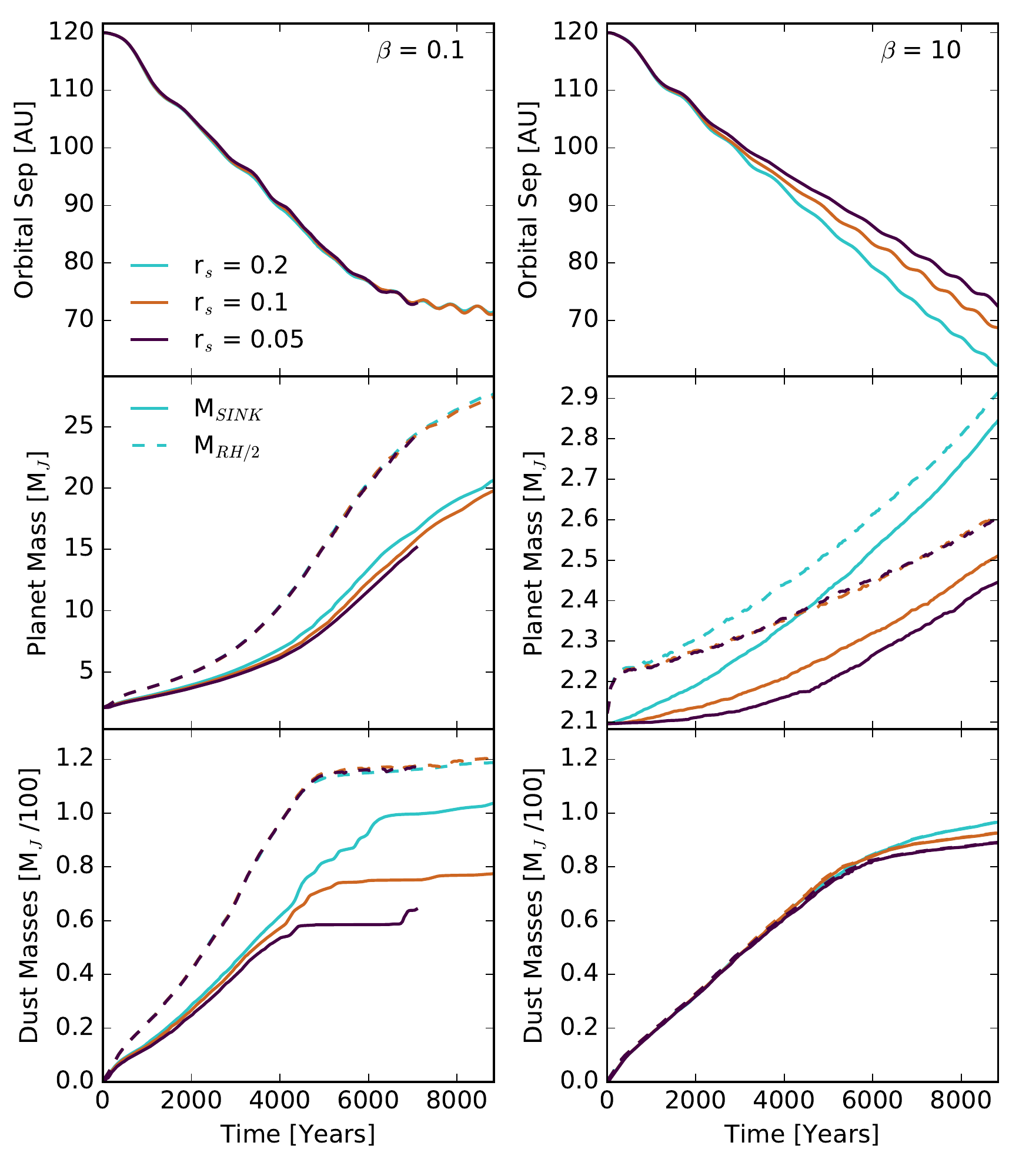}
\caption{Axes follow Figure \ref{fig:res_N}. In this case the number of SPH gas particles is set to 1 million while the sink particle radius ($r_{\rm s}$) is varied between 0.2, 0.1 \& 0.05 AU. }
\label{fig:res_sink}
\end{figure}

In this section we vary the SPH particle number (Figure \ref{fig:res_N}) and the sink particle radius (Figure \ref{fig:res_sink}) in order to investigate how these resolution limits affect our simulations. The top, middle and bottom panels of these figures show the planet separation, mass and the accreted pebble mass versus time. 
In the lower two rows of panels the solid curves correspond to the sink particle whilst the dashed curves show the total mass of the sink particle and the material inside the half Hill sphere of the planet. 
We fix the pebble radius at $a=1$ cm and unless specified the particle number and sink radius are set to 1 million \& 0.1 AU.
We run tests for the two opposite cooling limits, $\beta = 0.1$ and $\beta =10$, shown in the left and the right panels of each figure respectively. The physical implications of these cooling regimes will be discussed in Section \ref{sec:cooling}.

\subsubsection{Particle number tests}\label{sec:N_tests}
\label{sec:N_test}
It is important to ensure that the particle number in SPH simulations is high enough such that the length scales of relevant physical processes are well resolved. 
Wide orbit gas giants are initially expected to migrate in the Type I regime and will not open a gap unless they reach masses of $\sim 5\mj$ or more \citep[e.g.,][]{BaruteauEtal11,MalikEtal15,Nayakshin17a}.
In this regime, migration rates are governed by the Lindblad torques from the disc inside and outside of the planet \citep[e.g.,][]{LinPap79,Tanaka02,Bate03}. Low particle densities in the vicinity of the planet may lead to a miscalculation of these torques, hence a miscalculation of the migration rates. This effect will be more significant for the $\beta$ = 10 case as the gas in the vicinity of the planet is hotter, less dense and therefore less well sampled by the SPH.  

In Figure \ref{fig:res_N} the number of SPH particles is varied between 0.2, 1 \& 2 million.
Comparing the top four panels of Figure \ref{fig:res_N}, we find that the migration rate and planet mass have converged for SPH particle number $\geq$ 1 million.
From the bottom panels, we see that the dust mass accreted by the sink decreases with increasing particle number. 
In the low resolution case, the gas pressure maximum around the planet is less well resolved and so dust particles are able to accrete more rapidly onto the sink particle. 
This effect shows no sign of converging with increased particle number, but since we do not aim to fully model the deep interior of the planet this is not overly concerning.
Whilst the sink particle solves a numerical limitation it does not provide an accurate model for the internal structure of the migrating fragment.
We believe that the lack of convergence for $\beta=0.1$ is due to a failure to properly resolve the circumplanetary accretion disc at low resolution. When this dense gas disc is resolved it increases the stopping time of dust grains approaching the sink and effectively slows their rate of accretion, hence the variation in accreted dust mass in Figure \ref{fig:res_N}. 
The circumplanetary disc in these $\beta=10$ simulations is $\sim$ 10 $M_J$ whilst the total mass of trapped dust is at maximum 0.01 $M_J$. Since the gas mass dominates over the dust mass we are not worried that back reaction from the unresolved behaviour of this dust has a significant impact on the evolution of the simulation. Simulations with a fully resolved circumplanetary disc are too computationally expensive at this point so we must accept this limitation.

Importantly, the total gas and pebble mass inside the half Hill sphere do converge with increased particle number. 
This shows us that while we cannot be sure of the exact density profiles of material inside the half Hill sphere, we can be reasonably confident about the total gas and pebble mass in that region.
Disagreement between the curves of different particle number is far smaller than the differences in accretion rates due to varying $\beta$. This shows that numerical artefacts in the accretion rates are sufficiently small compared to the differences in the gas cooling physics that we aim to study.
As a note in passing, resolution requirements would have to be strongly increased for planets of mass much smaller than studied here, e.g., for Neptune-mass planets, because the Hill spheres of such planets are much smaller.

\subsubsection{Sink radius tests}
\label{sec:sink_tests}

We now test how the value of the sink radius used in our simulations may influence our results.
The runs in Figure \ref{fig:res_sink} have a fixed SPH particle number of 1 million and sink radii that are varied between 0.2, 0.1 \& 0.05 AU. Note that the radii of isolate pre-collapse GI planets is $\sim$ 0.05-0.5 AU ($10^3-10^4 R_J$) \citep{HelledEtalPP62014} depending on mass and central temperature. The sink radii used here are therefore comparable to the physical sizes of the objects that we aim to study.

In the left panels we show the $\beta$ = 0.1 case. The migration rates and planet masses are converged for all values of sink radius. The dust mass accreted onto the sink particle is not converged although we have discussed in Section \ref{sec:N_test} that this is not crucial since the `missing' pebbles are in the gas disc around the planet.

On the right we present the $\beta$ = 10 case. The choice of sink radius is important here since increasing the sink radius noticeably increases the rate of gas accretion onto the sink particle. 
This result agrees with previous simulations \citep{Cuadra06} which found that increasing the sink radius increases the rate of spurious gas accretion caused by the artificial vacuum of the sink particle. See especially the Appendix in \cite{Nayakshin17a}.
This effect becomes more significant in the inefficient cooling case as hot gas resists accretion onto the planet better due to a larger pressure gradient than cold gas and thus is more strongly affected by a change in the sink radius. 
Fortunately, the actual difference in accreted gas mass is small and in fact the mass inside of the half Hill sphere has converged at a sink radius of 0.1 AU. Given this convergence, we chose the sink radius of 0.1 AU for the simulations in this paper as a reasonable compromise between physical accuracy and numerical expense.

\subsubsection{Time step convergence}\label{sec:beta10_dt}
\begin{figure}
\includegraphics[width=0.99\columnwidth]{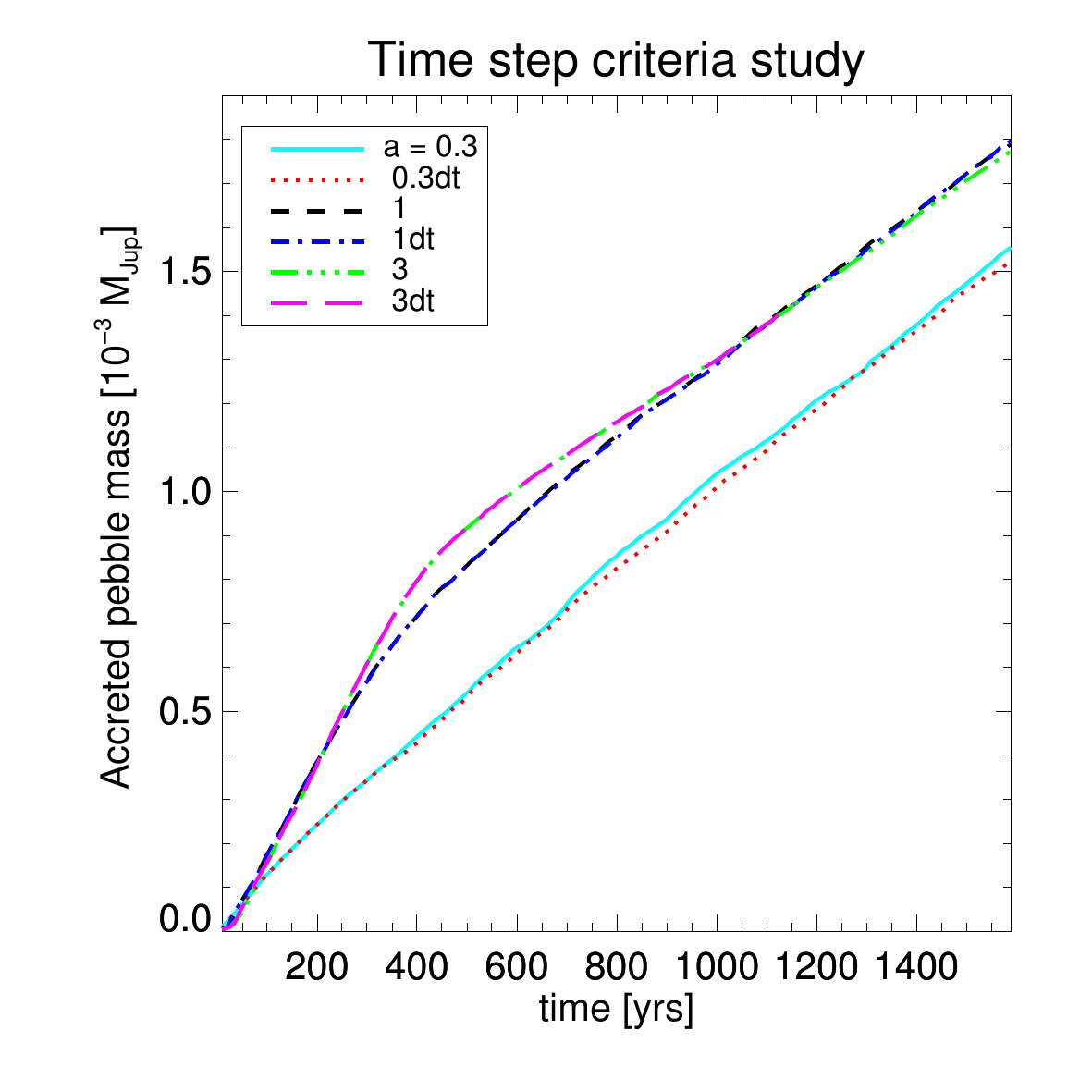}
 \caption{Convergence of grain accretion onto the planet for several grain sizes (marked in the legend) with respect to decreasing the dust particle time step by the factor of 2 (curves marked "dt" in the legend). }
\label{fig:dt}
\end{figure}

In Section \ref{sec:grain_dyn} we assumed that gas particle velocity changes little during the dust time step, which then allowed us to propose an approximate integration scheme for dust given in Equation \ref{vdust_dt1}. It is rather clear that if this assumption is wrong then varying dust time step will uncover numerical artefacts. Fig. \ref{fig:dt} shows how reducing time step by a factor of two affects the resulting pebble accretion rates for three different pebble sizes, $a = 0.3$, 1 and 3~cm.  All of these simulations are performed for $\beta =10$ models.The curves are labelled by the grain size in the legend. Curves labelled with "dt" are those calculated with twice smaller time step. It is apparent that changes in the results are minimal. The curves for $a=1$ cm are right on top of one another, for example. This indicates that our simulations are converged with regard to the time step and force calculation criteria.
 
\section{The Effect of Varying Physical Parameters on Disc Evolution}\label{sec:disc_evo}

\begin{table}
\centering
\begin{tabular}{ c c c } 
 Varied Parameter & Parameter Space & Figures \\
 \hline
 Gas Particle Number & 2$\times 10^5$, $10^6$  \& 2$\times 10^6$ & \ref{fig:res_N}, \ref{fig:SPH_t1}-\ref{fig:SPH_t1_tauQ} \\ 
 Sink Radius & 0.2, 0.1 \& 0.05 AU & \ref{fig:res_sink}\\ 
 Initial Planet Mass & 0.5, 1, 2, 4 \& 8 $\mj$ & \ref{fig:planet_mass0} \\
 Grain Size (a) & 0.1 \& 1 cm & \ref{fig:FF0} \& \ref{fig:grain_size_sigma} \\
 Pebble Fraction & 3$\times 10^{-4}$, $10^{-3}$ \& 3$\times 10^{-3}$ $M_{\odot}$ & \ref{fig:f_peb} \\
 Radiative Feedback & 0, 10$^{-3}$ \& 3$\times 10^{-4} \Lsun$  & \ref{fig:irr_feedback} \\
\hline
\end{tabular}
\caption{Summary of runs in this paper. Unless otherwise specified, N = 10$^6$, r$_{s}$ = 0.1 AU, $M_{P0}$ = 2 $\mj$, a = 1cm, pebble fraction = $10^{-3}$ and radiative feedback = 0. Figures \ref{fig:SPH_t1}-\ref{fig:SPH_t1_tauQ} correspond to the N = 2$\times 10^6$ runs.}
\label{table:runs_summary}
\end{table}

\begin{figure*}
\includegraphics[width=0.99\columnwidth]{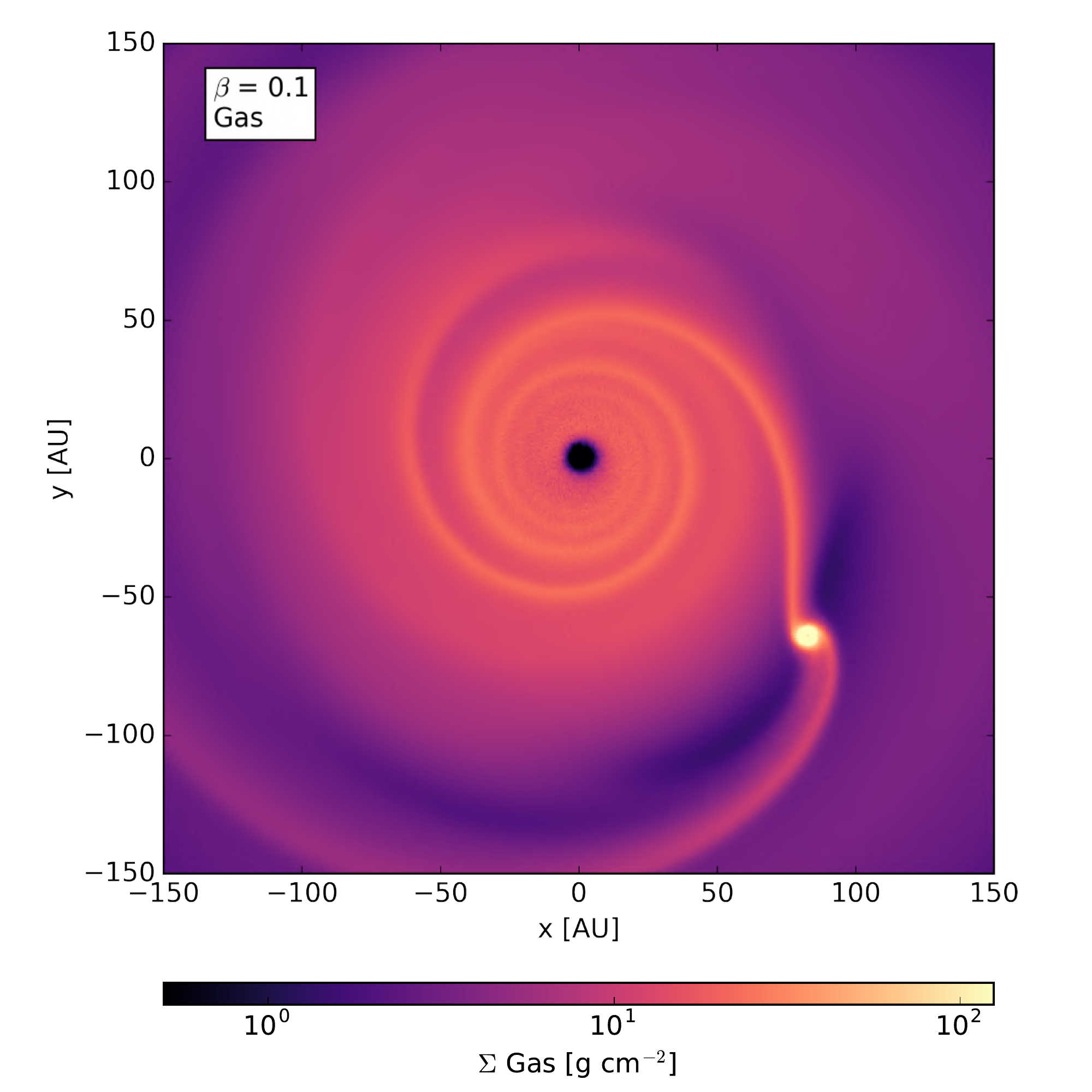}
\includegraphics[width=0.99\columnwidth]{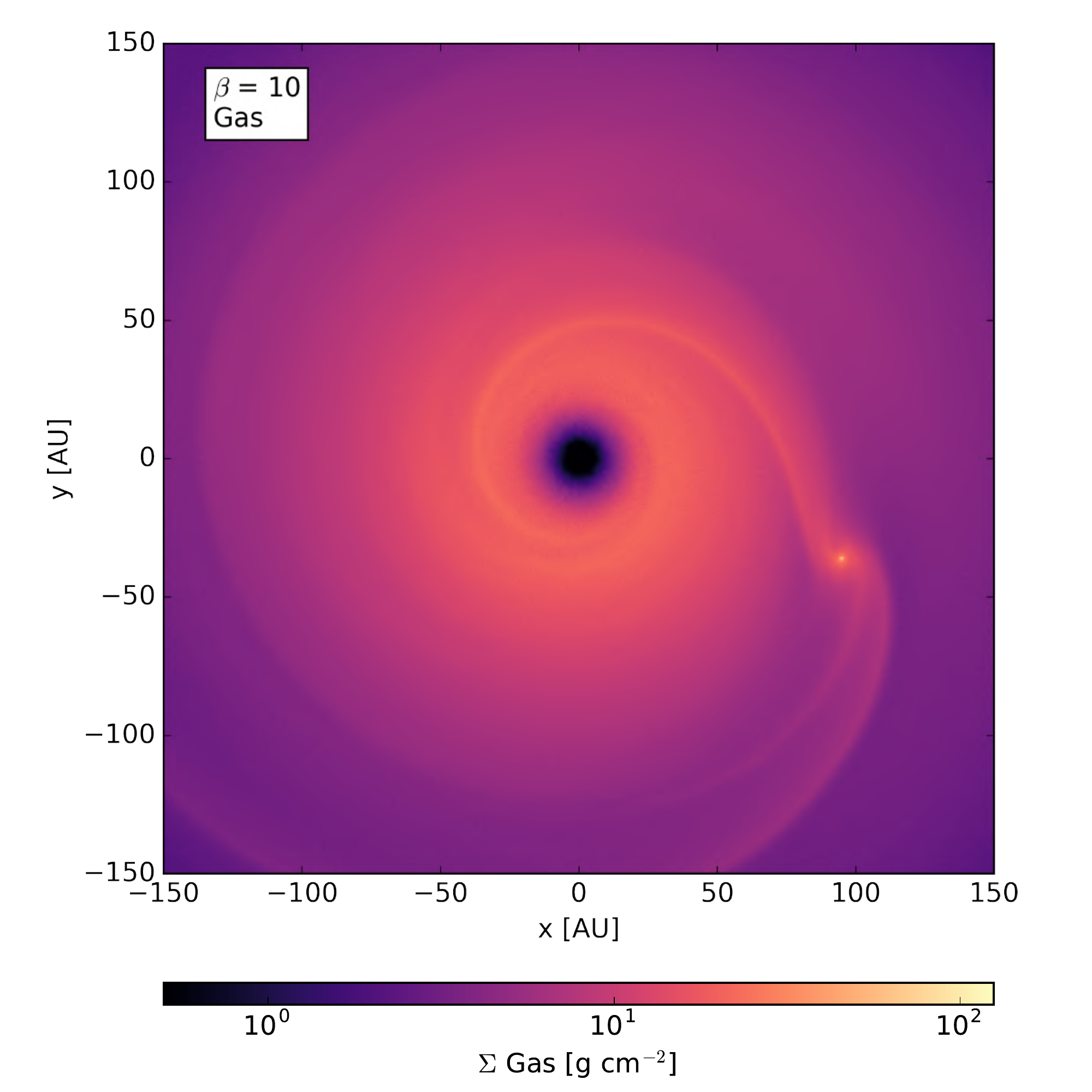}
\includegraphics[width=0.99\columnwidth]{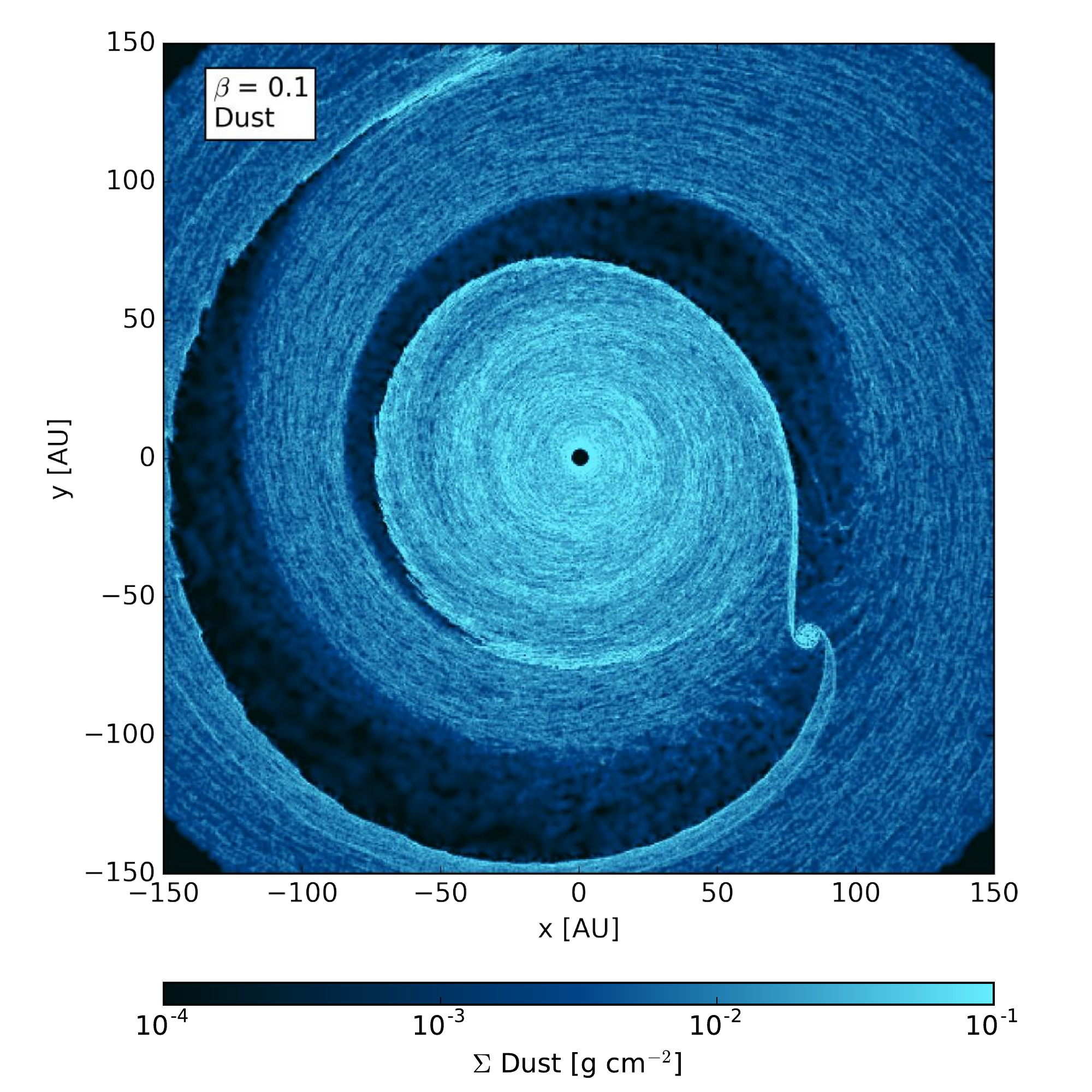}
\includegraphics[width=0.99\columnwidth]{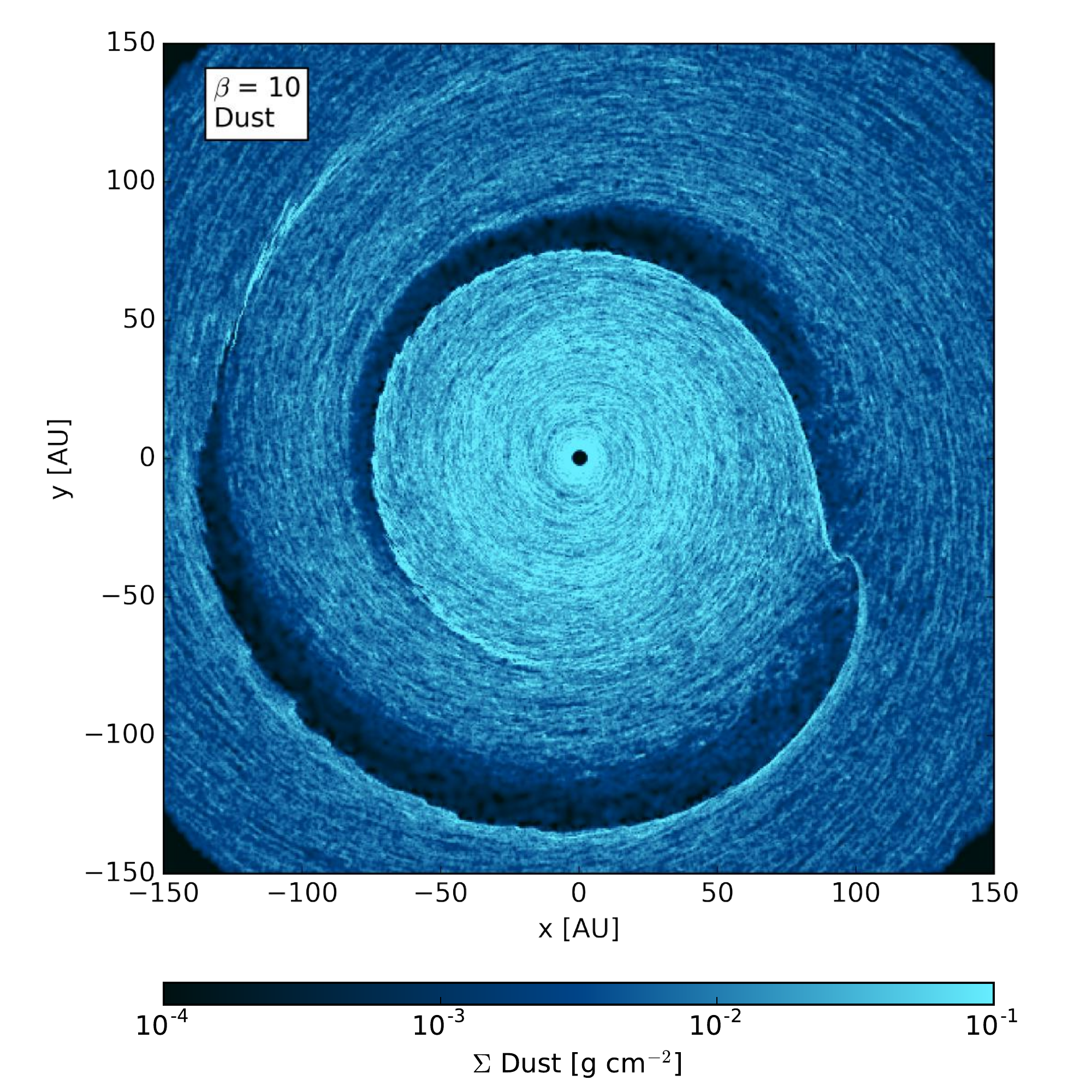}
\caption{A top down projection of the gas (top panels) and dust (bottom panels) surface densities in this disc at time $t = 2230$ years from the beginning of the simulation. The planet is located at the gas density peak in each figure at $(x,y) \sim (100, -50)$ AU. Left: $\beta = 0.1$, right: $\beta = 10$. From Figure \ref{fig:res_N} it can be seen that the planet has grown to $\sim 5 \mj$ in the efficient cooling case, this has caused it to open a much deeper gap. The corresponding grain Stokes numbers can be seen in the lower two panels of Figure \ref{fig:SPH_t1_tauQ}.}
\label{fig:SPH_t1}
\end{figure*}

\begin{figure*}
\includegraphics[width=0.99\columnwidth]{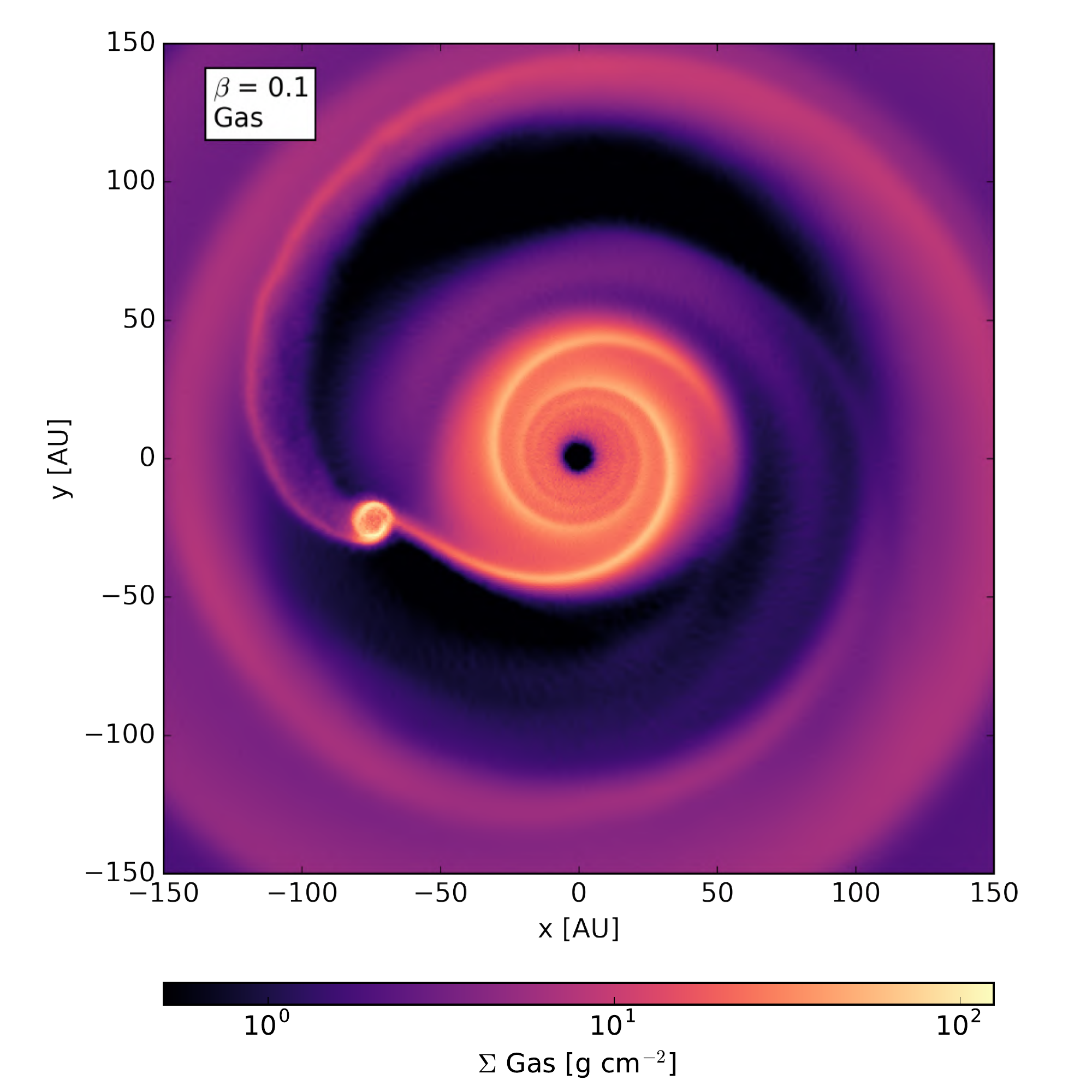}
\includegraphics[width=0.99\columnwidth]{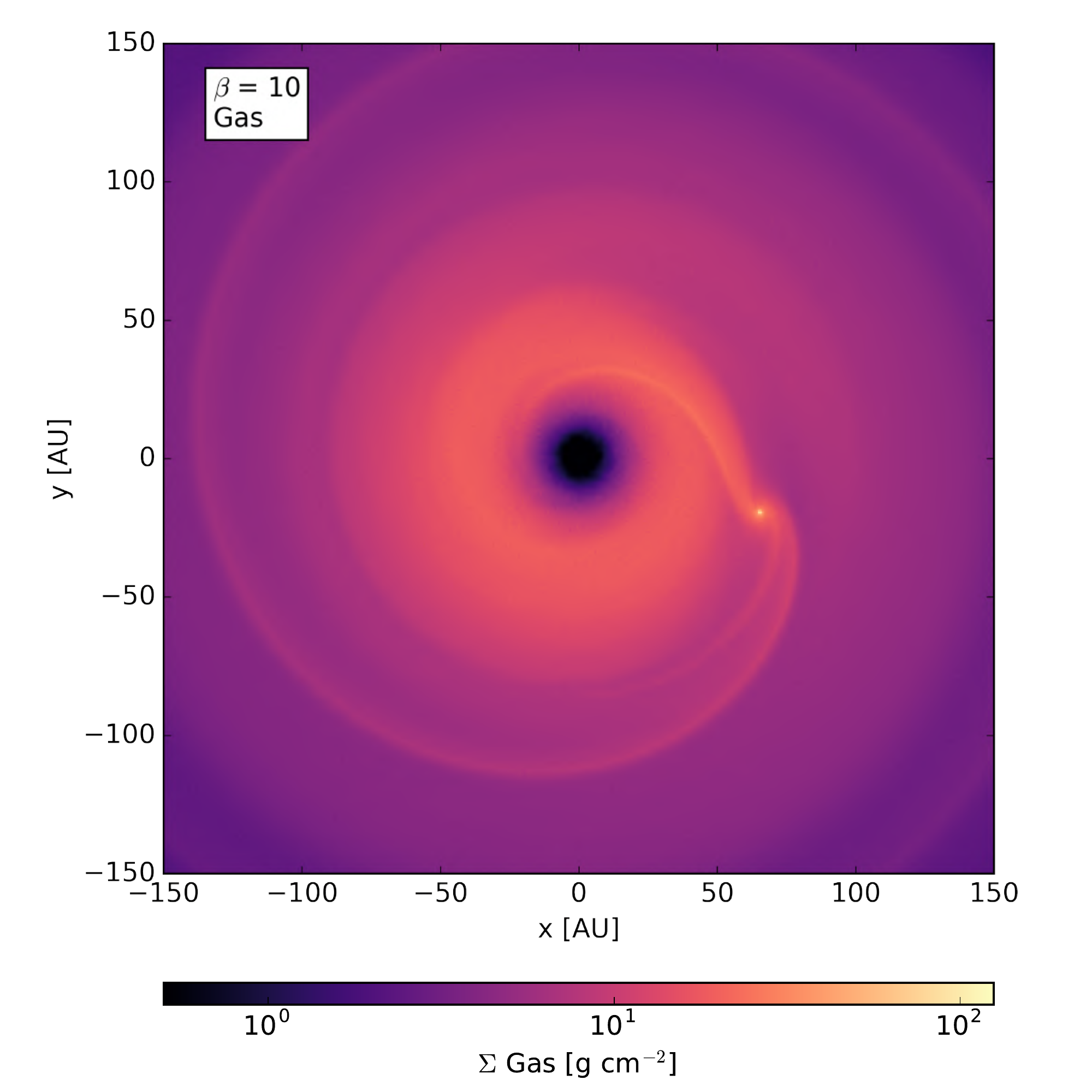}
\includegraphics[width=0.99\columnwidth]{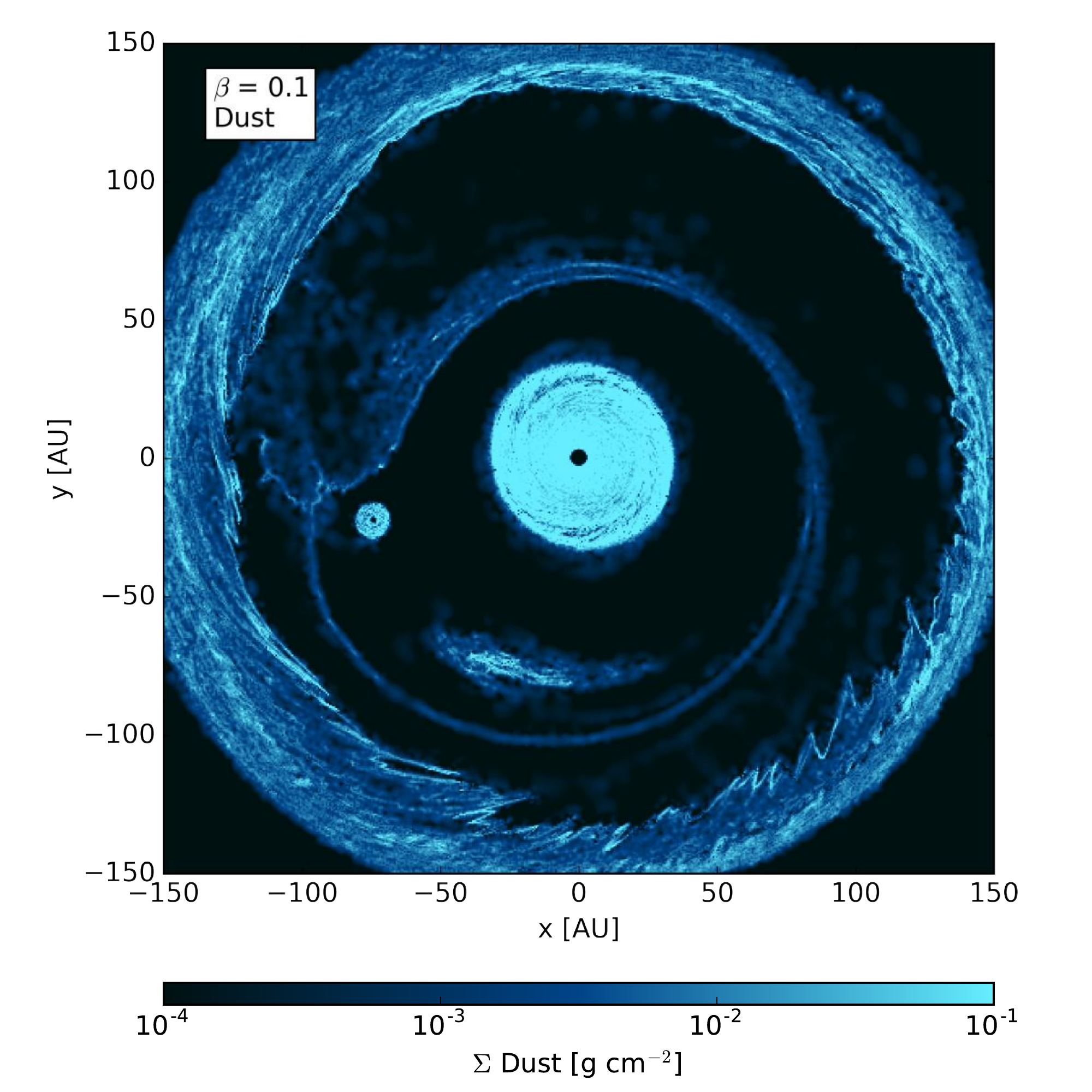}
\includegraphics[width=0.99\columnwidth]{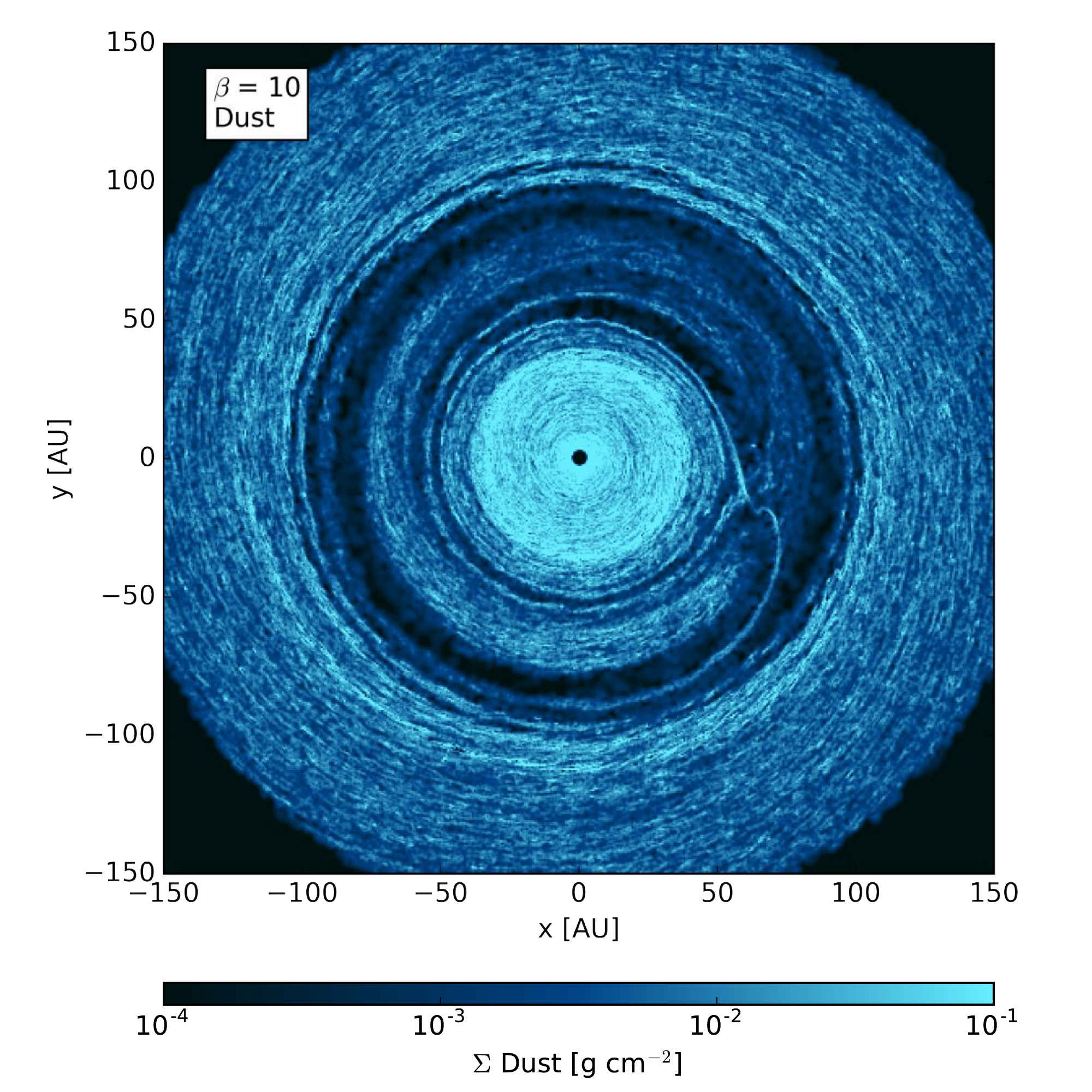}
\caption{Same as Figure \ref{fig:SPH_t1}, but at $t=6800$ years. In the efficient cooling disc, the planet has carved a very deep gas gap and accreted the majority of 1 cm pebbles. There is no gas gap in the inefficient cooling disc, but the pebble distribution has still been depleted.}
\label{fig:SPH_t2}
\end{figure*}

\begin{figure*}
\includegraphics[width=0.99\columnwidth]{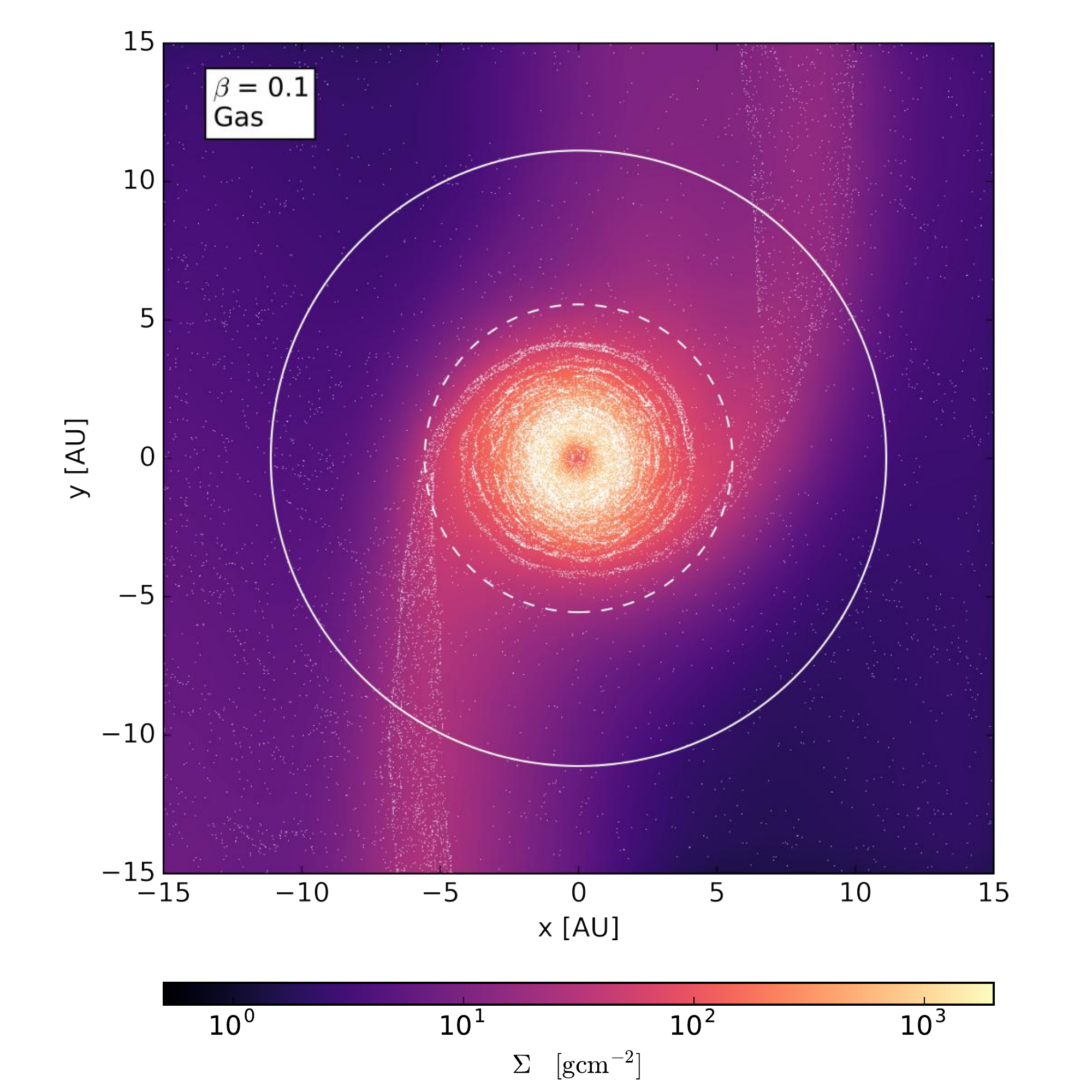}
\includegraphics[width=0.99\columnwidth]{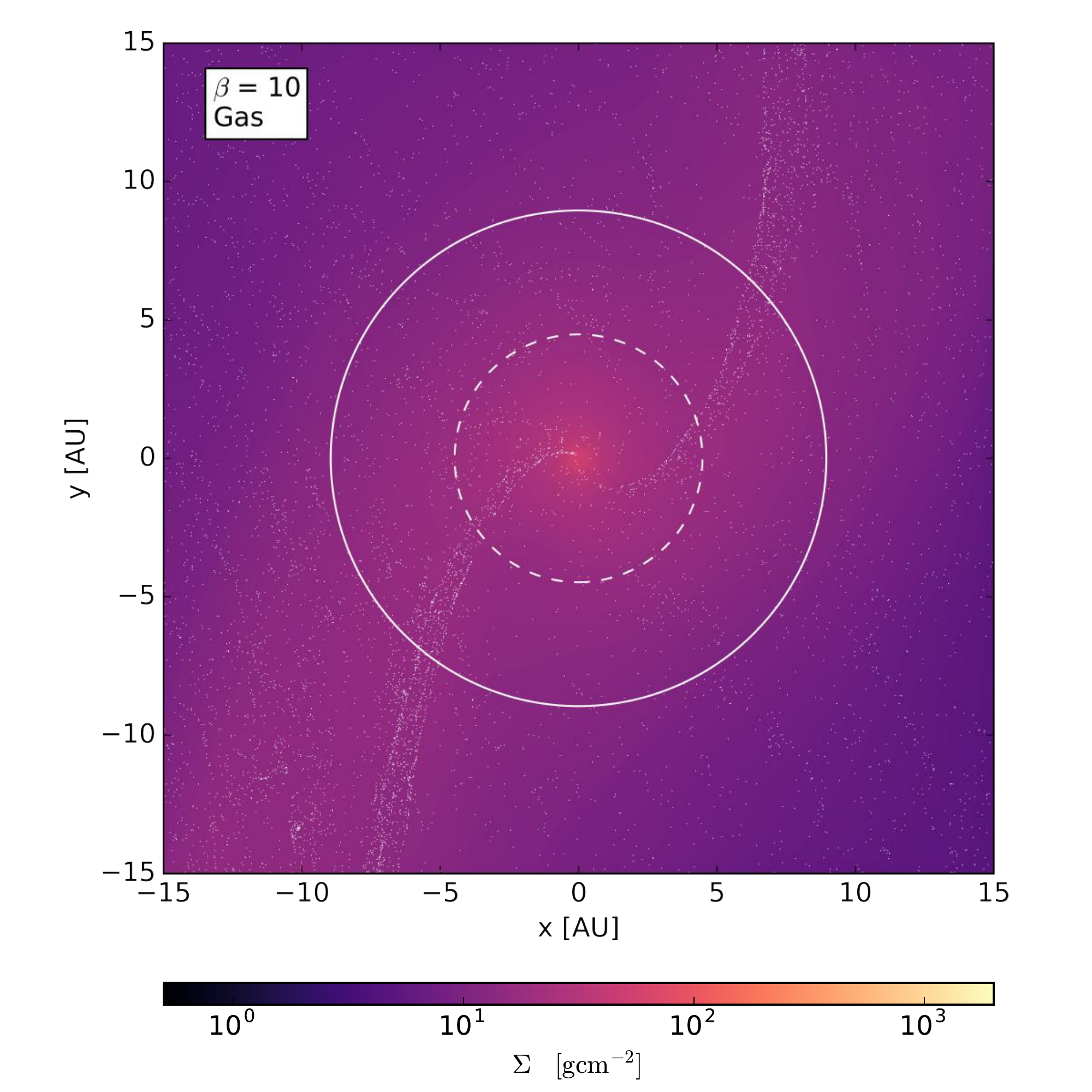}
\includegraphics[width=0.99\columnwidth]{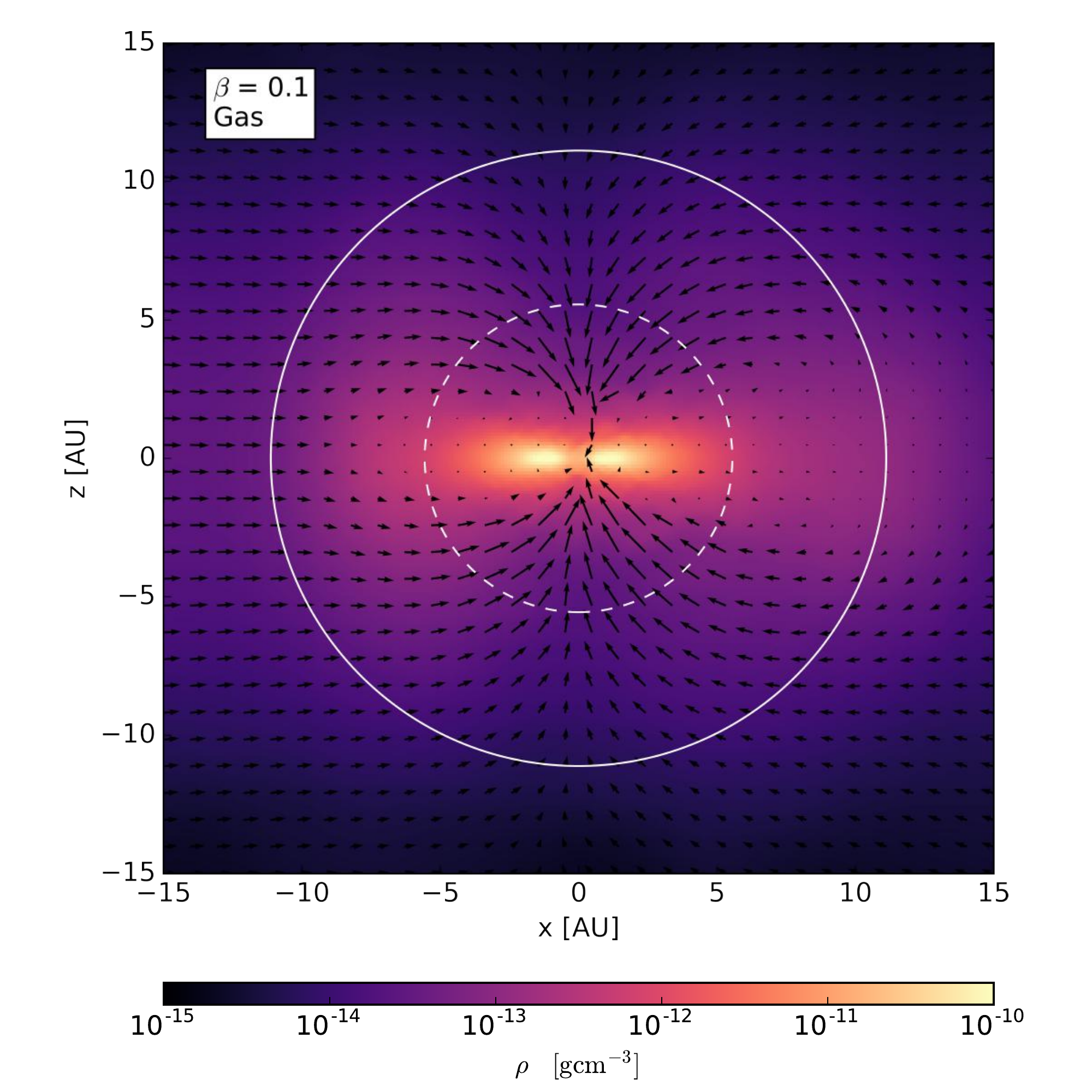}
\includegraphics[width=0.99\columnwidth]{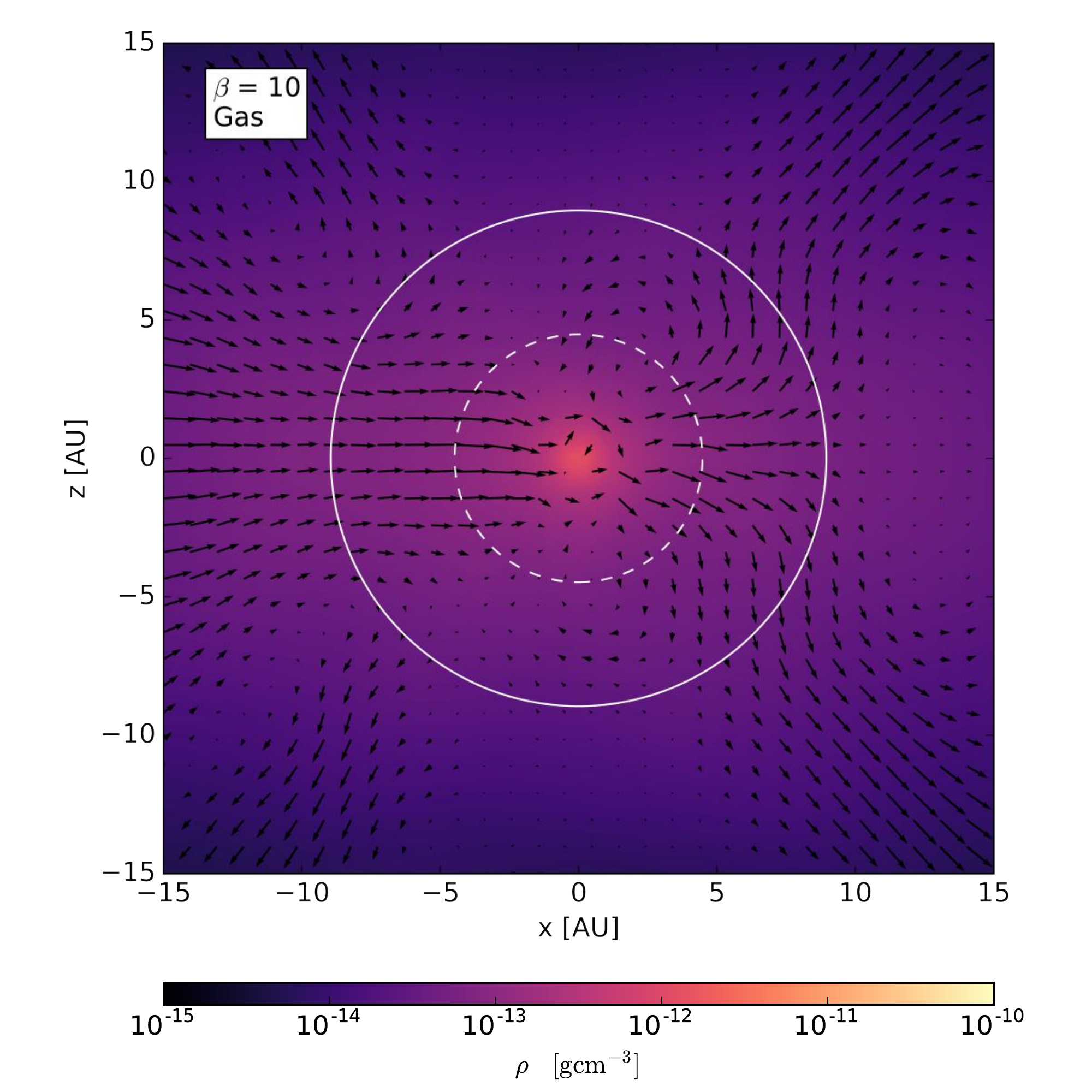}
\caption{Top panels: Gas surface density with SPH pebble positions over-plotted in white for $\beta$ = 0.1 (left) and $\beta$ = 10 (right) at $t = 2230$ years. The solid and dashed white lines show the full and half Hill radii. Lower panels: edge cuts of disc density. Velocity fields for the gas are over-plotted in black arrows and are relative to the motion of the planet. Note the compact disc structure in the $\beta$ = 0.1 case and the vertical inflow of gas. }
\label{fig:SPH_t1_zoom_over}
\end{figure*}

\begin{figure*}
\includegraphics[width=0.99\columnwidth]{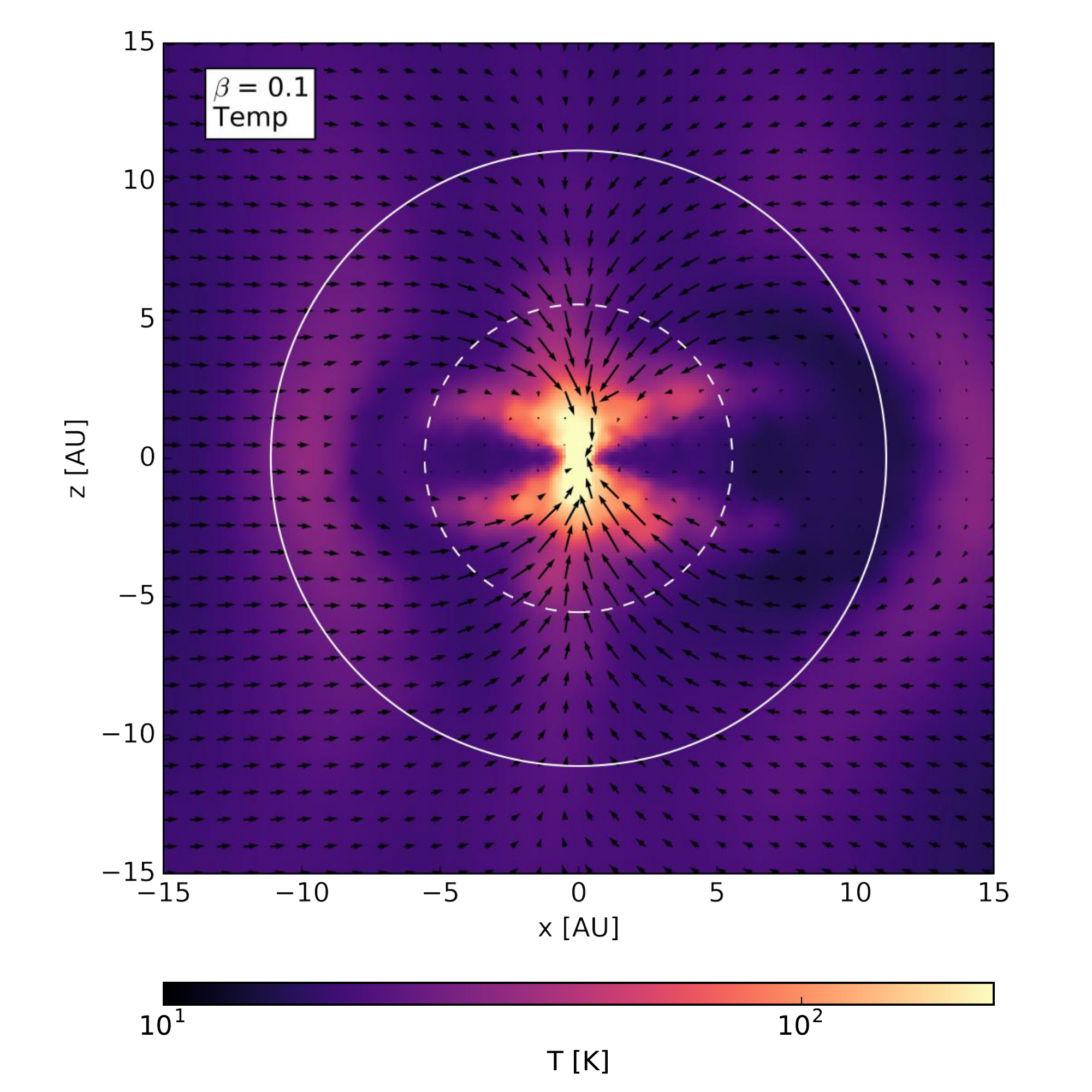}
\includegraphics[width=0.99\columnwidth]{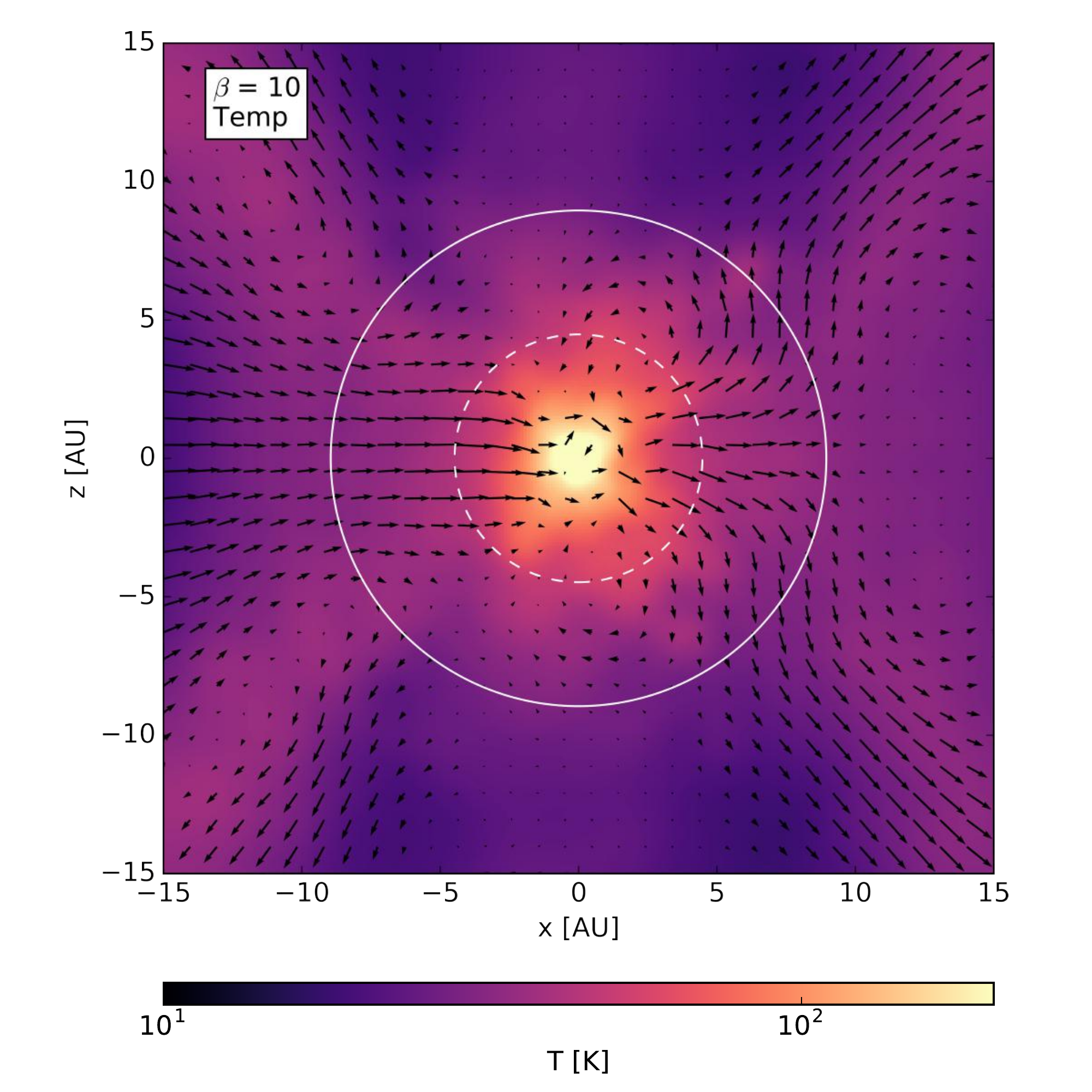}
\includegraphics[width=0.99\columnwidth]{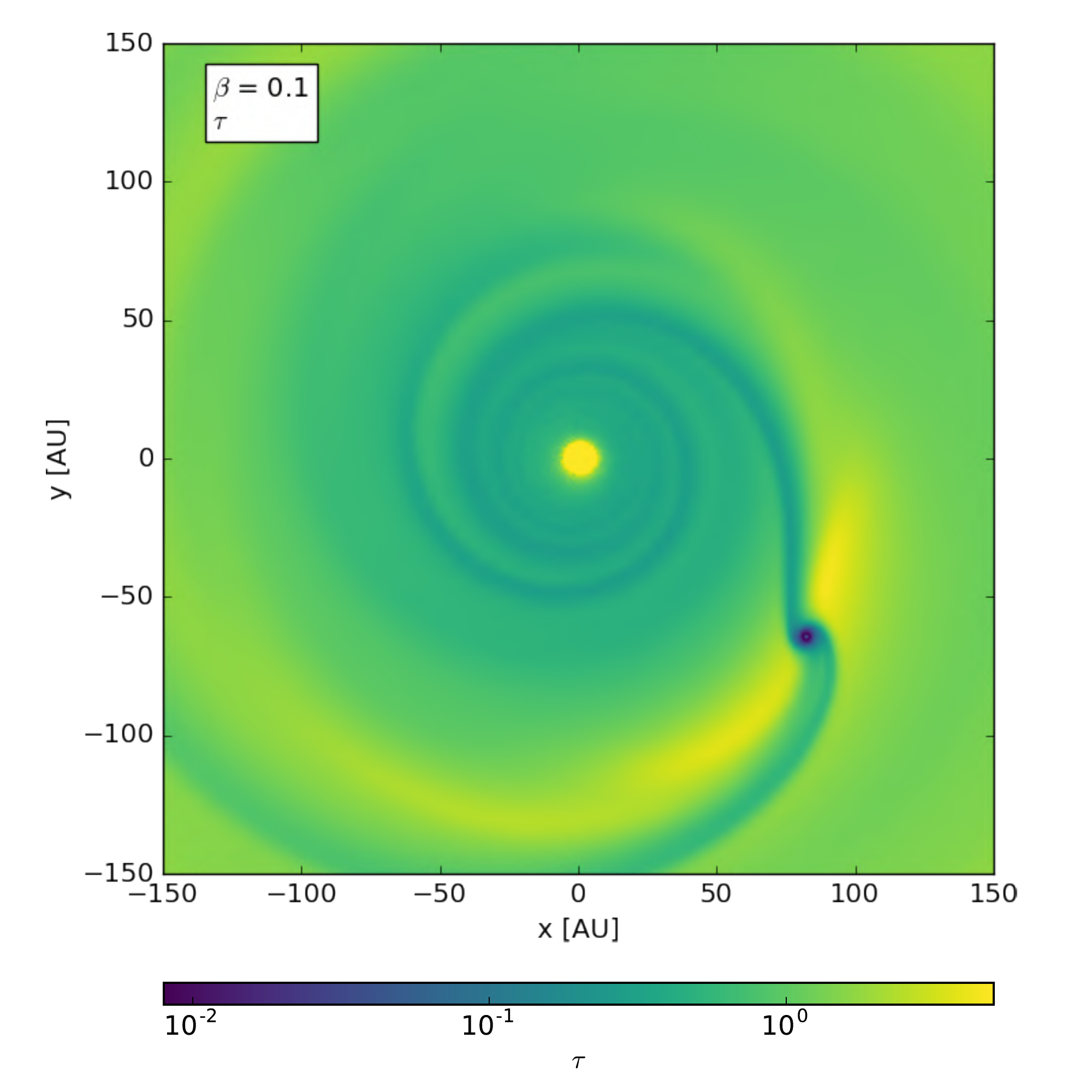}
\includegraphics[width=0.99\columnwidth]{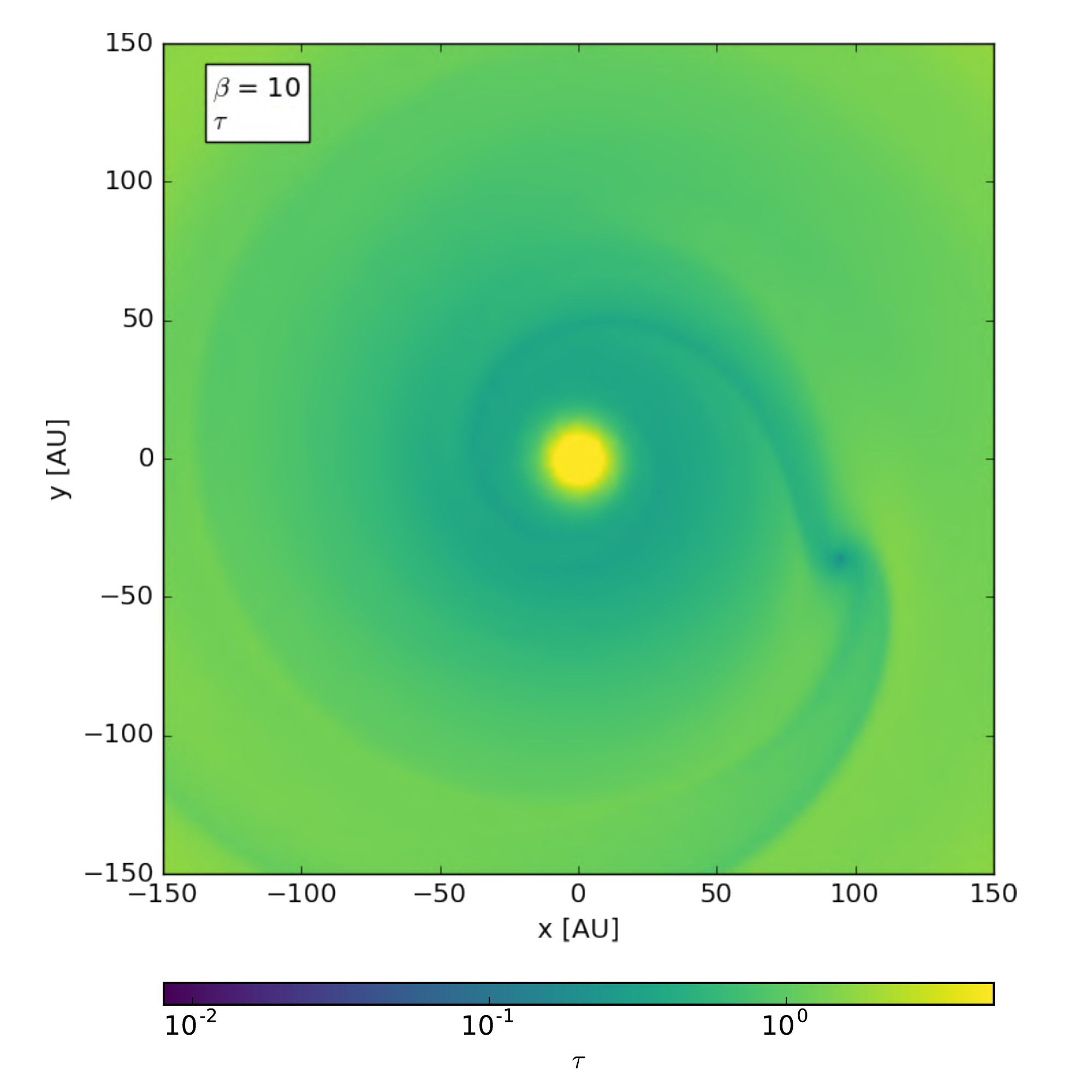}
\caption{Top panels: Edge on temperature plots for $\beta$ = 0.1 (left) and $\beta$ = 10 (right) at $t = 2230$ years. Note the cool accretion disc of gas on the left and the X morphology on the right due to opposing gas flow streams as the inefficiently cooled gas is ejected vertically away from the planet.
Bottom panels: Stokes number ($\tau$) across the whole disc for $\beta$=0.1 \& $\beta$=10 and 1 cm grains at $t = 2230$ years. Note that the perturbation in $\tau$ is much greater for the efficiently cooled disc.
}
\label{fig:SPH_t1_tauQ}
\end{figure*}

In the following section we outline the general evolution of the disc and planet system as well as the gas and pebble dynamics close to the planet. 
We then examine how varying the $\beta$ cooling rate, initial planet mass, grain size of the pebble distribution, and planet radiative feedback affect our simulation results. 
The purpose is to explore how each of these parameters impact the migration timescale and the rate of gas and dust accretion onto the planet. An overview of the varied simulation parameters and the corresponding figures can be seen in Table \ref{table:runs_summary}.
If not otherwise specified, in the following simulations the initial planet mass is set to $2 \mj$, the number of gas particles is 10$^6$, the sink radius is 0.1 AU, the grain size is 1 cm, $Z_0 = 10^{-3}$ and the number of dust particles is set as explained in Section \ref{sec:ic}.

\subsection{Disc evolution overview}
In the first few dynamical times (1/$\Omega_K$ = 160 years at R = 100 AU) the gas and pebble distributions remain close to their initial unperturbed states. After this, a gap is carved in the local pebble distribution under the action of planet torques, migration and accretion. For planets that grow above approximately $4 \mj$, the gas disc around the planet also becomes depleted. We take the simulations for $\beta$ = 0.1 \& 10, with $N = 2$ million SPH particles (plus approximately 1 million dust particles) and pebble size $a = 1$~cm to demonstrate the general features of this evolution.

\subsubsection{The early linear phase}
Figure \ref{fig:SPH_t1} shows the top view of the gas (left panels) and the pebble (right panels) surface densities in the disc for $\beta$ = 0.1 (top panels) and $\beta$ = 10 (bottom panels). These snapshots correspond to a time of $t=2230$~years after the start of the simulation, or around two orbits at 100 AU. The simulations for each $\beta$ are qualitatively similar, in both the planet has migrated from an orbital separation of 120 AU to about 100 AU in just over 2000 years. This corresponds to migration timescale of 12,000 years. The planet drives a spiral arm in the gas disc in both simulations although the density contrast is much higher in the efficiently cooled case. A gap in the gas density of around an order magnitude depth is starting to form, which can be seen in the top left panel.
By contrast, a much deeper $\sim$ 2 order of magnitude gap has been carved in the surface density of the pebbles in both simulations. The feature is a little wider for the efficiently cooled gas but the two cases are similar. 

We see that dust grains are concentrated into the spiral arms as discussed in Section \ref{sec:tau_images}.
It is worth noting that the spiral features in the dust distribution in Figure \ref{fig:SPH_t1} are transient and are entirely due to the presence of the planet.
The central cavity in the gas surface density distribution is caused by the accretion of SPH particles onto the central sink particle. 
As in the case with accretion of gas onto the planet, the sink particle prescription over-estimates the rate of gas accretion onto the sink, so that the gas disc becomes depleted out to $\sim 30$~AU already during its relaxation time. This implies that the simulations are strongly affected by the inner boundary condition when the planet migrates to about 30-40 AU, and hence our focus should be the phase before the planet arrives there \citep[for further discussion of this see Figure 8 and Section 3.2.6 in][]{Nayakshin17a}.

For the efficiently cooled disc in the left hand panels there is a density peak around the planet in both the gas and dust distributions. This is due to hydrostatic pressure support of gas against planet gravity. A compressed `atmosphere' is built around the planet inside the Hill sphere \citep[cf. ][]{Nayakshin17a}. 
The enhanced gas density increases the stopping time for dust particles close to the planet and supports them from sedimenting onto the sink particle. This explains the disparity between sink particle mass and half Hill sphere mass in Figures \ref{fig:res_N} and \ref{fig:res_sink}.
In the inefficiently cooled disc the gas atmosphere of the planet remains hot and relatively diffuse and pebbles are able to sediment onto the planet rapidly once they enter the Hill sphere. 

\subsubsection{The later non-linear phase}
Figure \ref{fig:SPH_t2} shows the gas and dust discs from Figure \ref{fig:SPH_t1} at $t=6800$ years. By this time the planets in both simulations have migrated to R $\sim 75 $~AU. From Figure \ref{fig:res_N} it can be seen that by this time the planet in the $\beta=0.1$ disc has reached a mass of 15 $\mj$ and entered the Brown Dwarf regime. It has opened a deep gap in the gas distribution and has begun to transition into the Type II migration regime. This massive object has almost completely accreted its local dust distribution. Pebbles on wider orbits are trapped in the pressure maximum at the outer edge of the gas gap.

In the $\beta=10$ disc, the planet has only grown to around 2.3 $\mj$ and is not massive enough to open a gap in the gas disc. Provided its mass remains low, this planet will continue to migrate in the Type I regime and will most likely reach the inner disc \citep{BaruteauEtal11}. Despite this, the planet has still been able to open an order of magnitude gap in the local dust distribution. This effect is well known \citep{Paardekooper04} and is possible because 1 cm dust grains are only weakly coupled to the gas disc. Since these grains have no pressure support and are only weakly coupled to the gas, the planet is able to easily perturb their distribution.

\subsubsection{Gas and pebble dynamics close to the planet}
\label{gas_dynamics}
The top two panels of Figure \ref{fig:SPH_t1_zoom_over} show the structure of the gas surface density and pebble distribution around the planet location at the same time as Figure \ref{fig:SPH_t1}. The continuous colours in the figure represent gas properties whereas pebbles are shown as white points. The full and half Hill radii are plotted as solid and dashed white lines respectively. 
In the $\beta$ = 0.1 case a much more compressed gas disc is able to form due to the efficient cooling which acts to trap dust grains in finely structured decaying orbits around the planet. Fine structure inside the Hill sphere is limited by the resolution of our simulations. The ratio of SPH smoothing length $h$ to Hill radius $R_H$ is $\sim$ 0.1 for the $\beta$ = 0.1 case but only $\sim $ 0.5 for $\beta$ = 10. The Stokes number decreases linearly with increasing gas density and so pebbles will become more tightly coupled to the gas. 
This suggests that for Stokes number $\simlt$ 1, the pebble accretion rate for grains of a fixed size will be higher if the gas density around the planet is lower. When the gas density is high, the grains will no longer decouple significantly from the gas and so will follow the flow of the gas field more closely.
It will take higher resolution simulations to reliably model this behaviour close to the planet inside the Hill sphere.

The lower two panels of Figure \ref{fig:SPH_t1_zoom_over} show edge-on density profiles around the planet. Notice that in the efficient cooling case a dense accretion disc has begun to form. Velocity profiles in the gas relative to the motion of the planet are over-plotted in black. Gas rapidly accretes through the poles of the accretion disc in the efficient cooling case but flows past the planet when it is kept hot by inefficient cooling. Negative $x$ is directed towards the star. The flow relative to the planet is caused by the rapid inwards Type I migration.

The top two panels of Figure \ref{fig:SPH_t1_tauQ} show the edge-on projections of the gas temperature in the disc at t = 2230 years and show that unsurprisingly the gas tends to be hotter where gas streams collide. On the length scales of our simulations gas does not exceed several hundred kelvin and so we are justified in neglecting grain sublimation.
In the $\beta$ = 0.1 case in the top left there is a clear difference between gas that has settled into an accretion disc and cooled and gas that is accreting onto the planet from the poles and being heated by shocks. Since the gas cooling time is short this gas is still able to rapidly cool and accrete onto the sink particle.
The $\beta$ = 10 simulations in the top right have an `X' like structure in the hotter gas. 
This is due to gas compression between the disc and the spherical, pressure supported hot gas surrounding the planet.
Gas entering the Hill sphere on the left (from the inner disc) is deflected by the hot gas around the planet and is forced out of the Hill sphere, causing the gas density profile to bulge vertically compared to the disc. Note that temperatures deep inside the Hill sphere reach around 200K. Sub-resolution temperatures in the core of the proto-planet may be even higher, but in the regions we simulate the gas temperature is not high enough for dust sublimation to be important.

\subsubsection{Dimensionless pebble stopping time across the disc}
\label{sec:tau_images}
The two lower panels of Figure \ref{fig:SPH_t1_tauQ} show the Stokes number ($\tau$) across the whole disc (defined in Equation \ref{eq:tau}) for 1 cm grains and for both values of $\beta$. Globally the stopping time is $\sim$ 1, a little less than the orbital time. 
The high density spiral arms induced by the planet act as dust traps for inwardly migrating dust grains. Since grain stopping time is inversely proportional to local gas density grains become tightly bound to the gas in these high density spirals.
This explains the dust streams in the vicinity of the planet in the top two panels of Figure \ref{fig:SPH_t1_zoom_over}. These streams are more pronounced in the $\beta$ = 0.1 case since the stopping time trap is deeper.

In the $\beta$ = 0.1 case the Stokes number for grains close to the planet falls by $\sim$ 2 orders of magnitude inside the half Hill sphere due to the high density circumplanetary gas disc. This causes pebbles to become trapped inside the half Hill sphere as the pebbles become tightly coupled to the gas. 
This effect can be seen in the lower panels of Figure \ref{fig:res_N} and explains why a large mass of dust is trapped inside the half Hill sphere of the planet and not accreted onto the sink particle in the efficiently cooled disc.
Pebble accretion in the $\beta = 10$ case remains efficient since the reduction in stopping time is a little less than an order of magnitude. 
The pebbles remain relatively decoupled from the gas which allows pebble accretion to proceed efficiently.

\subsection{Dependence on cooling efficiency}\label{sec:cooling}
From Figures \ref{fig:res_N} \& \ref{fig:res_sink} it can be seen that the value of $\beta$ in the disk cooling prescription initially has little effect on the inwards migration rate of the planet. After a few orbits, the migration begins to stall in the $\beta$ = 0.1 case due to runaway gas accretion which dramatically increases the planet mass. This causes the planet to open a gap in the gas disk and begin to enter the Type II migration regime. By contrast, in the $\beta$ = 10 case gas accretion is significantly less efficient and the lower mass planet can continue to migrate in the Type I regime. 
By comparison between the two cases, increasing the cooling timescale by a factor of 100 leads to around a factor of 30 decrease in accreted gas mass on a time scale $\sim$ 10$^4$ years. 

Notably, the pebble accretion rate is almost independent of the value of $\beta$. We will later conclude in Section \ref{sec:size} that this is due to pebbles accreting in the Hill regime \citep{LambrechtsJ12} in which essentially all pebbles entering the Hill sphere of the planet via Keplerian shear are captured by it. 
This result implies that planets migrating through inefficient cooled discs may become metal enriched due to pebble accretion. If gas cooling is efficient however, the degree of metal enrichment will be much weaker since gas accretion will dominate. These conclusions depend on a number of additional factors which we shall discuss later.

\subsection{Dependence on initial planet mass}
\label{sec:MP0}
\subsubsection{Migration rate}

\begin{figure}
\includegraphics[width=0.99\columnwidth]{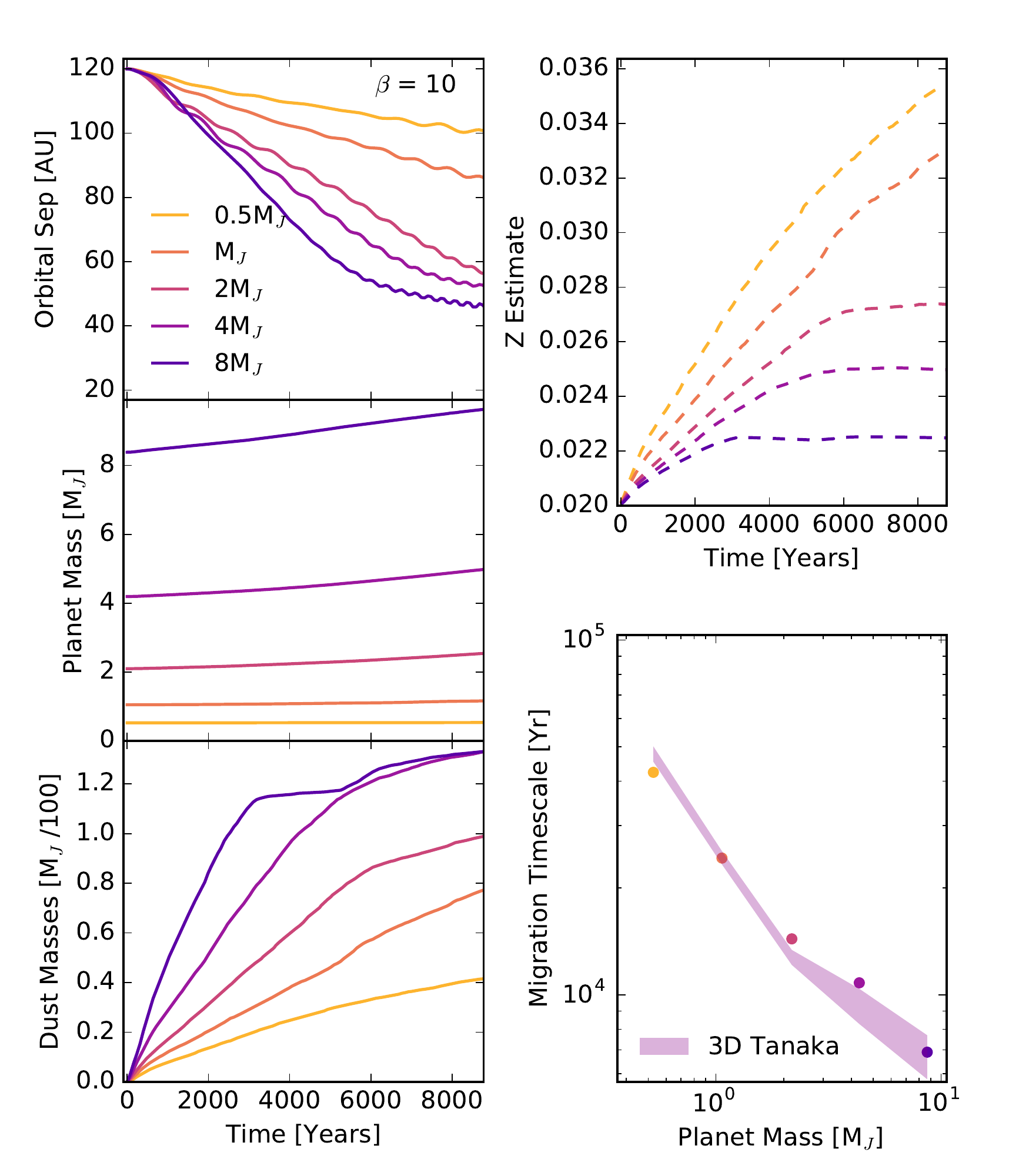}
\caption{Left: same as Figure \ref{fig:res_N}, but only for $\beta = 10$. Coloured curves correspond to different initial planet masses as shown in the legend. Right top: metallicity estimate, calculated with 1 cm grains as 10 \% of metals and assuming the other 90 \% is tightly coupled to the gas. Right bottom: Type I migration timescale from 1000-4000 years. Shaded region represents maximum and minimum analytic timescale \citep{Tanaka02} calculated from disc parameters during migration. Scatter points show the average migration timescale for each planet from our simulations. There is strong agreement, supporting the idea that our planets are migrating in the Type I regime.}
\label{fig:planet_mass0}
\end{figure}

Previous studies have demonstrated that gas giant migration through the outer disc is well described by the Type I planet migration regime \citep{BaruteauEtal11}, provided that the gas giant does not grow large enough to open a gap in the disc. It is well known that there is an underpopulated `desert' of gas giants beyond 20 AU, observations show that the occurrence rate of these planets in dynamically old systems is low, around 1-5 \% \citep{ViganEtal17}.
This suggests that if GI is a reliable planet formation mechanism, migration timescales must be short for a range of physically motivated initial fragment masses in order to efficiently remove planets from the outer disc. In Figure \ref{fig:planet_mass0} we explore how varying the initial planet mass affects migration rates, dust accretion efficiency and planet metallicity over the course of our simulations.

In this set of simulations only the $\beta$ = 10 case is included since for lower $\beta$ rapid gas accretion quickly dominates over the initial planet mass. All planets initially migrate inwards on short migration timescales: $t_{mig} = -R_{P} / \dot{R}_{P}$, where $R_P$ is the orbital separation of the planet from the star.
\cite{Tanaka02} derived an analytic Type I migration timescale for 3D disc planet resonant interactions via Lindblad torques as

\begin{equation}
t_{mig} = (2.7 + 1.1 \sigma)^{-1} \dfrac{M_*^2}{M_P} \dfrac{1}{\Sigma R_P^2 \Omega} \left( \dfrac{c_s}{R_P\Omega} \right)^2,
\label{eq:tanakamig}
\end{equation}

where $\sigma$ is the exponent in the disc surface density profile ($\Sigma \propto R^{-\sigma}; \sigma$ = 1 in our simulations). Whilst this result is valid for gas only discs, we expect it to hold for our simulations since the dust to gas ratio remains low in the bulk of our discs.

We compare this to the results from our simulations in the lower right panel of Figure \ref{fig:planet_mass0}. The scatter points represent the average migration timescales of our planets during the linear migration phase (taken to be between 1000 and 3000 years) whilst the shaded region shows the corresponding timescales calculated from Equation \ref{eq:tanakamig}. 
The spread in the shaded region is caused by the time variability of the local parameters over the course of the migration. We achieve good agreement with the analytic theory which confirms that our planets in the $\beta$ = 10 disc rapidly migrate in the Type I regime. The agreement also implies that our simulations correctly resolve the Lindblad torques acting on our planets. 

For the 4 \& 8 $\mj$ planets, the rate of migration slows after a few thousand years. In Section \ref{sec:mig_discussion} we discuss how this is caused by the simulation inner boundary and artifical gas accretion onto the central star.
Whilst these planets have started to open gaps in their local gas distributions, they continue to migrate during this transition. Ultimately, we expect the eventual end point of migration for each planet to be a strong function of the disc mass and the gas cooling prescription.

In this regard, the results of \cite{MalikEtal15} are very relevant. They find that massive planets may continue to migrate in the Type I regime even after they have grown above the gap opening mass. A commonly used formula for predicting the gap opening mass of planets is $M_T = 2M_* (H/R)^3$ \citep{Bate03}. \cite{CridaEtal06} gave a more refined condition that also takes into account the disc physical viscosity. However, \cite{MalikEtal15} show that this approach may be insufficient for very rapidly migrating planets because there is also the question of how quickly the planet will switch from the Type I to Type II regime. Our results support their conclusion that planets as massive as $8 \mj$ continue to migrate inwardd rapidly, despite formally satisfying the conditions for the transition in the Type II regime. This migration occurs at a rate more consistent with Type I rather than Type II migration, at least until the planets stall at the inner numerical resolution limit of the disc. Until recently, most population synthesis models discounted the gap opening transition timescale and switched to Type II migration immediately after triggering the transition mass criterion \citep[e.g.,][]{ForganRice13b, GalvagniMayer14,NayakshinFletcher15}. \cite{MullerEtal18} show that including a gap opening timescale argument increases the fraction of massive planets that migrate towards the inner disc, this result should be included in future gravitational instability population synthesis studies.

\subsubsection{Metal enrichment}\label{sec:Z_vs_m}

The relative rates of planet pebble and gas accretion is one of our key simulation outputs. In the upper right panel of Figure \ref{fig:planet_mass0} we use this data to estimate the metallicity of each migrating planet, defined as the ratio of the metal mass of the planet to its total mass. In order to make this calculation we take into account both the directly simulated large decoupled pebbles (scaled such that 10\% of metal mass is in pebbles) and also the population of grains too small to be decoupled from the gas. Assuming Solar metallicity for the initial disc and the planet, $Z_0 = Z_{\odot} = 0.02$, we specify that when the planet accretes gas mass $M_{\rm gas}$ it also gains $Z_0 (1 - f_{\rm peb}) \times M_{\rm gas}$ in microscopic dust mass, where $f_{\rm peb} = 0.1$ here is the mass fraction of metals locked in pebbles. The planet metallicity ($Z_{\rm P}$) is then equal to 

\begin{equation}
Z_{\rm P} = \dfrac{Z_0 M_{\rm P0} + M_{\rm peb} + Z_0 (1 - f_{\rm peb}) M_{\rm gas}}{M_{\rm P}}\;,
\end{equation}
where $M_{\rm P0}$ and $M_{\rm P}$ are the initial planet mass and the mass at time $t$ respectively. $M_{\rm peb}$ and $M_{\rm gas}$ are the pebble and gas masses accreted by the planet by time $t$, as plotted in the left panels of Figure \ref{fig:planet_mass0}.

The top right panel of Figure \ref{fig:planet_mass0} shows that planets of lower initial mass become more metal abundant due to pebble accretion compared to their more massive analogs. This trend is expected due to analytical scalings pointed out in Section \ref{sec:analytics}. In fact, all three of the factors discussed in Section \ref{sec:analytics} that lead to greater metal over-abundance by pebble accretion at lower planet masses can be seen in Figure \ref{fig:planet_mass0}.

Concentrating first on the linear part of the simulations, that is, the first few thousand years, we notice from the top right panel of Figure \ref{fig:planet_mass0} that planet metallicity increases more rapidly for lower mass planets. Next, focusing on the orbital separation versus time plot (top left panel), we see that lower mass planets migrate slower thus have more time to accrete pebbles. Finally, the bottom left panel shows that most massive planets may stop accreting pebbles  {\em before} they reach the inner boundary of our gas disc. We shall later see that this is caused by these planets opening gaps in the pebble disc sooner than a gap is opened in the gas disc. In other words, more massive objects not only migrate more rapidly, leaving them less time to collect pebbles, but also open gaps in either pebbles or both gas and pebbles sooner than their less massive analogs, terminating pebble accretion even sooner.

Note that the final metallicity of our objects depends on the structure of the inner disc, which is unfortunately adversely affected by the inner boundary condition we employ for numerical reasons. Nevertheless, it appears that pebble accretion does provide a qualitatively correct trend of negative correlation between metal over-abundance in giant planets versus their mass, as observed by \cite{MillerFortney11} and \cite{ThorngrenEtal15}.

\subsection{Dependence of accretion rate on pebble size}\label{sec:size}

\begin{figure}
\includegraphics[width=0.92\columnwidth]{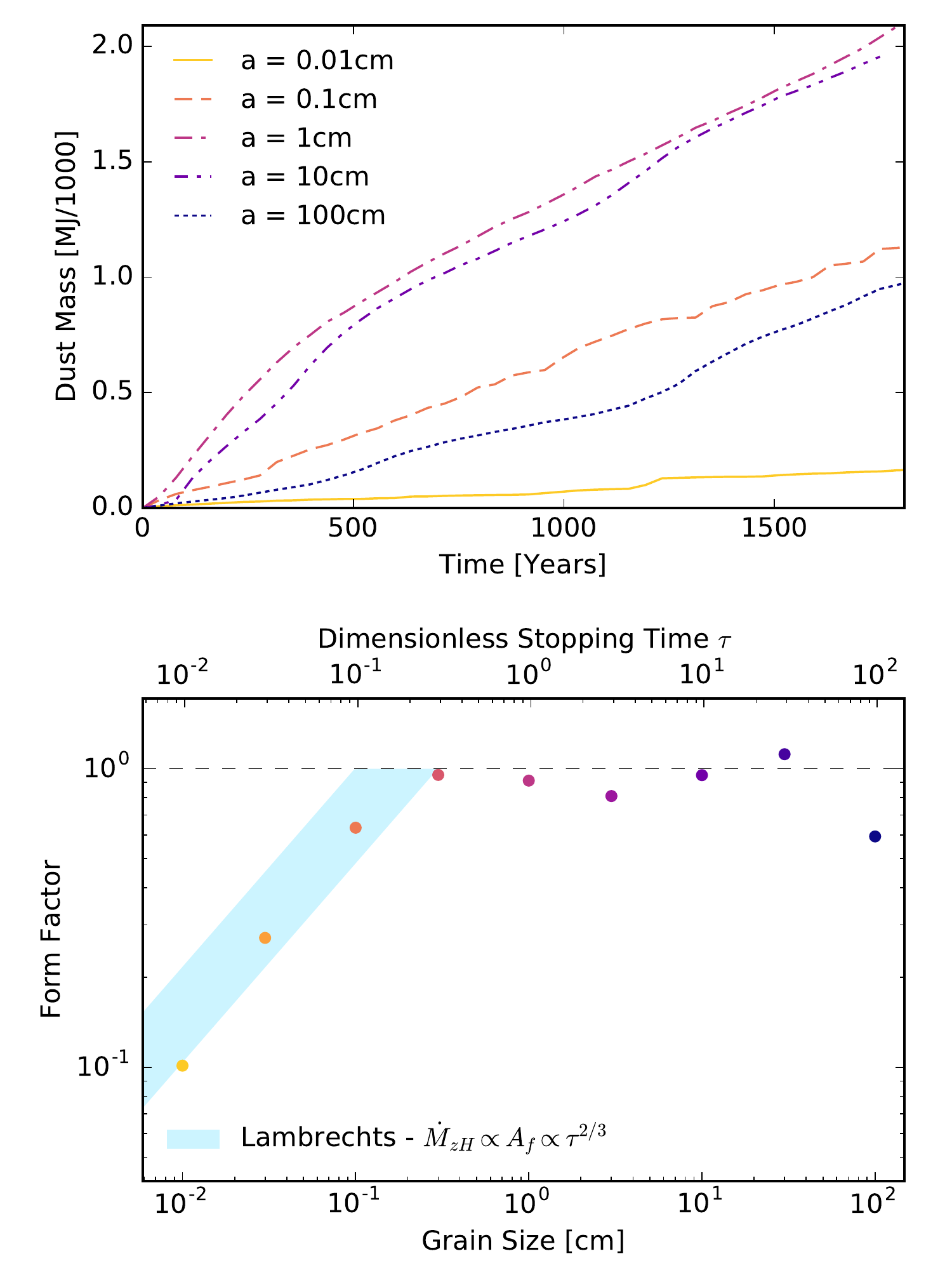}
\caption{Top: Accreted dust mass against time for different pebble sizes with $\beta$ = 10. Bottom: Accretion form factor (Equation \ref{FF0}) as a function of grain size (bottom) and Stokes number $\tau$ (top). Shaded in blue is the relation $\dot{M}_H \propto \tau^{2/3}$ expected by \protect\cite{LambrechtsJ12}. The region is set such that $A_{\rm f} =1$ accretion occurs at between $\tau=0.1$ (indicated by Lambrechts et al.) and $\tau=0.3$ (as found from our simulations). Thus, planets efficiently accrete pebbles in the range of stopping times $\tau$=[0.1-100].}
\label{fig:FF0}
\end{figure}

\begin{figure}
\includegraphics[width=0.99\columnwidth]{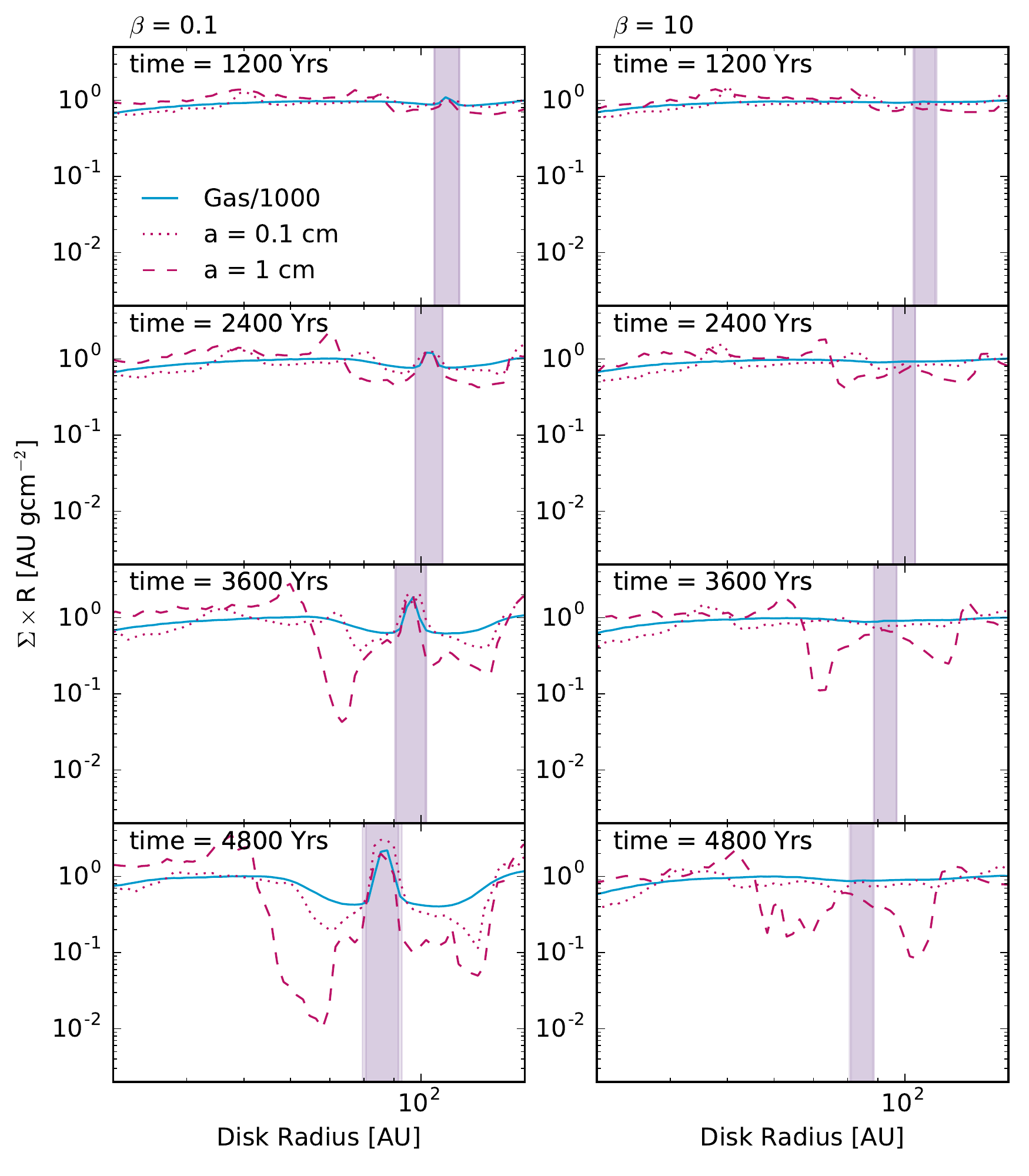}
\caption{These plots show how $\Sigma \times R$ against $R$ varies as planets migrate through the disc. The vertical shaded regions show the extent of the half Hill Radius of the planet. The solid and dashed profiles correspond to the azimuthally averaged gas and dust surface densities respectively. The two dust profiles are taken from separate simulations and overlaid for comparison. Vertical dashed lines represent the centre of mass of the dust over the range 30-150 AU.  Note how in the efficient cooling regime the planet is able to carve a much larger gap in the gas profile.}
\label{fig:grain_size_sigma}
\end{figure}

The simulations presented so far were carried out assuming a fixed grain size, $a=1$~cm. It is currently not quite clear how large grains can actually grow when large speed collisions, eroding larger grains into smaller grains, are taken into account \citep{DD05,Laibe2008, LP14, DrazkowskaEtal14,BoothClarke16}. It is hence worthwhile to explore how our results depend on the pebble size assumed.

Figure \ref{fig:FF0} shows how pebble accretion in the initial linear phase depends on the pebble size in our $\beta$ = 10 disc simulations. The top panel shows the total mass in pebbles ($M_{\rm z}$) accreted by the planet as a function of time for different grain sizes, from $a= 0.01$ cm to $a= 100$ cm. The panel below that shows a normalised pebble accretion rate as a function of the pebble grain size on the bottom, and also as a function of the Stokes number, $\tau$, on the top scale.

The dimensionless measure of the pebble accretion rate is defined through an `accretion form factor', $A_{\rm f}(a)$, to represent the efficiency with which pebbles entering the Hill sphere through Keplerian shear are accreted onto the planet. The pebble accretion rate measured in the linear regime from the simulation is defined as $\dot M_{\rm peb} = (M_{\rm z}(t_2) - M_{\rm z}(t_1))/(t_2 - t_1)$, where $t_1$, $t_2 = $ 500 and 1500 years respectively. The theoretically expected pebble accretion rate is found by considering the rate at which pebbles enter the Hill sphere through Keplerian shear ($\dot M_{\rm zH}$) is given analytically by Equation 38 in \cite{LambrechtsJ12} as
\begin{equation}
\dot M_{\rm peb, exp} = 2 R_{\rm H}^2 \Omega_{\rm K} \Sigma_{\rm peb}.
\label{eq:hill_rate}
\end{equation}
The accretion form factor is then defined as 
\begin{equation}
A_{\rm f} = \dfrac{\dot M_{\rm peb}}{\dot M_{\rm peb, exp}\;}.
\label{FF0}
\end{equation}
From the bottom panel of Fig. \ref{fig:FF0}, it appears that $A_{\rm f}(a) \approx 1$ for a broad range of Stokes numbers, $0.1 \simlt \tau \simlt 30$, which implies that the Hill accretion rate is a very good approximation to the pebble accretion rate in this `Hill regime' \citep[see][]{LambrechtsJ12}. 

Figure \ref{fig:FF0} demonstrates the well known result that pebbles with Stokes numbers of $\sim$ 1 are most efficiently accreted from the disc \citep{LambrechtsJ12}. This is due to a balance between their frictional stopping times and their Hill sphere crossing timescale. 
Grains with $\tau <<$  1 experience large frictional drag and are unable to decouple from the gas flow and enhance pebble accretion. They remain tightly bound to the gas and are swept past by Keplerian shear if the gas remains gravitationally unbound to the planet. For grains with $\tau >>$ 1 gas drag is insufficient to damp the relative Keplerian shear so they are unable to lose kinetic energy rapidly enough to become bound to the planet.
It is worth noting that $A_{\rm f}(a)$ never significantly increases above one. The planet is unable to capture grains from beyond the radius of the Hill sphere since at these separations Keplerian shear dominates over the local gravitational effect of the planet. This is not a surprise since the Hill sphere is defined as the region within which the gravitational influence of the planet dominates over the central object.

The plot is slightly misleading since the Stokes numbers is calculated for pebbles in the bulk of the disc. Inside the Hill sphere, gas density increases due to accretion onto the planet which acts to decrease the Stokes number of grains that enter this region. This explains why pebbles with Stokes number $\gg$ 1 seem to be accreted efficiently by the planet, a finding somewhat in disagreement with \cite{LambrechtsJ12}. 
\cite{LambrechtsJ12} use a constant gas density model which is appropriate for pebble accretion in the planetesimal regime. 
By contrast to \cite{LambrechtsJ12} we model pebbles with fixed physical sizes rather than fixed Stokes numbers. This allows us to self-consistently explore pebble dynamics in environments where the local gas density changes by orders of magnitude.
Higher gas density in our runs increases the frictional gas drag inside the Hill sphere and allows for more efficient capture of larger pebbles.

We caution that these results cover a short time (about one orbital time at the location of the planet) from $t=0$, during which the disc remains close to its initial unperturbed state. During later evolution the planet will sculpt both the gas and the pebble distributions in the disc and hence we should expect that in the non-linear regime there may be significant deviations of $\dot M_{\rm z}$ from the Hill accretion rate. We only expect this high pebble accretion regime to continue whilst the planet remains in the rapid Type I migration regime where it is constantly resupplied with pebbles during its migration.

\subsection{Dependence on pebble fraction}
\begin{figure}
\begin{centering}
\includegraphics[width=0.49\columnwidth]{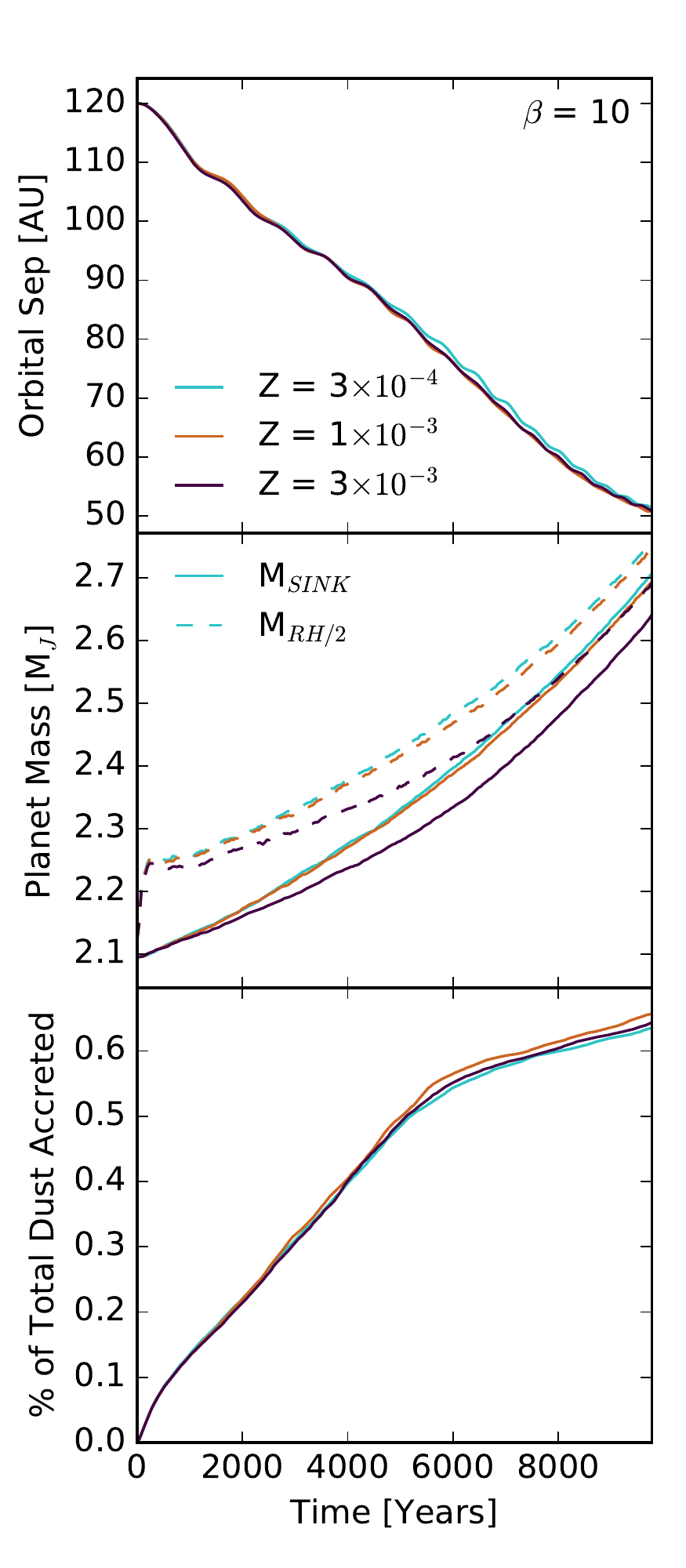}
\caption{Same as Figure \ref{fig:res_N} for $M_{P0} = 2 M_J$. This plot demonstrates that varying the initial pebble fraction has little impact on the large scale migration or gas accretion history of the planet. In the lower panel we have plotted the accreted pebble mass as a percentage of the total initial dust mass. This demonstrates that accreted dust mass increases linearly with pebble fraction in this regime.}
\label{fig:f_peb}
\end{centering}
\end{figure}

So far we have assumed that the fraction of metals locked into Stokes number $\sim$ 1 pebbles is 10 \% ($f_{peb}$=0.1).
We now choose to vary that parameter. In Figure \ref{fig:f_peb} we set f$_{peb}$ = 0.03, 0.1 \& 0.3 in order to explore how back-reaction from the pebbles onto the gas affects the accretion rates.

From the top panel of Figure \ref{fig:f_peb} we can see that the planet migration timescale remains unaffected by the variation in pebble fraction. This is unsurprising since planetary migration depends on the interaction of the planet with the local disc, which is dominated by the gas mass. Varying $f_{peb}$ will only lead to variations in the disc mass of fractions of 1\%.

Increasing the fraction of metals in pebbles acts to slightly decrease the planet mass. This may be due to back-reaction from the pebble distribution causing frictional heating of the gas disc. With more mass locked in pebbles, this back-reaction has a larger heating effect on the local gas, which remains hotter and therefore is prevented from accreting onto the planet. The effect is noticable but is small compared to the change in planet mass caused by altering the $\beta$ cooling factor.

In the lower panel of Figure \ref{fig:f_peb} we plot the percentage of the initial pebble disc mass that has been accreted onto the planet. These units were chosen in order to demonstrate that the rate of pebble accretion onto these large planets is linear with pebble fraction. In the parameter space of pebble fractions that we study here, the effect of back-reaction from the pebbles on the gas remains relatively small.

\subsection{Gaps opened by planets in pebble and gas discs}\label{sec:gaps}

Figure \ref{fig:grain_size_sigma} plots $\Sigma \times R$, where $R$ is in units of AU, against $R$ for both gas and pebbles at several different times. In the unperturbed disc these profiles are independent of radius since the initial disk surface density is set as $\Sigma \propto 1/R$. The gas surface density was divided by 1000 to compare with the pebble surface density on the same scale.
For the simulations presented in the figure, two pebble sizes are considered, $a = 1$ and $a=0.1$ cm, as well as two values of the $\beta$ cooling parameter. The vertical solid purple bar shows the extent of the half Hill sphere of the planet. 

Consider first the gas surface density profile. Checking back with Figure \ref{fig:res_N}, we see that the total mass within half Hill sphere grows from $2\mj$ to about $15 \mj$ by time $t=4800$~years for the $\beta = 0.1$ simulation. For $\beta=10$ this growth is much more modest, only to about $2.5\mj$. 
Due to this large difference in the planet mass, the $\beta=0.1$ planet starts to open a gap in the gas disc by the time of the last (bottom) snapshot shown in Figure \ref{fig:grain_size_sigma} while the $\beta=10$ planet is not able to affect the gas disc in any significant way. 

Note that the strong peak in the gas distribution inside the Hill radius around the $\beta=0.1$ planet is due to the circumplanetary accretion disc. As gaseous material is depleted close to the planet the Lindblad torques weaken and the migration rate of the planet transitions into the Type II regime.

Our planets are able to rapidly open deep gaps in the dust disc during their migration.
Additionally, these gaps are deeper in the 1 cm pebbles by comparison to the 1 mm pebbles as the former are less tightly coupled to the gas disc than the latter.
The gaps in the pebbles are however much weaker in the $\beta = 10$ case, which means that they may continue to accrete pebbles. This is due to the large difference in planet mass between $\beta =0.1$ and $\beta=10$ simulations. High mass planets cut themselves off from the pebble disc due to stronger gravitational torques that they impose on their surroundings. These results support our calculations in Section \ref{sec:analytics} that lower mass planets are more efficient at accreting pebbles.

\subsection{Including radiative feedback from the planet}
\begin{figure}
\includegraphics[width=0.99\columnwidth]{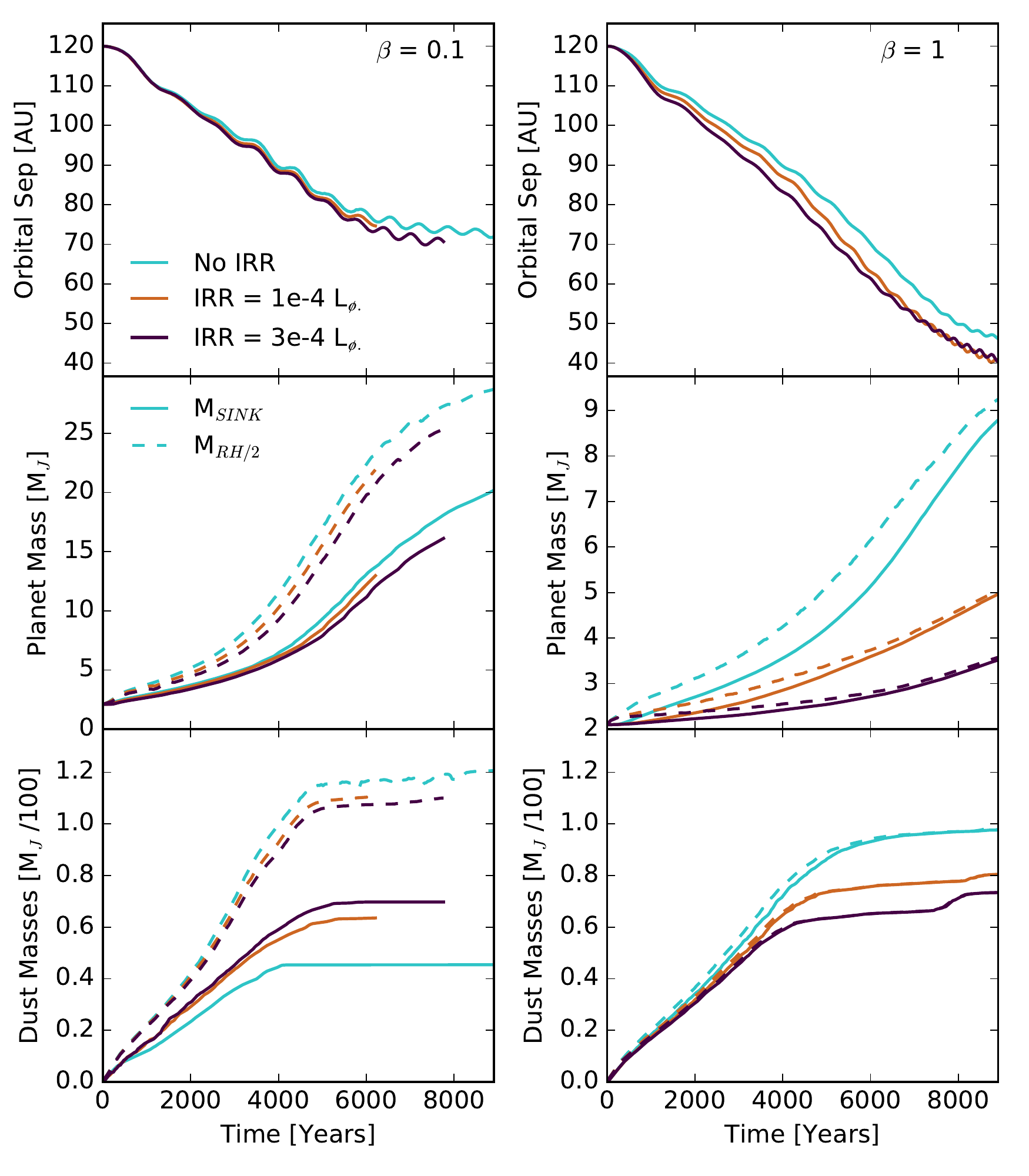}
\caption{Same as Figure \ref{fig:res_N} for $M_{P0} = 2 M_J$. The curves represent the case with no radiative feedback and with feedback = $10^{-4} \Lsun$ \& $3 \times 10^{-4} \Lsun$. Note that the right hand column now corresponds to $\beta = 1$. In the $\beta = 1$ case, including radiative feedback significantly limits gas accretion.}
\label{fig:irr_feedback}
\end{figure}
 
Considering the large impact that the $\beta$ cooling parameter has had on our results, it is clear that accurately modelling the local gas temperature around a migrating planet is crucial to understanding its accretion history. In light of this, we now investigate how radiative feedback from the planet modifies the results of our simulations. 

A proper model of radiative feedback would include 3D radiative transfer calculations which is beyond the scope of our paper. Instead we use a simple prescription for the effects of planetary feedback as described in \cite{NayakshinCha13}. In this scenario, the planet luminosity is given by its contraction luminosity (see below). Note that accretion luminosity of the planet in this scenario is much smaller than the contraction luminosity because the planet radius is very large compared to Core Accretion planets considered by \cite{Stamatellos15}. 
The exact value for planet luminosity, $L_{\rm fb}$, depends on the evolution of the internal structure of the planet, which is not modelled in our simulations. We explore two values of feedback luminosity, $L_{\rm fb} = 10^{-4} \Lsun$ \& $3 \times 10^{-4} \Lsun$, which are of a similar order to those obtained by \cite{HelledEtal11} for pre-collapse gas giants in fully self-consistent stellar evolution calculations of planet contraction. In this regime the radius of the planet is $\sim 10^3-10^4 R_J$.

The feedback onto the gas surrounding the planet is implemented by introducing the minimum gas temperature $T_{\rm min}(r)$, where $r$ is the distance to the planet. If gas temperature $T$ falls below $T_{\rm min}$ then we reset it to $T_{\rm min}$. The planet feedback is usually important only at distances $r \simlt R_{\rm H}$ from the planet. We acknowledge that our feedback prescription is a simple scheme though the benefits of developing a more computationally expensive solution are limited whilst there are such large uncertainties in the local gas cooling rates.

Figure \ref{fig:irr_feedback} demonstrates the impact of radiative feedback. We chose two low values of the $\beta$ cooling parameter for these tests, $\beta = 0.1$ and $\beta = 1$ as physically interesting. Radiative feedback for a $\beta \gg 1$ cases is of a very minor importance as the gas is already hot around the planet location. From the middle panels of Figure \ref{fig:irr_feedback} it can be seen that radiative feedback acts to suppress gas accretion onto the planet, as is known from the previous work. However, the effect is rather meagre for the $\beta = 0.1$ case; the accreted gas and dust masses vary by only about 10\% for the cases with and without feedback from the planet.

This effect is much more dramatic in the $\beta$ = 1 case, decreasing the gas accretion rate onto the planet by about an order of magnitude in the high luminosity case. 
The rate of pebble accretion is not strongly affected by the slowdown in gas accretion, pebbles are able to decouple from the hot gas flow and accrete onto the planet.

A general conclusion from these numerical experiments with radiative feedback from the planet is that the extra heating from the planet may take a moderately low $\beta$-parameter case into the realm of an effective high $\beta \simgt 10$ case. 
When the combination of cooling timescale and radiative feedback prevent gas around the planet from cooling, the gas accretion rate is reduced by the pebble accretion rate is affected less strongly.
Thus, gas giant planets formed by gravitational disc instability may be efficiently metal enriched even if $\beta\sim 1$ when the radiative feedback from the contracting planet is taken into account. This is crucial as in order to make a gas giant planet by disc fragmentation, the disc cooling parameter $\beta$ may have to be as small as $\sim 3$ \citep{DengEtal17}.

\section{Discussion}\label{sec:discussion}

In this paper we performed 3D hydrodynamical simulations, modelling accretion of large grains -- pebbles -- onto gas giant planets and brown dwarf mass objects embedded into large scale massive gas discs around Solar type stars. 
Pebble accretion alters chemical composition of the objects, and this may then be used to distinguish between theories of gas giant and brown dwarf formation. 

\subsection{Main results}\label{sec:main}

Recent work \citep{Nayakshin17a} indicated an interesting dichotomy in the fate of gas clumps embedded into massive protoplanetary discs at separations of tens to $\gtrsim 100$~AU. Clumps embedded into inefficiently cooling discs, i.e., $\beta\gtrsim 10$ in terms of the dimensionless cooling time, tend to migrate towards the inner $\sim 10$AU disc at nearly a constant gas mass on timescales as short as $10^4$ years. In the opposite limit when the disc cools rapidly, i.e., $\beta \lesssim 1$, there is a runaway gas accretion onto the clumps, with clumps quickly becoming massive brown dwarfs or even low mass stellar companions. These much more massive objects open deep gaps in their discs and migrate in the much slower Type II regime. 

In contrast to gas accretion, our simulations show that pebble accretion is hardly influenced by how quickly the gas cools. We find that pebbles accrete onto the embedded objects close to the expected maximum `Hill rate' (Equation \ref{eq:hill_rate}) provided that their corresponding Stokes numbers are greater than about 0.1. This corresponds to grains of size about 0.1-1 cm for reasonable disc parameters. Pebble accretion is however strongly suppressed when the object opens a deep gap in the disc.

The rate of pebble accretion found in our simulations is physically large and may produce planets that are significantly over-abundant in their metal composition compared to their host disc and star. 
This result depends on the fraction of total metal mass locked into pebbles in the disc (in Figure \ref{fig:planet_mass0} we take this value to be 10\%).

Furthermore, while pebble accretion is not very sensitive to the gas cooling rate, it is strongly dependent on the planet mass and this yields observationally testable predictions. In particular, both simulations (Section \ref{sec:Z_vs_m}) and simple analytical arguments (Section \ref{sec:analytics}) show that metal enrichment via pebble accretion should anti-correlate with the planet mass.

Finally, we considered what happens with clumps embedded in rapidly cooling discs, $\beta \le 1$, when the radiative feedback from the contracting object is taken into account. It was found that for very rapid cooling, $\beta = 0.1$, radiative feedback has a negligible impact on our results since it is efficiently reprocessed through the disc via rapid gas cooling.  
For $\beta$ = 1 discs however, our radiative feedback prescription strongly suppresses gas accretion onto the planet. This shows that realistic self-gravitating discs, which are expected to have moderate values of the cooling parameter, $3\le \beta \lesssim 10$ \citep{Gammie01,Rice05,MeruBate11a,DengEtal17}, may nevertheless be in the effectively inefficient cooling regime due to planet feedback on the surrounding gas.

\subsection{Observational connection}\label{sec:obs}

\subsubsection{Current observations and theoretical ideas}\label{sec:cur_obs}

Planet bulk metallicity determines how dense the planet is for a given mass. Using appropriate stellar evolution modelling, it is therefore possible to invert the observed planet mass-radius data to infer the planet metal budget \citep{Guillot05}, although there are some uncertainties to do with whether the metals are all in a solid core of the planet or uniformly distributed in the envelope \citep{MillerFortney11}.

Nevertheless, observations show convincingly that the overabundance of metals in giant planets with respect to their host stars  decreases with increasing mass of the planet, approximately as $Z_{\rm P}/Z_* \propto M_{\rm p}^{-0.45 \pm 0.09}$ \citep{ThorngrenEtal15}. This result can be interpreted in terms of the Core Accretion theory which predicts a similar but slightly steeper relation \citep{MordasiniEtal14b}. Also, \cite{ThorngrenEtal15} find that a model with a constant Toomre parameter, $Q$, for the disc at the planet growth location could explain the planet metal over-abundance if the planet accretes all the metals within its `feeding' zone defined by a few Hill radii. 

The high rates of dust accretion in our simulations lead to gaps in the dust distribution (and gas distribution for low $\beta$) which appear similar to gaps observed in Class I \& II discs with ALMA \citep{BroganEtal15}. These gaps were analysed and fitted extensively by \cite{DipierroEtal15,DipierroEtal16a,DipierroEtal17}, but in the case for a low mass disc. These authors ran longer time scale simulations in order to make predictions of gap width based on planet mass. 
Although our simulations may be visually similar, in practice they are different since our planets experience very rapid migration, faster than the radial drift velocity of even Stokes number $\sim$1 grains. 
The early gap features we observe would vary on orbital timescales as the planet migrates through the dust population. If these perturbations in the dust field exist due to migrating GI planets, they may be seen in ALMA observations of early systems. Further high resolution simulations will be needed in order to determine the longevity of these features as well as their relative widths and dependence on grain size.

\subsubsection{Implications of our simulations}

The results of this paper suggest that planets and brown dwarfs created by gravitational fragmentation of massive cold discs at tens to hundreds of AU may provide an alternative explanation for the observed metallicity over-abundance of gas giants \citep{ThorngrenEtal15}. Here we did not model planet migration inwards of about 30 AU due to numerical limitations. From previous work, however, we know that planets and brown dwarfs may continue to migrate down to the sub-AU region and even perish in the star \citep[e.g.,][]{VB06,BoleyEtal10,NayakshinLodato12,GalvagniMayer14,NayakshinFletcher15}. They can also be scattered into the sub $\sim 1$~AU region by N-body interactions and then circularise there by tidal interactions with the star \citep{RiceEtal15}. Hence the planets we study here at tens of AU may well be relevant for the planets observed at $\sim 0.1$~AU.

In Section \ref{sec:Z_vs_m} we found that planets with lower initial masses became more metal overabundant during their migration phase by comparison to planets with higher initial masses. This is expected based on simple analytical arguments spelled out in Section \ref{sec:analytics}.
In brief, in the Type I regime lower mass planets migrate at a slower rate and are therefore able to more efficiently clear the pebble distribution at each annuli while they migrate inwards. Secondly, they also remain the Type I regime for longer. Higher mass planets open deep gaps in the pebble and gas discs sooner, leaving the Type I regime, and hence stop accreting pebbles. From our discussion of the accretion form factor in Section \ref{sec:dust_discussion} we can see that gravitational instability planets will become very metal overabundant, but only if a large fraction of the disc metal mass is in the size range $\sim [0.1-100]$  cm. In our simulations we have taken the fraction of metal mass in large grains to be 10\%. In order to reproduce the giant planet metallicities found by \cite{ThorngrenEtal15} this fraction would need to be in the range 30-50\%.

These conclusions are however qualitative. While the downward trend in the pebble enrichment of planets with mass $M_{\rm p}$ is clear from our simulations, much more work is necessary to determine the exact dependence of $Z_{\rm P}$ on planet mass and other parameters, such as mass of the star, $M_*$. 
If GI planets form frequently, efficiently migrate through the disc and survive inside 10 AU then they must produce a population of planets that agree with observational data for gas giant planets in this regime.

Previous work in this field has had some success, population synthesis of gravitational instability planets by \cite{NayakshinFletcher15} (see their Figure 17) reproduced the observed correlations in the earlier sample by \cite{MillerFortney11}. The work presented here justifies the Hill capture rate formula for pebble accretion, which was assumed by \cite{NayakshinFletcher15}, at least in the linear regime.
We also note that gravitational instability planets can also be enhanced in metals at birth \citep{BoleyDurisen10, BoleyEtal11a} and later by stripping the outer layers of the envelope \citep{BoleyEtal10,Nayakshin10c,HelledEtalPP62014}. These effects have not been taken into account here.

\subsection{Caveats and uncertainties}\label{sec:caveats}

While we are confident that our main results are robust, there are a range of physical uncertainties in the problem that make it difficult to make a direct comparison of the model with observations.

\subsubsection{Rapid Type I migration}
\label{sec:mig_discussion}

Migration rates and formation frequency via GI are not independent, both depend heavily on disc mass. It is therefore important to consider the interplay between these two phenomena.
Investigating disc fragmentation without a reliable migration model will lead to an overestimated number of planets left in the outer disc whilst studying migration without gravitational instability leaves large uncertainties in the choice of initial disc profile and planet  mass. 
Our model makes this second simplification by assuming that a gravitationally bound planet has already formed. This approach allows us to make progress towards understanding migration at higher resolution, but formation and migration theories must be combined in order to build towards an accurate population synthesis of GI planets.

Our choice of a laminar gas disc ensures that the planet has an initially gas rich local environment. 
This may lead to rapid early migration; Type I migration operates efficiently when there are high local gas densities at the location of the Lindblad torques.
It is reasonable to question whether planets forming through a natural fragmentation mechanism would satisfy this condition as their local environments would by necessity be less homogeneous. On the other hand, \cite{BaruteauEtal11} found that disc fragmentation stochastically enhances migration rates. Regardless, fragmentation will certainly lead to divergence from the analytic Type I migration rates.

In several of our runs planet migration stalls at around 40 AU in the disc. This is most notable in Figure \ref{fig:planet_mass0} for the 4 \& 8 $M_J$ planets and is caused by low numerical resolution in the inner disc. Our effective computational inner boundary moves outwards during the simulations due to the relaxation of the gas disc surface density through unphysical accretion of gas particles onto the stellar sink particle. This effect is discussed at length in Section 3.2 of \cite{Nayakshin17a} for gas only simulations in which he varied the sink radius of the star and found that this directly altered the inner stalling location of migrating planets.
We have therefore decided to limit discussion in this paper to the Type I phase of the migration. We do not dispute the findings of \cite{BaruteauEtal11} who demonstrate rapid and efficient migration from 100 to 10 AU for the 5 $M_J$ planets in their simulations.

\subsubsection{The effect of gas cooling on accretion and migration}

In this paper we have used the $\beta$ cooling prescription which is a well researched but ultimately simple approximation of cooling in protoplanetary discs. 
More realistic cooling prescriptions are limited by large (order unity) uncertainties in opacity due to a lack of information about grain size and growth rates in protoplanetary dust discs \citep{SemenovEtal03}.  
By varying the $\beta$ parameter we have shown that differences in gas cooling prescriptions may lead to wildly different gas accretion histories and migration paths. This highlights the importance of treating gas cooling properly when conducting global simulations of migrating gravitational instability gas giants. 

In addition to the global assumptions about cooling rates made by the $\beta$ prescription, our models also discount changes in cooling that would take place inside the extended planetary atmosphere. Enhanced metallicity inside the planet may increase the cooling timescale, whilst efficient grain sedimentation may in fact decrease the cooling timescale by preferentially transporting metals deep into the planet core \citep{HelledEtalPP62014}. Self-consistently modelling dense planetary atmospheres is computationally expensive but will be crucial for accurately determining gas accretion rates onto planets in future simulations. 

Despite these possible improvements, it is important to consider the usefulness of such computationally expensive efforts given the degeneracies involved. There is limited benefit to improving local and global cooling prescriptions whilst large uncertainties remain in the rates of planetary radiative feedback and in the size and mass distribution of grains throughout the disc.

\subsubsection{The negative planet mass-metallicity correlation}
\label{sec:M-Z}
\label{sec:dust_discussion}
We have examined how developments in the pebble accretion field may lead to metal enrichment during the migration of GI planets through the outer disc. The timescale for metallicity enhancement is similar to the migration timescale in our simulations, and so it is important to consider how the initial dust distribution might affect grain accretion.
In this work we have considered a simple laminar dust distribution in order to isolate the effects of gas cooling on the dust accretion. In reality, fragments in GI discs form in high surface density spiral waves. Dust is likely to be concentrated in these spiral structures and especially in overdense fragmenting clumps at around a factor two density enhancement \citep{BoleyDurisen10}. 

The dust distribution may also have been perturbed and depleted by the migration of previous GI planets which may lead to the depletion of grains available for capture during the early migration phase. Additionally, there are large uncertainties involved in the distribution of dust mass as a function of grain size as well as in the replenishment rates of the large dust grains that we consider in these simulations \citep{DD05}. It is hard to guess how more complex dust distributions would affect pebble accretion rates for migrating planets, this poses an interesting question for future lines of research.

We have also neglected further evolution of the planet metallicity after the Type I migration phase. Later radial migration of dust grains could enrich these planets during subsequent evolution of the disc. Tidal stripping of planet atmospheres may also enhance the metallicity of these planets once they migrate to the inner disc \citep{Nayakshin17a}. Accreted grains are likely to sediment towards the planet core and so atmosphere stripping will preferentially remove lower metallicity material from the planet surface \citep{HelledEtalPP62014}. Combining all of these physical mechanisms into a unified fragmentation model with a realistic dust size distribution will be necessary in order to discover whether GI simulations can accurately predict the negative exponent in the mass-metallicity power law.

\section{Conclusions}

In this paper we have used 3D SPH simulations to examine the relative rates of gas and dust accretion onto gas giants migrating through the outer disc.
We find that order Jupiter mass planets in massive discs migrate from 100 AU in the Type I regime in $\sim$ $10^4$ years with migration timescales inversely proportional to their masses. This is in agreement with \cite{BaruteauEtal11} who state that `fast inwards migration should be a generic expectation for planets formed by gravitational instability'. 

In our simulations inefficient gas cooling rates and radiative feedback from migrating planets acted degenerately to suppress gas accretion. We examined two extreme cooling rates of $\beta$ = 0.1 \& $\beta$ = 10 as well as a range of physically motivated feedback luminosities. An irradiative temperature floor from the central star prevented fragmentation in our discs.
In the efficient gas cooling case our planets entered a runaway accretion phase and rapidly grew into brown dwarfs. When we considered inefficiently cooled discs and included planetary feedback the rate of gas accretion was suppressed and our planets remained close to their initial $\sim M_J$ masses.

We found that dust grains with Stokes numbers $\sim$ [0.1-100] were efficiently accreted even when gas accretion was strongly suppressed. These accreted grains were able to decouple from gas that flowed into the Hill sphere of our planets even when the gas streams remained too hot to become gravitationally bound to the planet. This decoupling enhanced the metallicity of our planets. We also found that this metallicity enhancement was anti-correlated with planet mass; lower mass planets became more metal overabundant by comparison to higher mass planets.
This result may allow GI planets to reproduce the negative mass-metallicity correlation for gas giants observed in the inner disc. These findings are preliminary and will need to be extended in order to discover whether GI planets are able to correctly predict the metallicity enhancement of hot and warm Jupiters. 
The degree of metallicity enhancement depends on the initial distribution of metals across different grain sizes and on the spatial distribution of metals in gravitationally unstable proto-planetary discs. We leave further study of this metallicity enhancement for future work.

\section{Acknowledgements}
We acknowledge support from STFC grants ST/K001000/1 and ST/N504117/1, as well as the ALICE High Performance Computing Facility at the University of Leicester, and the STFC DiRAC HPC Facility (grant ST/H00856X/1 and ST/K000373/1). DiRAC is part of the National E-Infrastructure. We are also grateful to the anonymous referee for their insightful comments on the paper, specifically concerning dust and gas coupling tests.

\section*{Appendix}
\renewcommand{\thesubsection}{\Alph{subsection}}

\subsection{SPH smoothing length resolution}
\begin{figure}
\includegraphics[width=0.99\columnwidth]{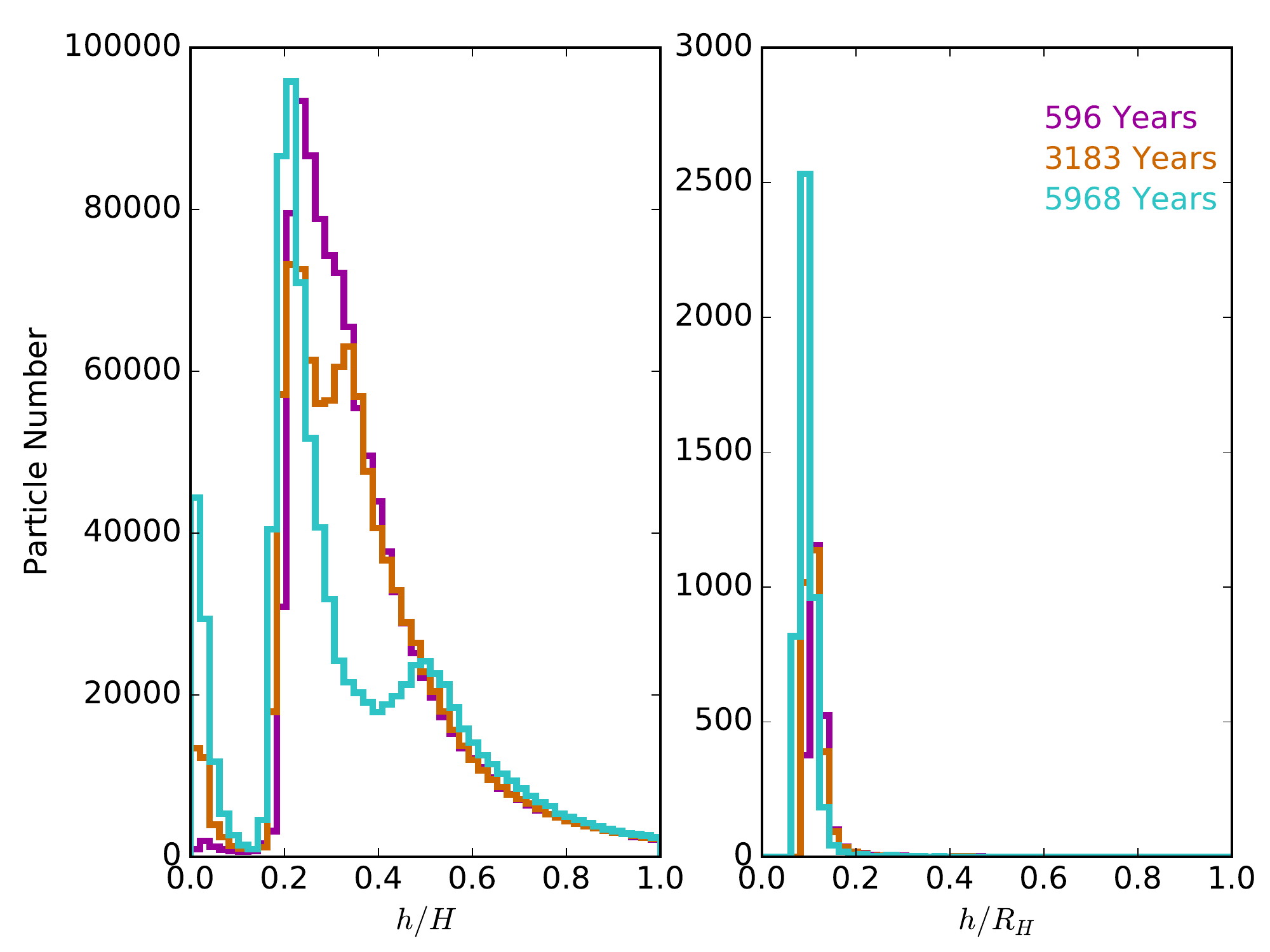}
\includegraphics[width=0.99\columnwidth]{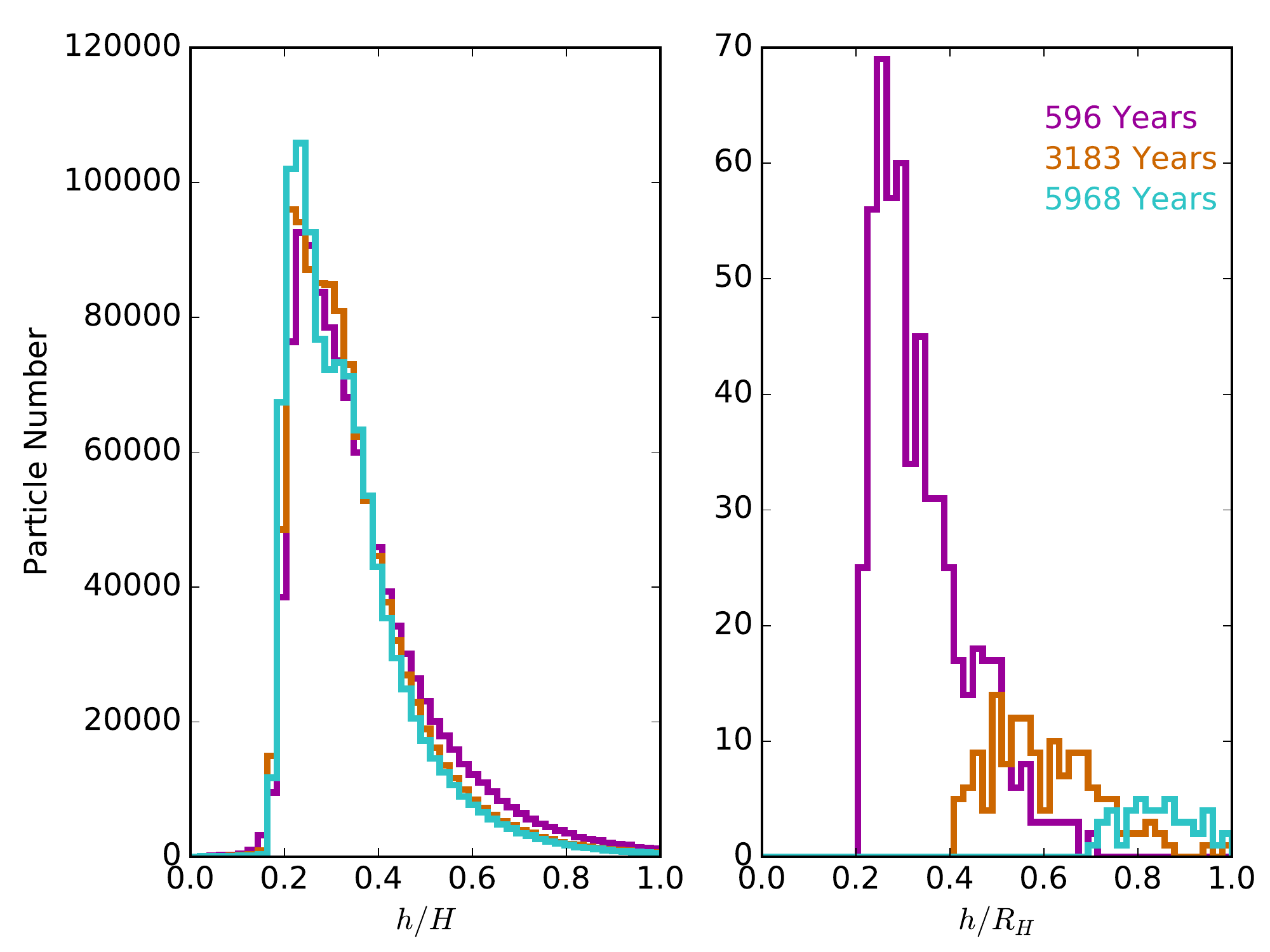}
\caption{Plots to show the ratio SPH smoothing length ($h$) against disc scale height ($H$) as well as Hill sphere radius ($R_H$). Top: $\beta=0.1$, n=$2\times10^6$, bottom:  $\beta=10$, n=$2\times10^6$.}
\label{fig:SPH_res}
\end{figure}

SPH simulations become dominated by particle noise at low resolutions. This is especially important in regimes where accretion onto sink particles is depleting the total number of SPH particles over the lifetime of the simulation. Figure \ref{fig:SPH_res} plots the histograms of particle smoothing lengths as ratios against disc scale height in the global disc and as a ratio against the Hill radius for particles inside the Hill sphere of the planet. 

For both values of $\beta$ the wider disc remains well resolved for the duration of the simulation with around three SPH smoothing lengths per scale height. This supports the particles number studies that we completed in Section \ref{sec:N_test}. Since the planetary migration rates matched Type I very well (see Figure \ref{fig:planet_mass0}) we expected the global disc resolution to remain reasonably high.

In the top panels is can be seen that for the $\beta=0.1$ case a considerable number of SPH particles are used to describe the compact circumplanetary accretion disc. This disc is resolved to about a tenth of the Hill radius, which is a similar scale to the fine structured accretion streams seen in Figure \ref{fig:SPH_t1_zoom_over}. This suggests that the specific structure may be defined by the resolution limit but grains in this regime would still remain supported from accretion by the high gas densities and their corresponding short stopping times. A detailed understanding of the internal structure would require a higher resolution study. 

For $\beta=10$, however, the resolution inside the Hill sphere is much lower. This represents the SPH particle density decreasing due to heating caused by the inefficient cooling rate in the disc.
Errors due to the low resolution may explain the visible difference that particle number has on the gas accretion rate in Figure \ref{sec:N_test}. As discussed in that section though, the relative gas accretion rate is small compared to the total planet mass so for the purpose of this study the error is acceptable. It is reassuring that in Figure \ref{sec:N_test} the dust accretion rate remains insensitive to the particle number. This suggests that the SPH gas resolution inside the Hill sphere in the inefficient cooling case is high enough to accurately simulate the drag on dust particles accreting onto these planets.

Whilst having a higher resolution inside the Hill sphere of the planet is desirable, we have discussed in the paper that once dust particles have entered the Hill sphere they are efficiently accreted at close to the Hill rate. 
Pebbles in the range $\tau \sim [0.1-100]$ are sufficiently decoupled from the gas disc that inside the Hill sphere they will generally remain gravitationally bound to the planet regardless of the precise gas dynamics. Therefore we feel that our simulations are successful in capturing the bulk change in metallicity for these migrating planets due to the accretion of large grains.

\subsection{Dust velocity dispersion}
\label{app:v_disp}

\begin{figure}
\includegraphics[width=1.02\columnwidth]{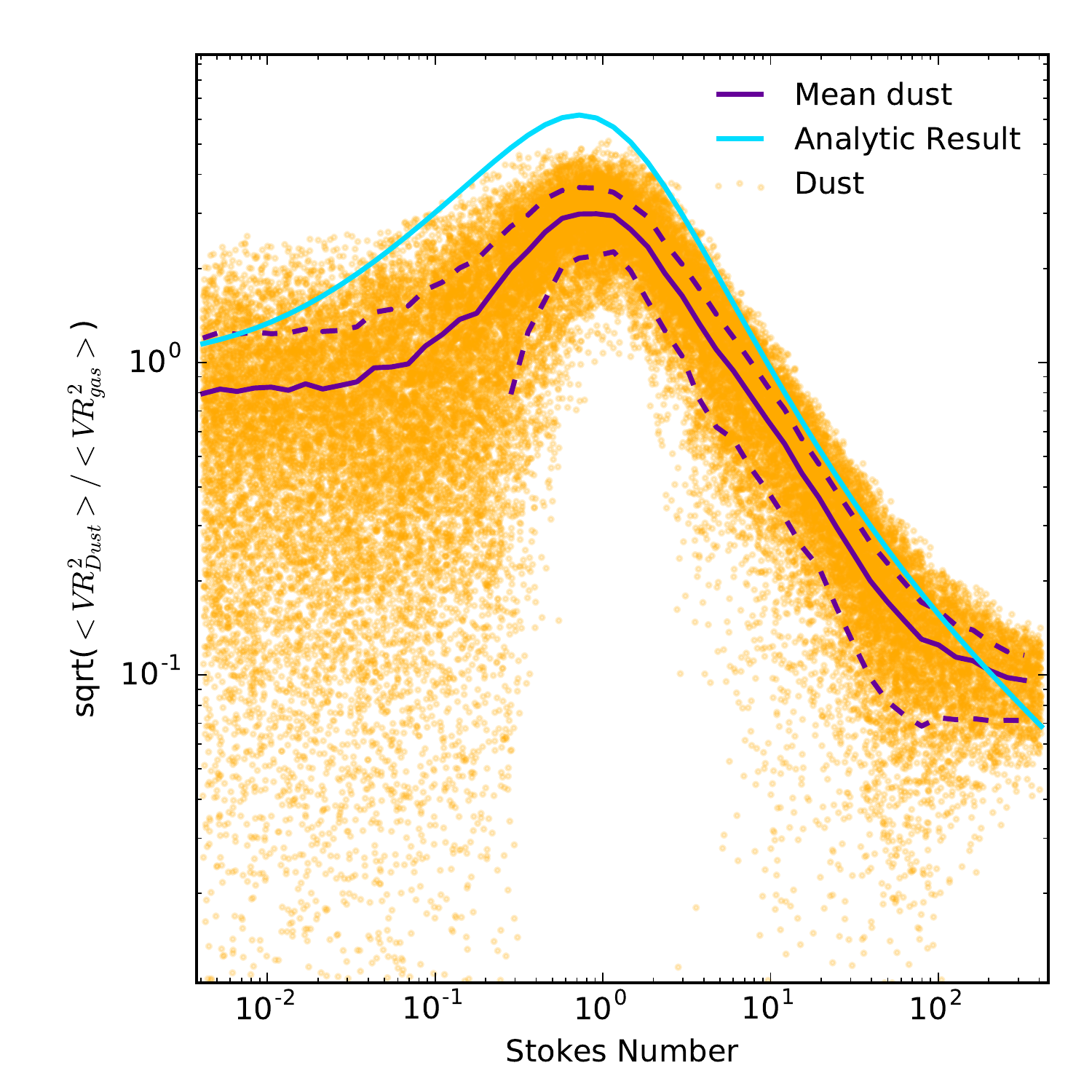}
\caption{The orange scatter show the RMS dust radial velocities as a fraction of the local RMS gas radial velocity. The purple solid/dashed purple curves show the mean/deviation for this data. The blue line is the analytic formula (Equation 33a in \protect\cite{YoudinLithwick07}). We see an order of magnitude agreement between the two results suggesting that velocity dispersion in the SPH field is indeed driving dispersion in the dust velocity.}
\label{fig:vr_disp}
\end{figure}

\begin{figure}
\includegraphics[width=1.02\columnwidth]{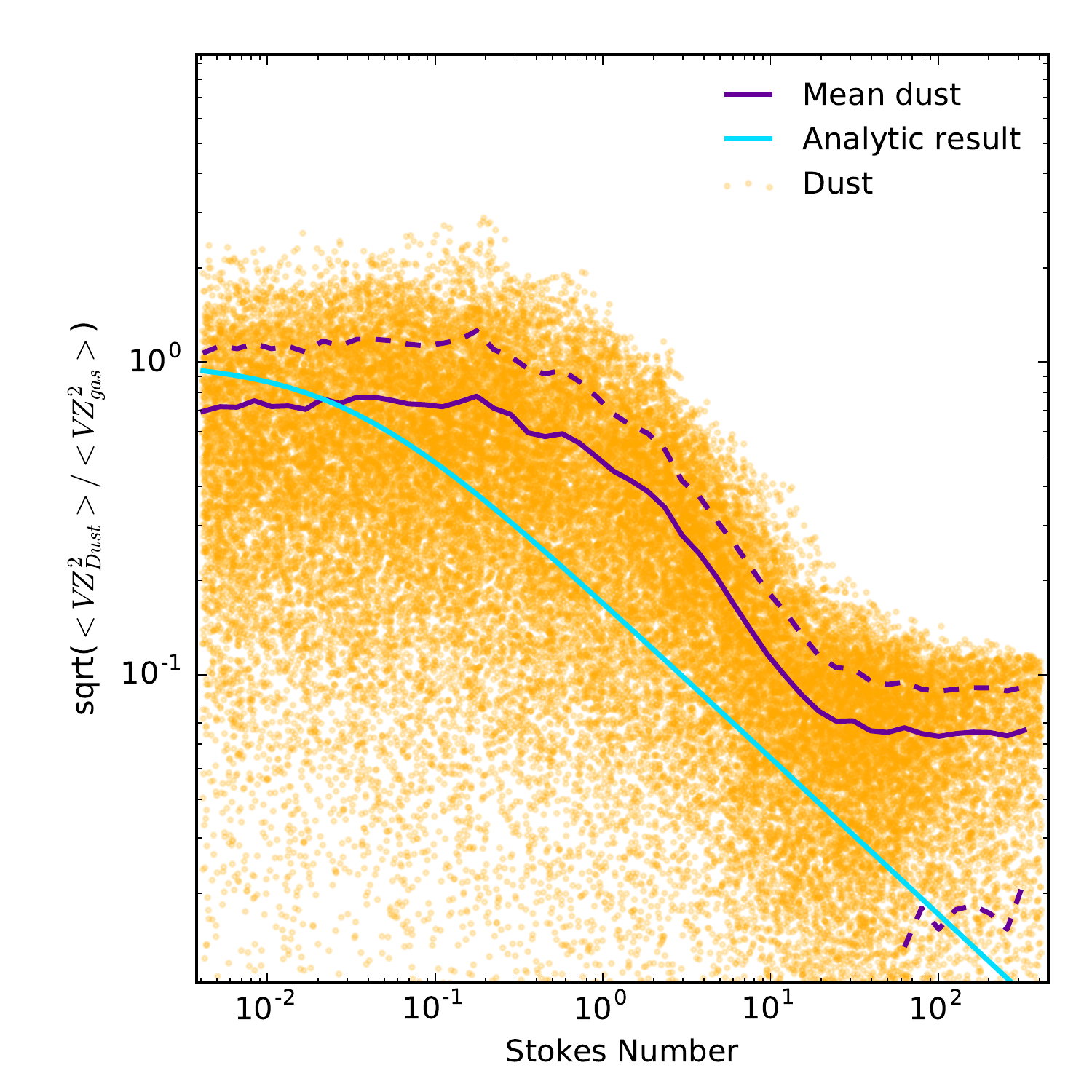}
\caption{The orange scatter show the RMS dust vertical velocities as a fraction of the local RMS gas vertical velocity. The purple solid/dashed purple curves show the mean/deviation for this data. The blue line is the analytic formula (Equation 20 in \protect\cite{YoudinLithwick07}).}
\label{fig:vz_disp}
\end{figure}

In order to explore the reason for dust dispersion in Figure \ref{fig:vrad} we have consulted \cite{YoudinLithwick07} who present analytic predictions for the dispersion velocities of dust particles suspended in gas. The general idea is to treat the SPH gas distribution as a turbulent field that stirs up the dust. They make the assumption that the viscous transport $\alpha$ is a good approximation for the $\alpha$ that governs the size of turbulent cells. A physical value for $\alpha$ can be found from the numerical value of $\alpha_{AV}$ in \textsc{gadget-3} using the method detailed in \cite{LodatoPrice10},

\begin{equation}
\alpha =  \alpha_{AV} \dfrac{1}{10}\dfrac{\langle h \rangle}{2H}
\end{equation}

where $\langle h \rangle$ is the mean smoothing length in the disc and $H$ is the disc scale height. We introduce the factor of 2 to acknowledge that the smoothing length is defined differently in \textsc{gadget-3} to the majority of SPH codes. Having made these assumptions, the diffusion constant $D$ for dust can be recast in terms of disc viscosity ($\nu$) as

\begin{equation}
D \sim \nu = \dfrac{\alpha c_s^2}{\Omega_K} = \langle v_G^2 \rangle t_{\rm eddy}
\end{equation}

where $\langle v_G^2 \rangle$ is the kernel weighted mean square gas velocity measured over nearest neighbours in the SPH field and $t_{\rm eddy}$ is some characteristic timescale for turbulent eddies. In the simplest case this expression can then be used to calculate the ratio of velocity dispersion in dust ($\langle v_D^2 \rangle$) compared to gas in the form

\begin{equation}
\dfrac{\langle v_D^2 \rangle}{ \langle v_G^2 \rangle} = \dfrac{1}{1 + \dfrac{\tau}{\tau_{\rm eddy}}} = \dfrac{1}{1 + \dfrac{\tau \Omega_K \langle v_G^2 \rangle}{\alpha c_s^2}}
\end{equation}

where $\tau_{\rm eddy}$ = $t_{\rm eddy} \Omega_K$ and $\tau$ is the Stokes number of the dust. This idea is extended by \cite{YoudinLithwick07} who perform a more in depth analysis including differential sheer and the different regimes of coupling between gas and dust particles. We compare against their Equations 33a and 20 in Figures \ref{fig:vr_disp} and \ref{fig:vz_disp}. These Figures are based on the simulation test for the Weidenschilling radial velocity drift presented in Figure \ref{fig:vrad}.

The intrinsic velocity dispersion in the SPH field generates the scatter in the dust field that appears in Figures \ref{fig:vrad} and \ref{fig:dust_settle}, especially for low values of $\tau$. From Figure \ref{fig:vr_disp} we see that the radial velocity dispersion in our simulations agrees reasonably well with the analytic treatment of \cite{YoudinLithwick07}. This demonstrates that it is indeed the intrinsic noise of the SPH particles that is driving this velocity dispersion. 

From Figure \ref{fig:vz_disp} we see that the simulation vertical velocity dispersion matches the analytic prediction to the correct order of magnitude, but does not decrease as rapidly with particle Stokes number. The simulation dispersion is therefore higher than the analytic prediction for larger grain sizes. We are not overly concerned with this result since the run from which the figure is derived was not designed to test this analytical prediction. We started all the dust particles at 0.01 of the gas particle positions. It is possible that the deviations are caused by the pebble vertical distribution not having reached equilibrium for larger Stokes number particles. The deviations in the vertical velocity dispersion of dust particles are unlikely to affect the main conclusions of our paper. We experimented with different initial geometrical thickness of the dust layer and found that dust accretion rates onto the planets were very similar.

\subsection{Additional dust settling tests}
\begin{figure}
\includegraphics[width=0.49\columnwidth]{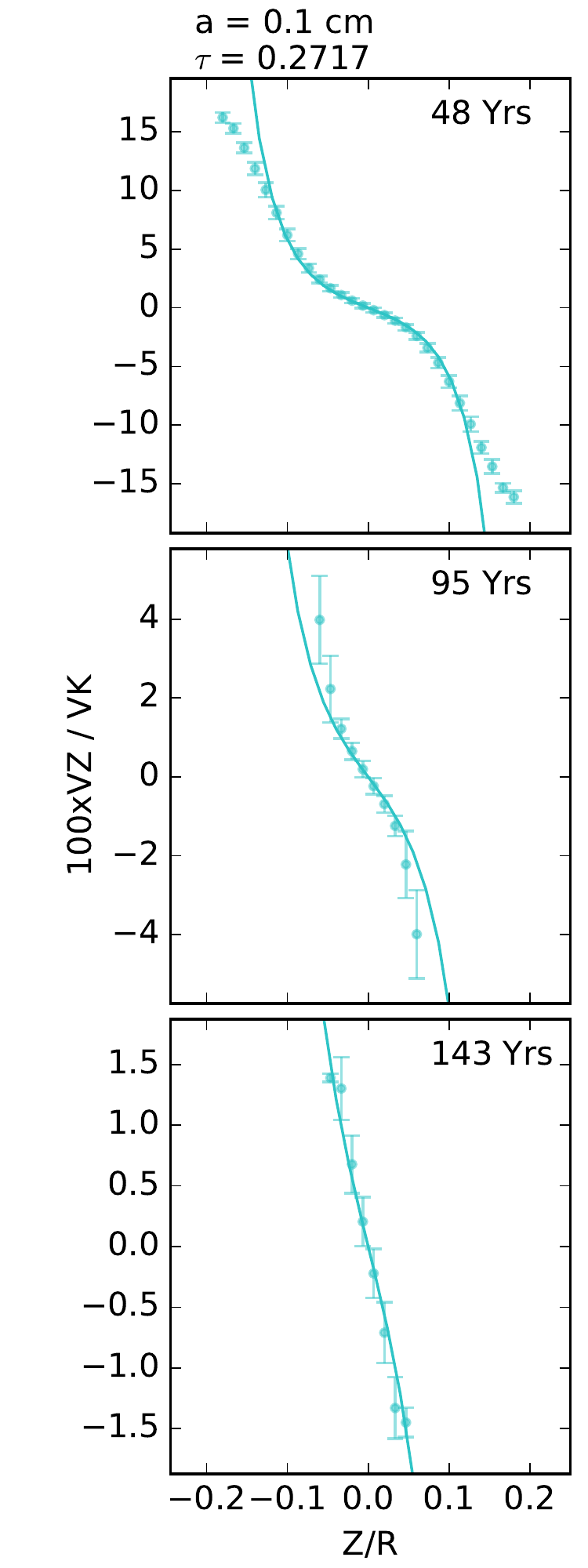}
\includegraphics[width=0.49\columnwidth]{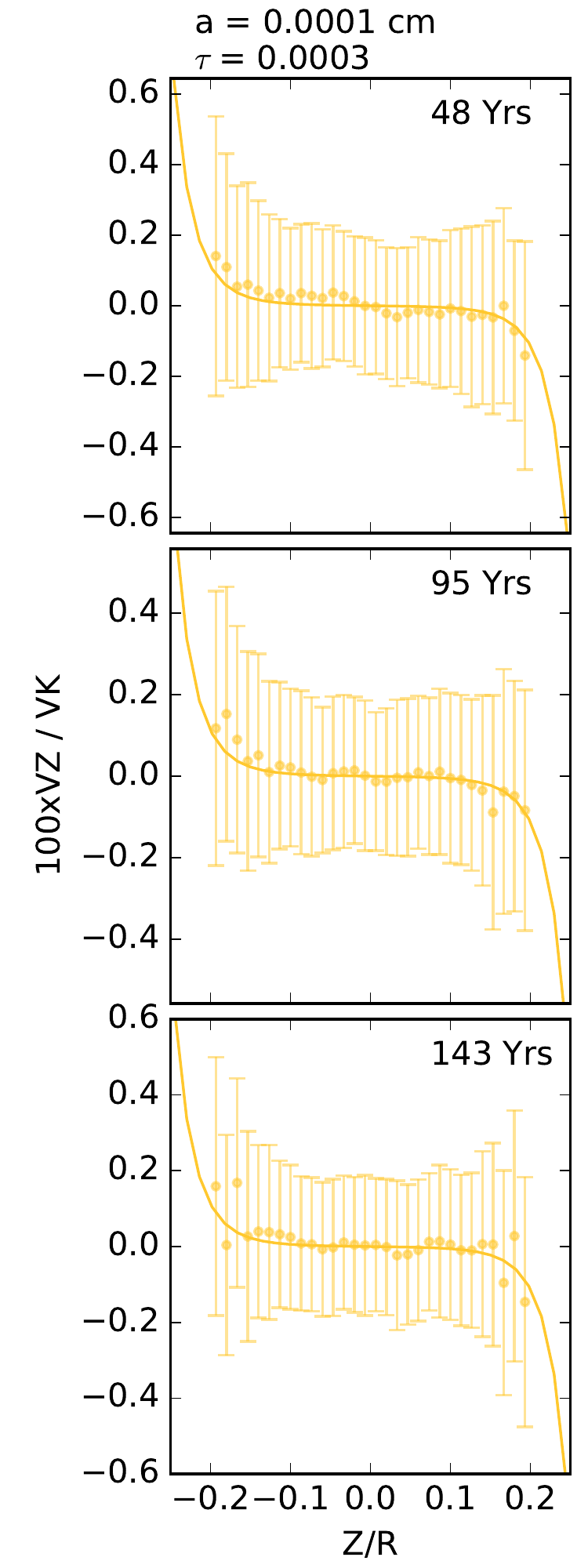}
\caption{Dust settling tests for $a$=0.1cm \& 0.0001cm in a 0.01M$_\odot$ gas disc at 50 AU. The solid line marks the analytic settling solution. Filled circles and error bars represent binned particle locations and the 1$\sigma$ spread within each bin. Notice that the $a$=0.1cm grains have a high Stokes number and so overshoot the expected settling solution.}
\label{fig:dust_settleAP}
\end{figure}
Further to Section \ref{sec:dust_settle}, in Figure \ref{fig:dust_settleAP} we present settling tests for grain sizes $a$=0.1cm \& 0.0001cm. 

Notice from the left hand plots that the $a$=0.1cm grains overshoot the analytic settling velocity. They settle very rapidly and oscillate around the disc midplane as described for the $K_S^E / \hat{m}_D$ = 1 ($\tau$ = 0.6) grains with low coupling to the gas in \cite{LAandBate14}. From the bottom left panel it can be seen that after this initial oscillation phase is damped these grains match very tightly to the settling solution.
In the right hand plots the grains have a very low Stokes number ($\tau$=0.0003) and so are very tightly coupled to the gas. Whilst the large spread reflects the very large coupling to the SPH velocity noise field (See Appendix \ref{app:v_disp}), the mean settling velocity is again in very good agreement with the analytic expectation.

\subsection{Validity of the Epstein regime}
\label{app:eps}
\begin{figure}
\includegraphics[width=0.99\columnwidth]{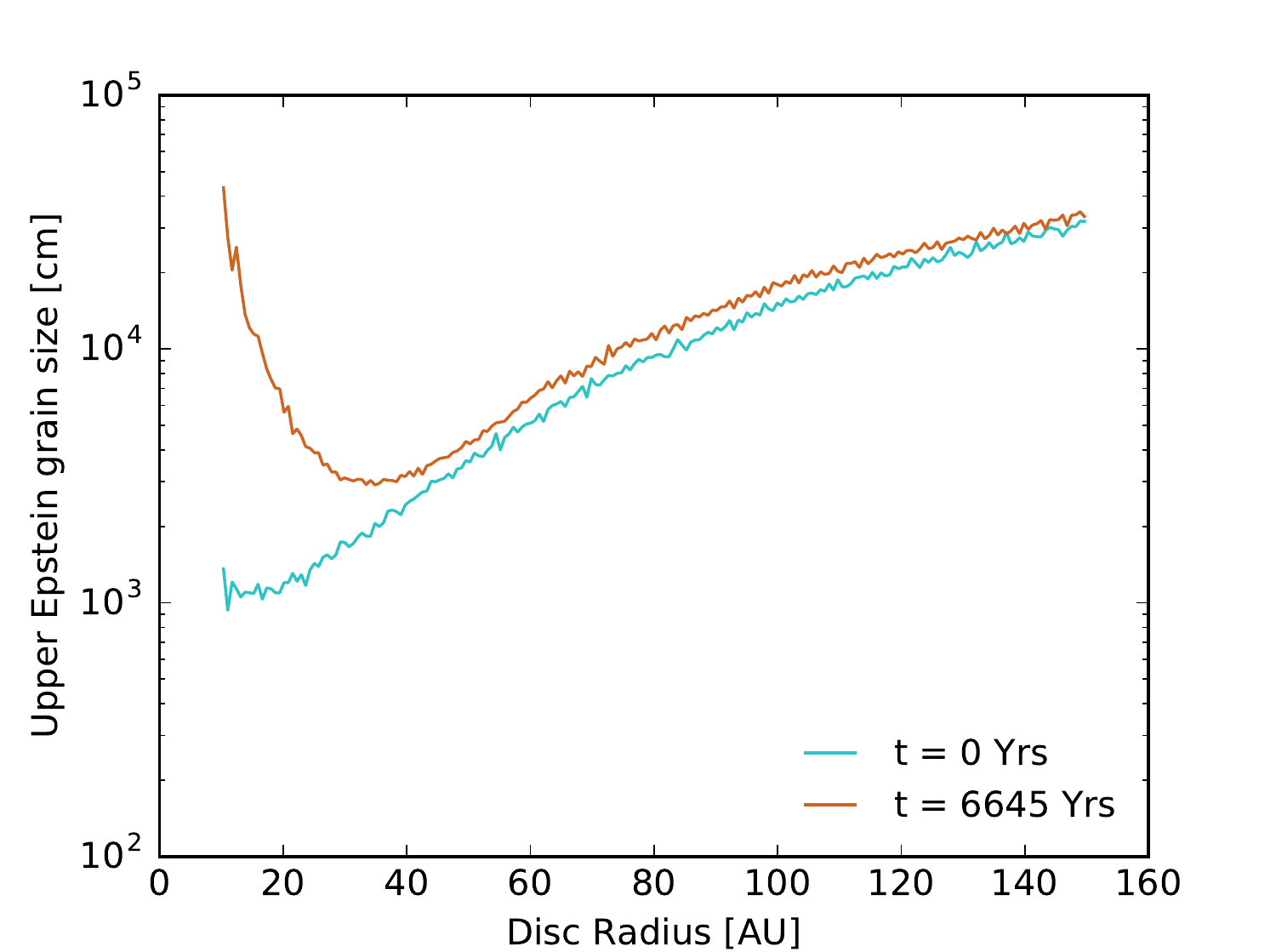}
\caption{Plots for the validity regime for Epstein drag based on Equation \ref{eq:eps}, grain sizes below the lines are well described in the Epstein regime.}
\label{fig:eps}
\end{figure}

In Figure \ref{fig:eps} we plot Equation \ref{eq:eps} that describes the regime in which the Epstein drag law applies. We find that the grains in our simulations fall within the Epstein regime by more than two orders of magnitude across the disc.
The turn over at $\sim$ 40 AU over the course of the simulation reflects the artificial depletion of surface density in the inner disc due to unphysical accretion at the central boundary.

\subsection{Conservation of angular momentum}
\begin{figure}
\includegraphics[width=0.99\columnwidth]{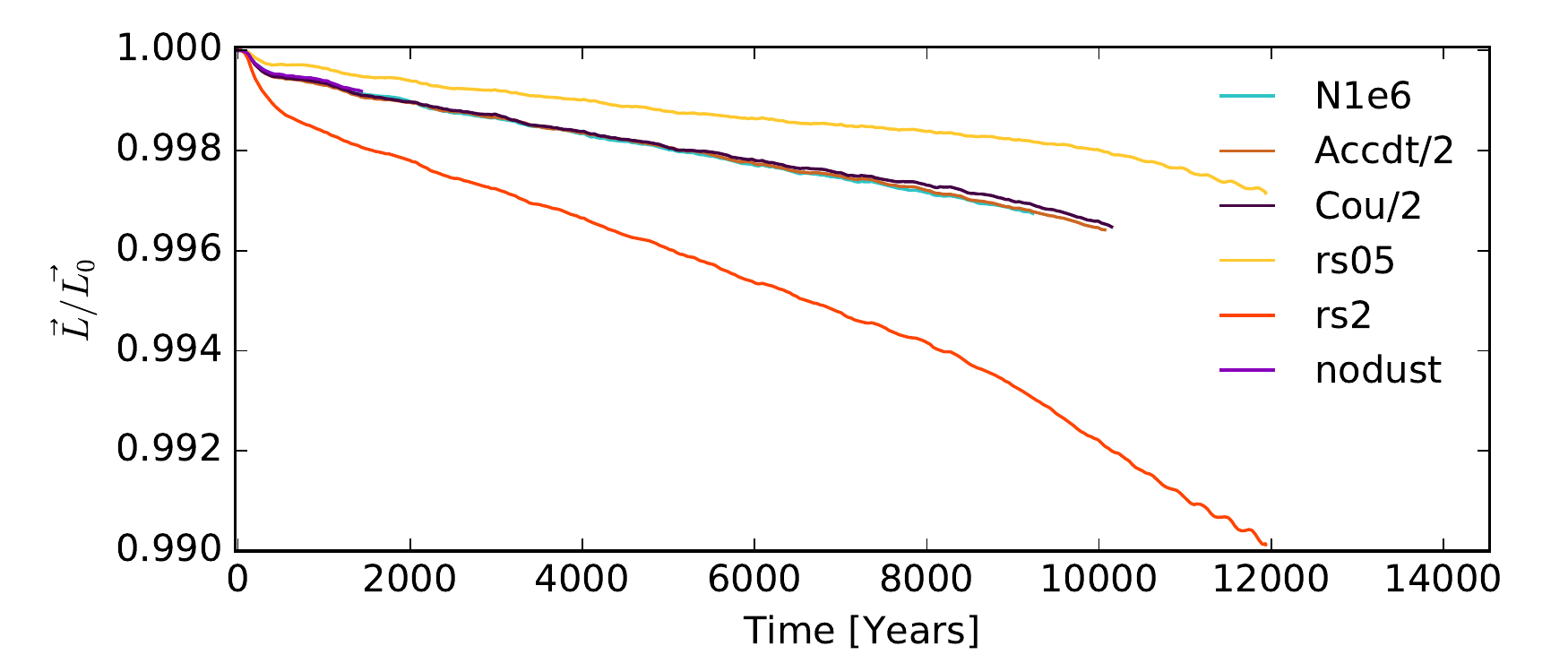}
\caption{Plot of total angular momentum as a fraction of the initial total angular momentum for $N=10^6$, $\beta=10$ run. In this suite of tests the simulations typically lose 0.3\% angular momentum over 10,000 years. Note that only changes in the planet sink radii causes an appreciable change in angular momentum conservation. This suggests that the loss of angular momentum is not due to the implicit dust scheme.}
\label{fig:L_cons}
\end{figure}

In Figure \ref{fig:L_cons} we plot the total angular momentum of the system (star, planet, gas and dust) as a fraction of the initial total angular momentum for the $N=10^6$, $\beta=10$ set up. The total angular momentum typically drops by 0.3\% over 10,000 years which is a very acceptable tolerance given the other uncertainties in the paper. Varying the SPH Courant parameter (Cou/2), the minimum timestep at the stellar sink radius (Accdt/2) and running the simulation without dust (nodust) have little to no impact on the change of total angular momentum. 

The main result from Figure \ref{fig:L_cons} is that angular momentum loss is a strong function of planetary sink radius.
In this version of \textsc{gadget-3} data about the spin of sink particles is not recorded. This means that during accretion linear momentum is conserved but information about angular momentum is lost. 
This can be seen from Figure \ref{fig:L_cons} in which altering the sink radius of the particle dramatically changes the accuracy of angular momentum conservation during the run. These results demonstrate that the loss of angular momentum over the simulation is not caused by our implicit dust scheme. 
In future work we will aim to record the angular momentum information that is lost during accretion in order to better understand the angular momentum conservation of the system. 

\bibliographystyle{mnras}
\bibliography{humphries} 



\bsp	
\label{lastpage}
\end{document}